%% file: main-edsm.tex
\documentclass[times,preprint]{elsarticle}
\input{preamble.tex}
\usepackage{color,soul}

\begin{document}

\begin{frontmatter}
	\title{Deep encoder-decoder hierarchical convolutional neural networks for conjugate heat transfer surrogate modeling}%

	\author[1]{Takiah Ebbs-Picken}
	\author[1]{David A. Romero}
	\author[1]{Carlos M. Da Silva}
	\author[1,2]{Cristina H. Amon\texorpdfstring{\corref{cor1}}{}}
    \cortext[cor1]{Corresponding author.}
	\ead{cristina.amon@utoronto.ca}
	
	\address[1]{Department of Mechanical and Industrial Engineering, ATOMS Laboratory, University of Toronto, 5 King's College Road, Toronto, ON M5S 3G8, Canada}
	\address[2]{Chemical Engineering and Applied Chemistry, University of Toronto, 5 King`s College Road, Toronto, ON M5S 3G8, Canada}
	
	
	\begin{abstract}
	\sloppy
	Conjugate heat transfer (CHT) analyses are vital for the design of many energy systems.
	However, high-fidelity CHT numerical simulations are computationally intensive, which limits their applications such as design optimization, where hundreds to thousands of evaluations are required.
	In this work, we develop a modular deep encoder-decoder hierarchical (DeepEDH) convolutional neural network, a novel deep-learning-based surrogate modeling methodology for computationally intensive CHT analyses. 
	Leveraging convective temperature dependencies, we propose a two-stage temperature prediction architecture that couples velocity and temperature fields. 
	The proposed DeepEDH methodology is demonstrated by modeling the pressure, velocity, and temperature fields for a liquid-cooled cold-plate-based battery thermal management system with variable channel geometry.
	A computational mesh and CHT formulation of the cold plate is created and solved using the finite element method (FEM), generating a dataset of 1,500 simulations.
	The FEM results are transformed and scaled from unstructured to structured, image-like meshes to create training and test datasets for DeepEDH models.
	The DeepEDH architecture's performance is examined in relation to data scaling, training dataset size, and network depth. 
	Our performance analysis covers the impact of the novel architecture, separate DeepEDH models for each field, output geometry masks, multi-stage temperature field predictions, and optimizations of the hyperparameters and architecture. 
	Furthermore, we quantify the influence of the CHT analysis' thermal boundary conditions on surrogate model performance, highlighting improved temperature model performance with higher heat fluxes. 
	Compared to other deep learning neural network surrogate models, such as U-Net and DenseED, the proposed DeepEDH architecture for CHT analyses exhibits up to a 65\% enhancement in the coefficient of determination $R^{2}$.
	\end{abstract}

	\begin{keyword}
		Surrogate modeling \sep machine learning \sep deep-learning \sep deep encoder-decoder hierarchical (DeepEDH) convolutional neural networks \sep conjugate heat transfer \sep battery thermal management
	\end{keyword}
\end{frontmatter}

\section*{Highlights}
\addcontentsline{toc}{section}{Highlights}%
\begin{itemize}
	\item Modular deep encoder-decoder convolutional neural networks with skip connections for conjugate heat transfer surrogate modeling.
	\item Hierarchical model architecture to couple velocity and temperature fields.
	\item Output geometry masks for improved flow model performance.
	\item Illustration of the surrogate modeling methodology considering a liquid-cooled cold-plate-based battery thermal management system. 
	\item Comprehensive surrogate model performance characterization.
\end{itemize}

\bigskip

\addcontentsline{toc}{section}{\protect\numberline{Nomenclature}}
\input{nomenclature-edsm.tex}

\begin{scriptsize}
\begin{multicols}{2}
	\printnomenclature
\end{multicols}
\end{scriptsize}

\bigskip

\clearpage
\tableofcontents 

\clearpage
\vfill

\input{main_body_short.tex}

\input{backmatter.tex}

\addcontentsline{toc}{section}{References}
\begin{singlespace}
	\bibliographystyle{ieeetr}
	\bibliography{references-edsm}
\end{singlespace}

\end{document}

%% file: preamble.tex
\usepackage{framed}

\usepackage{amsmath} 
\usepackage{amssymb} 
\usepackage{latexsym}
\usepackage{amsthm} 
\usepackage{mathtools} 
\usepackage{bm} 
\usepackage{mathrsfs} 
\usepackage{calrsfs}  

\usepackage{graphicx} 
\usepackage{float} 
\usepackage{adjustbox} 
\usepackage{array} 
\usepackage[subrefformat=parens,labelformat=parens]{subcaption} 
\usepackage{booktabs, multicol, multirow}
\usepackage{hhline} 
\usepackage{caption} 

\usepackage{setspace} 
\usepackage{parskip} 
\usepackage{lscape} 
\usepackage{titlesec} 
\usepackage{paralist} 
\usepackage{color} 
\usepackage[colorlinks,linktocpage=true]{hyperref} 
\usepackage[capitalise, noabbrev, nameinlink]{cleveref}
\usepackage{balance}                
\usepackage[margin=2.5cm]{geometry}

\usepackage{calc}

\usepackage{enumerate} 
\usepackage{listings} 
\usepackage[toc,page]{appendix} 
\usepackage{siunitx} 

\DeclarePairedDelimiterX{\norm}[1]{\lVert}{\rVert}{#1}
\DeclareMathAlphabet{\pazocal}{OMS}{zplm}{m}{n}  

\usepackage{nomencl}
\makenomenclature

\usepackage{etoolbox}
\renewcommand\nomgroup[1]{%
	\item[\bfseries
	\ifstrequal{#1}{S}{Symbols}{%
	\ifstrequal{#1}{A}{Acronyms}{%
	\ifstrequal{#1}{C}{Other Symbols}{}}}%
]}

\makeatletter
\DeclareRobustCommand{\volume}{\text{\volumedash}V}
\newcommand{\volumedash}{%
  \makebox[0pt][l]{%
    \ooalign{\hfil\hphantom{$\m@th V$}\hfil\cr\kern0.08em--\hfil\cr}%
  }%
}
\makeatother

\newlength\mylenA
\newlength\mylenB
\newlength\mylenC

\DeclareSIUnit\pixel{px}

%% file: nomenclature-edsm.tex
%
\nomenclature[S]{$A$}{area}
\nomenclature[S]{$C_{p}$}{specific heat capacity}
\nomenclature[S]{$d_{bc}$}{boundary condition thickness}
%
\nomenclature[S]{$\bm{F}$}{body force}
\nomenclature[S]{$\Delta H$}{enthalpy difference}
\nomenclature[S]{$h$}{convective heat transfer coefficient}
\nomenclature[S]{$\bm{I}$}{identity tensor}
\nomenclature[S]{$I$}{current}
\nomenclature[S]{$\bm{K}$}{stress tensor}
\nomenclature[S]{$k$}{thermal conductivity}
\nomenclature[S]{$\dot{m}$}{mass flow rate}
\nomenclature[S]{$\bm{n}$}{unit normal vector}
\nomenclature[S]{$p$}{pressure}
\nomenclature[S]{$p_{0}$}{outlet pressure}
\nomenclature[S]{$P_{0}$}{power source}
\nomenclature[S]{$\rho$}{density}
\nomenclature[S]{$Q$}{heat source}
\nomenclature[S]{$\bm{q}$}{heat flux}
\nomenclature[S]{$R^{2}$}{coefficient of determination}
\nomenclature[S]{$T$}{temperature}
\nomenclature[S]{$T_{ext}$}{external temperature}
\nomenclature[S]{$t$}{time}
\nomenclature[S]{$\bm{u}$}{velocity}
\nomenclature[S]{$\mu$}{dynamic viscosity}
\nomenclature[S]{$\volume$}{volume}
\nomenclature[S]{$V$}{voltage}

%
%
\nomenclature[A]{ANN}{artificial neural network}
\nomenclature[A]{CFD}{computational fluid dynamics}
\nomenclature[A]{CHT}{conjugate heat transfer}
\nomenclature[A]{CNN}{convolutional neural network}
\nomenclature[A]{DeepEDH}{deep encoder-decoder hierarchical}
\nomenclature[A]{FCN}{fully convolutional network}
\nomenclature[A]{FEM}{finite element method}
\nomenclature[A]{FPN}{feature pyramid network}
\nomenclature[A]{GAN}{generative adversarial network}
%
\nomenclature[A]{LSTM}{long short-term memory}
\nomenclature[A]{MSE}{mean squared error}
%
\nomenclature[A]{PDE}{partial differential equation}
%
\nomenclature[A]{RMSE}{root mean square error}
\nomenclature[A]{SCC}{Spearman's rank correlation coefficient}
\nomenclature[A]{SOC}{state of charge}
%

%% file: main_body_short.tex
\section{Introduction}
\label{sec:intro}
Heat transfer analyses are vital for designing many energy systems, with applications ranging from batteries and fuel cells to electric vehicles and power generation.
For many of these systems, temperature directly impacts their performance, requiring effective thermal management.
Designing and optimizing thermal management systems requires accurately predicting their heat transfer behavior.
Many of these applications use convection-based thermal management systems and require the consideration of heat transfer in both solid and fluid domains simultaneously;
such analysis is known as conjugate heat transfer (CHT).
Different governing partial differential equations (PDEs), the momentum, continuity, and energy equations, exist for the solid and fluid domains and are coupled at the solid-fluid interface.
Solving these equations in each domain requires a dynamic update of the interface boundary condition. 
Analytical solutions for CHT problems are limited and often rely on simplifications, limiting their practical applicability \cite{John_Senthilkumar_Sadasivan_2019}.
Instead, numerical approaches are commonly used to solve CHT problems \cite{John_Senthilkumar_Sadasivan_2019}, but these approaches can be computationally expensive, especially for high-fidelity three-dimensional  solutions.
Directly using high-fidelity models is unfeasible in many practical applications due to time and computational constraints.
For example, design optimization typically requires thousands of model evaluations, whereas control and digital twin applications often have limited computational resources \cite{EbbsPicken_Silva_Amon_2023,Hachem_Ghraieb_Viquerat_Larcher_Meliga_2021,Naseri_Gil_Barbu_Cetkin_Yarimca_Jensen_Larsen_Gomes_2023}.
To address these challenges, reduced-order surrogate models can approximate the behavior of the system at a decreased computational cost, enabling their use in applications like design optimization \cite{EbbsPicken_Silva_Amon_2023}, control \cite{Hachem_Ghraieb_Viquerat_Larcher_Meliga_2021}, and digital twin systems \cite{Naseri_Gil_Barbu_Cetkin_Yarimca_Jensen_Larsen_Gomes_2023}, where high-fidelity models are impractical.

Data-driven surrogate modeling methodologies have been proven effective in approximating the behavior of complex engineering systems. 
Surrogate modeling requires a regression task where the input needs to be mapped non-linearly to the output.
These approaches rely on input-output data observations to construct an approximation for the mapping or relationship of interest.
While some works incorporate physics-based constraints into the surrogate model \cite{Raissi_Perdikaris_Karniadakis_2019,Meng_Karniadakis_2020,Jin_Cai_Li_Karniadakis_2021,Natale_Svetozarevic_Heer_Jones_2022}, most data-driven approaches are purely data-based, requiring no prior knowledge of the system.
Common approaches include polynomial regression, symbolic regression, Kriging, radial basis functions, and artificial neural networks (ANNs).
Many studies have applied these approaches to predict specific values, such as extreme temperatures, velocities, or pressures, without reconstructing entire physical fields \cite{EbbsPicken_Silva_Amon_2023}.
For instance, polynomial response surface and Kriging methods have been used for shape optimization of turbine blades from temperature and pressure predictions \cite{Yu_Yang_Yue_2011}, boiler superheaters have been optimized using polynomial basis functions to predict heat transfer rates \cite{Maakala_Jarvinen_Vuorinen_2018}, and symbolic regressions have been applied in building energy demand response optimization \cite{Ren_Gao_Ma_Zhang_Sun_2024}.
Artificial neural networks have been used for various predictions, including closure coefficients for turbulent heat flux models in forced convection flows~\cite{Fiore_Koloszar_Mendez_Duponcheel_Bartosiewicz_2022}, internal fridge temperatures \cite{Antonio_Afonso_2011}, maximum chip temperatures~\cite{Ozsunar_Arcaklioglu_Dur_2009}, and electronic component temperatures and flow velocities in nanofluid filled enclosures~\cite{Kargar_Ghasemi_Aminossadati_2011, Ben-Nakhi_Mahmoud_Mahmoud_2008, Varol_Avci_Koca_Oztop_2007, Mahmoud_Ben-Nakhi_2007}.
While these traditional data-driven methodologies are effective for specific value predictions, they are typically incapable of reconstructing full physical fields and handling high-dimensional or permutation-invariant inputs without significant computational overhead, limiting their application to low-dimensional regression problems.

Alternatively, deep neural network data-based surrogate modeling approaches have been effective for applications involving regression of complex and non-linear behaviors.
These approaches are characterized by their numerous layers, making them distinct from shallow ANNs and other data-based methods.
Comparatively, deep neural networks can accurately reconstruct full physical field results, yielding similar accuracy for specific values derived from these fields \cite{Wang_Zhou_Yang_Huang_2022}. 
Common deep neural network architectures applied for surrogate modeling include recurrent neural networks \mbox{\cite{Jiang_Durlofsky_2023}}, long short-term memory (LSTM) neural networks \mbox{\cite{ZHANG2020115552}}, generative adversarial networks (GANs) \mbox{\cite{Wang_Zhou_Yang_Huang_2022,ZHANG2022121747}}, and convolutional neural networks (CNNs). 
Recurrent and LSTM neural networks are designed for sequential or time series predictions and typically require larger models and more training time than CNNs \mbox{\cite{Bai_Kolter_Koltun_2018,tc-16-1447-2022}}. 
Further, CNNs can offer similar or improved accuracy when the fields to be predicted show strong spatial dependence \mbox{\cite{Bai_Kolter_Koltun_2018,tc-16-1447-2022,DEHGHANI2023102119}}, such as for CHT-governed systems.
Generative adversarial networks are effective when limited training data is available; however, they require more complex models and longer training times \mbox{\cite{Wang_Zhou_Yang_Huang_2022}} than CNNs.
Hence, most research for surrogate modeling of engineering systems has focused on CNNs.

Initially designed for image classification, CNNs have proven effective for high-dimensional surrogate modeling.
Levering a combination of linear convolution operations and non-linear activations, CNNs extract features from inputs to predict relevant outputs.
Convolutional neural networks have proven effective for surrogate modeling of engineering systems in various applications. 
For example, they have been used to reconstruct cooling effectiveness fields for transpiration cooling~\cite{Yang_Min_Yue_Rao_Chyu_2019}, predict velocity fields around a cylinder~\cite{Jin_Cheng_Chen_Li_2018}, predict battery lifetime~\cite{He_Wang_Lu_Chai_Yang_2024}, and estimate electric vehicle spatial distributions for electricity grid planning~\cite{Tikka_Haapaniemi_Raisanen_Honkapuro_2022}.
Extending CNNs, encoder-decoder architectures use a sequence of convolutional layers to reduce the input size to a low-dimensional latent space, or code dimension, followed by a series of deconvolutional layers to reconstruct the output.
Compared to traditional CNNs, encoder-decoder architectures extract multi-scale features from the input more effectively.
Initially developed for image reconstruction, these architectures have recently been applied as surrogate models for various engineering systems.
For example, encoder-decoder CNN models have been used to predict flow fields around different objects~\cite{Guo_Li_Iorio_2016} and airfoils~\cite{Bhatnagar_Afshar_Pan_Duraisamy_Kaushik_2019}, shale gas production~\cite{Zhou_Li_Qi_Zhao_Yi_2024}, as well as CHT temperature and velocity fields \cite{Peng_Liu_Xia_Aubry_Chen_Wu_2021}.

To enhance information propagation within encoder-decoder networks, especially in cases with a direct input-output dependency, the U-Net architecture \cite{Ronneberger_Fischer_Brox_2015} introduced skip connections to link the encoder and decoder sections.
U-Net architectures have since been used to reconstruct temperature and velocity fields in various applications, such as nanofluid-filled finned absorber tubes for natural and forced convection~\cite{Hua_Yu_Peng_Wu_He_Zhou_2022} and film cooling in rocket combustors~\cite{Ma_Zhang_Haidn_Thuerey_Hu_2020}.
Dense architectures, which introduce shortcut connections between layers, aim to further improve information propagation and reduce the number of parameters in CNNs.
Examples include ResNet \cite{He_Zhang_Ren_Sun_2016}, Highway Networks \cite{NIPS2015_215a71a1}, and DenseNet \cite{Huang_Liu_Maaten_Weinberger_2016}.
These architectures enable more efficient training of deeper networks with fewer parameters and have been proven effective in image classification and segmentation tasks.
In the context of surrogate modeling, these networks have been applied to various engineering systems, including nanofluid-filled finned absorber tubes \cite{Hua_Yu_Peng_Wu_He_Zhou_2022, Hua_Yu_Zhao_Li_Wu_Wu_2023}, channelized subsurface flow systems \cite{Tang_Liu_Durlofsky_2020, 10.2118/203924-MS, Jiang_Durlofsky_2023}, and CO$_{2}$ plume migration \cite{Wen_Tang_Benson_2021}.
Building on these architectures, Zhu and Zabaras \cite{Zhu_Zabaras_2018} extended the fully convolutional DenseNet \cite{Jegou_2017_CVPR_Workshops} to develop the DenseED architecture for high-dimensional surrogate modeling.
DenseED demonstrated strong performance in surrogate modeling and uncertainty quantification for single-phase flow in heterogeneous media with a \num{4225}-dimensional input and multi-phase flow fields with a \num{2500}-dimensional input \cite{Mo_Zhu_Zabaras_Shi_Wu_2019}.
Notably, DenseED showed excellent performance with limited training data and fewer network parameters than non-dense architectures. 
Romero \textit{et al.} \cite{Romero_Hasanpoor_Antonini_Amon_2024} extended DenseED with the DeepWFLO architecture to predict turbine wake fields for wind farm layout optimization, demonstrating the architecture's effectiveness for velocity field predictions. 

In this work, we develop a new modular deep encoder-decoder hierarchical (DeepEDH) convolutional neural network architecture based on image-to-image regression, which builds on previous DenseED \cite{Zhu_Zabaras_2018} and fully convolutional DenseNet \cite{Jegou_2017_CVPR_Workshops} architectures. Our DeepEDH architecture is tailored for surrogate modeling of CHT problems.
While several studies have developed surrogate models for CHT analysis, most are limited to predicting specific values of interest, with few models reconstructing full pressure, velocity, and temperature fields. 
Notably, none of these studies leverage the coupled nature of flow and heat transfer in CHT problems, instead predicting each field separately. 
This work’s end-to-end surrogate modeling approach effectively addresses these research gaps. 
Our new DeepEDH convolutional neural network architecture leverages specific physics considerations. It introduces several novel aspects, including output geometry masks, separate models for each field, and a two-stage temperature prediction methodology.
Separate models predict individual physical fields, serving as surrogate models for different governing PDEs. 
We apply an output geometry mask after the decoder to ensure the validity of reconstructed flow fields only within the fluid domain. 
Additionally, for temperature fields, we use two-stage models where velocity and temperature fields are linked, leveraging the coupled behavior of CHT problems. 
Combining the DeepEDH architecture, output geometry masks, field-specific models, and the connected two-stage temperature methodology, we address critical research gaps in previous works and demonstrate the ability to reconstruct complete fields for CHT analysis with improved accuracy and efficiency, requiring fewer network parameters than previous models.

Our work introduces several other aspects not explored in previous investigations:
\begin{enumerate}
	\item A method for translating unstructured CHT simulation results into structured meshes, allowing the creation of convolutional surrogate model training databases from such results.
	\item A comprehensive assessment of numerous factors influencing model performance, including architecture modifications, separate DeepEDH models for each field, output geometry masks, multi-stage temperature architecture, and hyperparameter and architecture optimization.
	\item A thorough quantification of the effects of the neural network depth, data resolution, dataset size, and the magnitude of heat flux boundary conditions on surrogate model performance, allowing physical insights into the behavior of the CHT system.
\end{enumerate}
Our findings and evaluations of the proposed DeepEDH architecture and end-to-end surrogate modeling methodology provide rich scientific insights to advance future research on surrogate modeling for CHT analyses of complex engineering systems.

The remainder of the paper is structured as follows. In \cref{sec:ge_and_dp}, we present the governing equations for CHT analysis and outline the procedures for processing unstructured mesh data. 
\Cref{sec:methodology} defines surrogate modeling in the context of image-to-image regression neural networks and introduces the model architecture, training process, regularization techniques, and loss function.
In \cref{sec:implementation}, we provide details of an implementation of the surrogate modeling methodology, focusing on an application to a liquid-cooled cold-plate-based battery thermal management system. 
This includes information on the generated datasets, model specifications, and the training process.
\Cref{sec:results} includes a comprehensive analysis and characterization of the model's performance.
Finally, in \cref{sec:concl_sm}, we offer a summary of our findings and present our conclusions.

\section{Governing equations and data processing procedures}
\label{sec:ge_and_dp}
This section describes the governing equations for CHT analyses and the data processing procedures used to generate the training and validation datasets from unstructured numerical simulation results.

\subsection{Governing equations and boundary conditions}
\label{sec:gov_eqn}
Conjugate heat transfer analyses are crucial for designing and optimizing thermal management systems, for example, those providing indirect liquid cooling through metal cold plates, as the one depicted in the case study used in this work (\cref{sec:implementation_csp}), featuring a cold plate with fluid channels and solid pin-fins and walls. Additional application illustrations are available in our previous work on the design optimization of pin-fin cold plates for electric vehicle battery packs \cite{EbbsPicken_Silva_Amon_2024,EbbsPicken_Silva_Amon_2024_ate}. 
In these cold-plate-based systems, the fluid enters the channel with velocity $\bm{u}$ and heat is transferred to the fluid from the cold plate surface heated by heat flux $\bm{q}$.
Conjugate heat transfer is used to analyze such a system and determine the velocity and temperature fields in the fluid and the temperature field in the solid. 
This CHT analysis is governed by three equations: conservation of mass, momentum, and energy.
Specifically, for three-dimensional, steady, and laminar flow with constant fluid properties within the fluid domain, the governing equations are momentum conservation \cref{eqn:ns} and mass continuity \cref{eqn:continuity}.
\begin{equation}
	\begin{aligned}
		\label{eqn:ns}
		& \rho \left(\bm{u} \cdot \nabla\right)\bm{u} = \nabla \cdot \left[-p\bm{I} + \bm{K} \right] + \bm{F}, \\
		& \bm{K} = \mu \left(\nabla \bm{u} + \left(\nabla \bm{u}\right)^{T}\right),
	\end{aligned}
\end{equation}
\begin{equation}
	\rho \nabla \cdot \bm{u} = 0 \label{eqn:continuity},
\end{equation}
where $\bm{u}$ is the velocity vector, $p$ is the pressure, $\rho$ is the density, $\mu$ is the dynamic viscosity, $\bm{K}$ is the stress tensor, $\bm{F}$ is the body force, and $\bm{I}$ is the identity tensor.
We define boundary conditions for the fluid domain as no-slip walls \cref{eqn:no_slip}, mass flow inlet \cref{eqn:mass_flow}, pressure outlet \cref{eqn:p_out}, and symmetry \cref{eqn:sym}.
\begin{align}
    & \bm{u} = 0, \label{eqn:no_slip}\\
	& -\int_{\partial \Omega} \rho(\bm{u} \cdot \bm{n})d_{bc}dS = \dot{m}, \label{eqn:mass_flow}\\
	& \left[-\rho \bm{I} + \bm{K}\right]\bm{n} = -p_{0}\bm{n}, \label{eqn:p_out}\\
	&\begin{aligned}
		\label{eqn:sym}
		& \bm{u} \cdot \bm{n} = 0, \\
		& \bm{K_{n}} - \left[\bm{K_{n}} \cdot \bm{n}\right] = 0, \qquad \bm{K_{n}}=\bm{Kn},
	\end{aligned}
\end{align}
here $\bm{n}$ is the unit normal vector, $d_{bc}$ is the boundary condition thickness, $\dot{m}$ is the mass flow rate, and $p_{0}$ is the outlet pressure.
Energy conservation, governed by \mbox{\cref{eqn:energy}}, is applied in both the fluid and solid domains.
These domains are coupled at the solid-fluid interface. 
\mbox{\Cref{eqn:energy}} is solved homologously for the solid and fluid regions to ensure the continuity of temperature and heat flux across the interface.
\begin{equation}
	\begin{aligned}
		\label{eqn:energy}
		& \rho C_{p}\frac{\partial T}{\partial t} + \rho C_{p} \bm{u} \cdot \nabla T + \nabla \cdot \bm{q} = Q,\\
		& \bm{q} = -k \nabla T,
	\end{aligned}
\end{equation}
where $T$ is the temperature, $C_{p}$ is the specific heat capacity, $t$ is time, $\bm{q}$ is the heat flux, $Q$ is the volumetric heat generation rate, and $k$ is the thermal conductivity, with boundary conditions consisting of zero heat flux \cref{eqn:ht_insulated}, convective heat flux \cref{eqn:ht_insulated} with $q_{0} = h(T_{ext} - T)$, inflow \cref{eqn:ht_inflow}, outflow \cref{eqn:ht_outflow}, and symmetry \cref{eqn:ht_outflow}:
\begin{align}
    & -\bm{n} \cdot \bm{q} = q_{0}, \label{eqn:ht_insulated}\\
	&\begin{aligned}
		\label{eqn:ht_inflow}
		& \bm{n} \cdot \bm{q} = \rho \Delta H \bm{u} \cdot \bm{n}, \\
		& \Delta H = \int^{T}_{T_{ustr}} C_{p}dT,
	\end{aligned} \\
	& -\bm{n} \cdot \bm{q} = 0. \label{eqn:ht_outflow}
\end{align}
For \Crefrange{eqn:ht_insulated}{eqn:ht_outflow}, $q_{0}$ is the heat flux, $h$ is the convective heat transfer coefficient, $T_{ext}$ is the external temperature, $P_{0}$ is the power source, $\volume$ is the volume, and $\Delta H$ is the enthalpy difference. In this work's case study (\cref{sec:implementation}), the heat source is modeled based on battery heat generation, considering the battery temperature and state-of-charge (SOC) independent Ohmic heating \cref{eqn:ohmic_heat} \cite{Al-Zareer_Silva_Amon_2021}.
\begin{equation}
	\label{eqn:ohmic_heat}
	\dot{Q}_{\text{gen}} = |V-VOC(SOC)|\times I,
\end{equation}
where $V$ is the battery voltage, $VOC$ is the open circuit voltage, and $I$ is the current.

The heat flux, generated by a battery module in this work's case study (\cref{sec:implementation}),  is utilized as the cold plate's boundary condition for the energy \cref{eqn:energy}.
The governing \cref{eqn:ns,eqn:continuity,eqn:energy} with appropriate boundary conditions are then solved with the finite element method (FEM).
To obtain accurate results across the fluid and solid domains, the mesh is refined at the fluid-solid interface to capture the high temperature and velocity gradients.
Further, a conformal mesh at this interface ensures continuity of the temperature and heat flux from the solid to fluid domain, allowing the conjugate heat transfer to be accurately captured.

\subsection{Data structuring and scaling procedures}
\label{sec:data_struct}
Convolutional neural networks were generally designed for image-related tasks, such as recognition or segmentation, where image-like data is used for input and training.
For CHT surrogate modeling, previous literature primarily used structured meshes composed of square cells, which resemble image pixels and are suitable for CNNs.
However, for solving CHT problems with complex geometries, it is often advantageous to use unstructured meshes composed of irregular polygons without a specific structure.
To adapt these numerical solutions for surrogate modeling, we transform the unstructured result data to structured data post-solver.
By mapping to a structured grid after solving, we can represent the results as images with lower resolution than the unstructured solver mesh, leading to smaller models with faster and more efficient training.

To transform two-dimensional unstructured mesh data (\cref{sub@fig:unstruct_grid}) into structured mesh data (\cref{sub@fig:struct_grid}), we define a structured mesh as a grid of square cells with $n_{y}$ rows and $n_{x}$ columns. 
For each $c_{\text{struct}}$ structured cell, we calculate an area-weighted average of the values from overlapping $c_{j_{\text{unstruct}}}$ unstructured cells, repeating this process for all cells in the new mesh. 
The area-weighted average is calculated according to \cref{eqn:struct_grid}.
\begin{equation}
	\label{eqn:struct_grid}
	\Psi(c_{i_{\text{struct}}}) = \frac{\sum_{j=0}^{n}A(c_{j_{\text{unstruct}}}) \cdot \Psi(c_{j_{\text{unstruct}}})}{\sum_{j=0}^{n}A_(c_{j_{\text{unstruct}}})} \qquad \qquad \forall c_{j_{\text{unstruct}}} \in dom(c_{i_{\text{struct}}})
\end{equation} 
where $\Psi$ is the field value of interest, such as pressure, velocity, or temperature, $A$ is the area of the cell, and $dom$ is the domain of a cell.

\begin{figure}[H]
	\centering
	\begin{subfigure}{0.3\textwidth}
		\centering
		\includegraphics[width=0.7\textwidth]{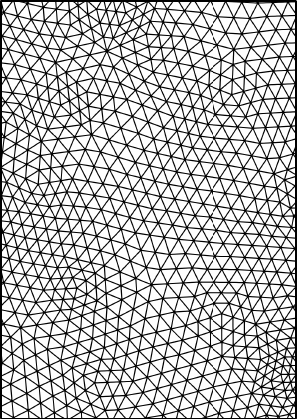}
		\caption{\label{fig:unstruct_grid}}
	\end{subfigure}
	\qquad
	\begin{subfigure}{0.3\textwidth}
		\centering
		\includegraphics[width=0.7\textwidth]{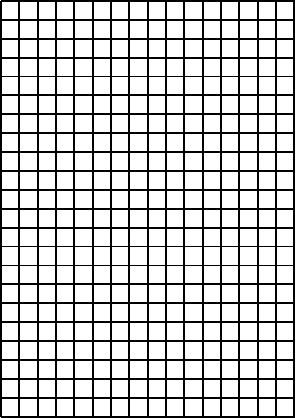}
		\caption{\label{fig:struct_grid}}
	\end{subfigure}
	\caption[Unstructured and structured meshes]{Illustration of the application of \cref{eqn:struct_grid} to temperature, velocity, or pressure field results, transforming  (\ref{sub@fig:unstruct_grid}) an unstructured mesh onto  (\ref{sub@fig:struct_grid}) an image-like structured mesh where each cell is representative of an image pixel.\label{fig:grid_structuring}}
\end{figure}

\section{Surrogate modeling methodology}
\label{sec:methodology}
\subsection{Data-driven surrogate modeling as image-to-image regression}
The thermal management systems under consideration in this work are governed by the CHT equations and boundary conditions detailed in \cref{sec:gov_eqn}.
The solution process computes steady-state results, including pressure and velocity fields, as well as transient temperature results, using simulations based on the input geometry and predefined boundary conditions.
Assuming fixed boundary conditions and variable geometry, we can view this simulation process as a black-box function, denoted as $f$, that maps the geometry model $\bm{X_s}$ to physical fields $\bm{y}$, as expressed in \cref{eqn:sim_model}.
\begin{equation}
	\label{eqn:sim_model}
	\bm{y} = (P, \bm{u}, T) = f(\bm{x}, \bm{X_s})
\end{equation}
here $\bm{y}$ are the pressure ($P$), velocity ($\bm{u}$), and temperature ($T$) at each spatial location $\bm{x}$ in the domain.

Data-based surrogate modeling aims to replace the computationally expensive simulation process $f$ with a more efficient model.
This surrogate model is developed or trained using a limited dataset consisting of $n$ simulation inputs and outputs: $\left\{\bm{x}_{i}, \bm{y}_{i}\right\}_{i=1}^{n}$.
By treating the simulation inputs and outputs as images, the surrogate model can be formulated as an image-to-image regression model, similar to established CNN architectures.
Both the input and output are considered as images with pixel dimensions of $n_y$ rows and $n_x$ columns, where each pixel is a value to be predicted by the regression model. 
Given that the simulation results from process $f$ are defined on unstructured meshes, the data processing methods established in \cref{sec:data_struct} are used to transform $\bm{y}$ results to structured image-like data.
The regression models use the high-dimensional input to predict each pixel in the output, modeling a highly non-linear relationship to approximate the solution of the underlying governing PDE.
We can define the surrogate model as a function, denoted as $\hat{f}$, which approximates $f$ as outlined in \cref{eqn:surr_model}.
\begin{equation}
	\label{eqn:surr_model}
	\hat{\bm{y}} = (\hat{P}, \hat{\bm{u}}, \hat{T}) = \hat{f}(\bm{x}, \bm{X_s}, \bm{\theta})
\end{equation}
the $\hat{\bm{y}}$ predictions consist of pressure ($\hat{P}$), velocity ($\hat{\bm{u}}$), and temperature ($\hat{T}$) results, approximating the true $\bm{y}$ values.
The predictions are determined using the surrogate models with parameters $\bm{\theta}$, learned from the $n$ training simulations: $\left\{\bm{x}_{i}, \bm{y}_{i}\right\}_{i=1}^{n}$.

\subsection{Modular deep encoder-decoder hierarchical (DeepEDH) convolutional neural network architecture}
In this section, we briefly introduce the techniques used in developing the DeepEDH architecture presented in this work.
\subsubsection{Dense and fully convolutional encoder-decoder neural networks}
\begin{figure}[H]
	\centering
	\includegraphics[width=0.95\textwidth]{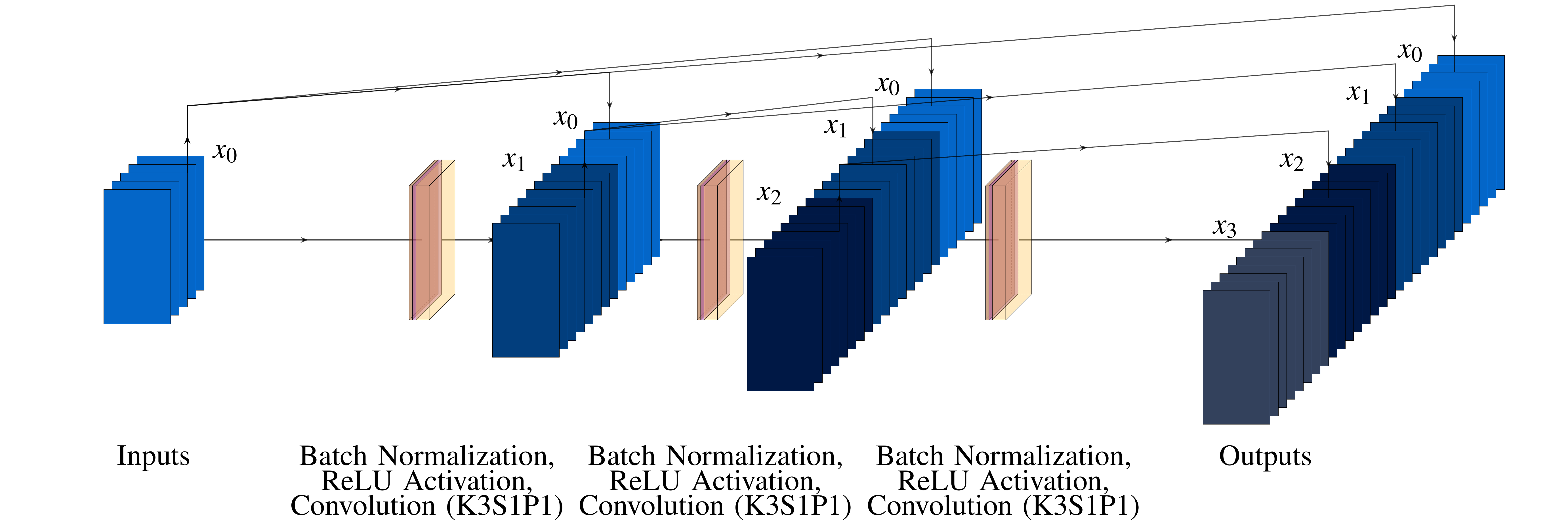}
	\caption[Neural network dense block]{Dense block based on the DenseNet \cite{Huang_Liu_Maaten_Weinberger_2016} architecture with 3 layers ($L=3$) and a growth rate of 8 ($K=8$). The previous feature maps, represented as blue-hued planes in the figure, are appended to the output from each block of batch normalization, ReLU activations, and convolution using a kernel with kernel size (K) of 3, stride (S) of 1, and padding (P) of 1. The feature map sizes remain constant through the dense block, while the number of feature maps grows by $K$ through each layer. \label{fig:dnet_arch}}
\end{figure}

Training deep convolutional neural networks can be challenging due to issues like vanishing gradients \cite{Hochreiter_1998} and accuracy degradation \cite{He_Zhang_Ren_Sun_2016}.
To address these problems, various techniques have been introduced, including normalized initialization and intermediate initialization layers \cite{He_Zhang_Ren_Sun_2016, Ioffe_Szegedy_2015, Glorot2010UnderstandingTD}, and shortcut connections, as seen in architectures like ResNet \cite{He_Zhang_Ren_Sun_2016} and Highway Networks \cite{NIPS2015_215a71a1}.
Shortcut connections are a fundamental feature of dense convolutional neural networks and enable more efficient training of deeper networks, allowing the output of a layer to be combined with the input to a layer further down the network.
DenseNet \cite{Huang_Liu_Maaten_Weinberger_2016} extended this concept, introducing shortcut connections to all subsequent layers.
For network layer $L$, there will be $K \cdot L$ feature maps added based on the growth factor $K$.
\Cref{fig:dnet_arch} shows an example of a dense block based on DenseNet with $L=3$ and $K=8$.

Fully convolutional networks (FCNs) \cite{Long_Shelhamer_Darrell_2015} are a class of neural networks that have been used for image-to-image regression tasks. 
These networks allow for pixel-wise predictions by replacing the fully connected layers with convolutional layers and including up-sampling (decoder) layers to recover the input size.
The use of skip connections between down-sampling (encoder) sections and decoder sections enables the preservation of information from the encoding process.
Recent works like U-Net \cite{Ronneberger_Fischer_Brox_2015}, Segnet \cite{Badrinarayanan_Kendall_Cipolla_2015}, and FPNs \cite{Lin_Dollar_Girshick_He_Hariharan_Belongie_2017} have demonstrated the effectiveness of FCNs.
By using both encoder and decoder sections, FCNs can learn both low-level and high-level features, while skip connections facilitate the flow of information across multiple scales.

\begin{figure}[H]
	\centering
	\begin{subfigure}{0.7985\textwidth}
		\centering
		\includegraphics[width=0.9\textwidth]{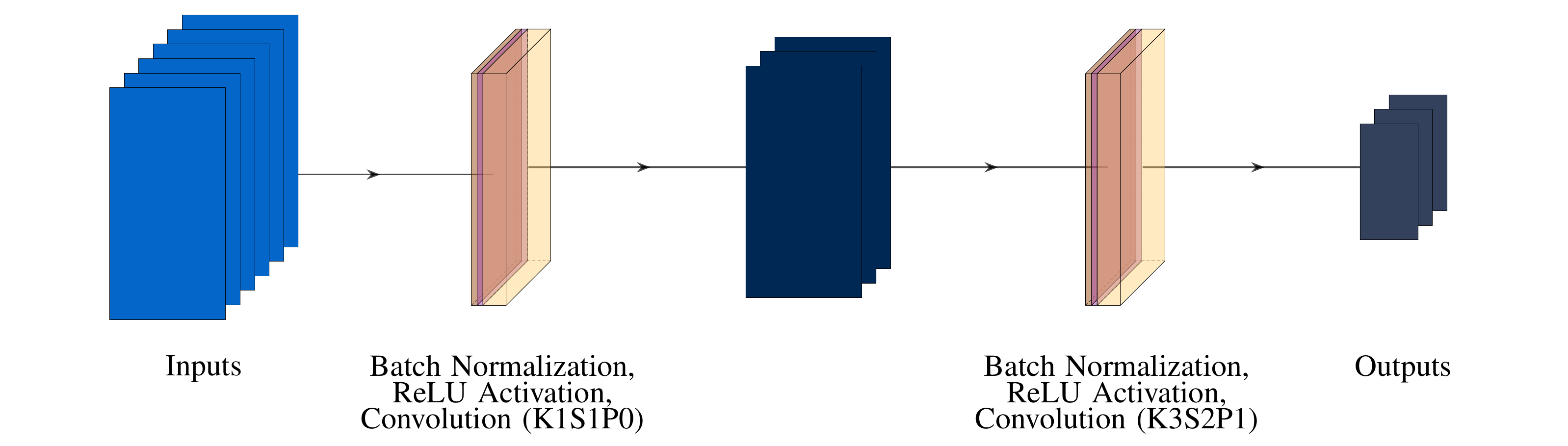}
		\caption{\label{fig:encode_layer}}
	\end{subfigure}
	\begin{subfigure}{0.925\textwidth}
		\centering
		\includegraphics[width=0.9\textwidth]{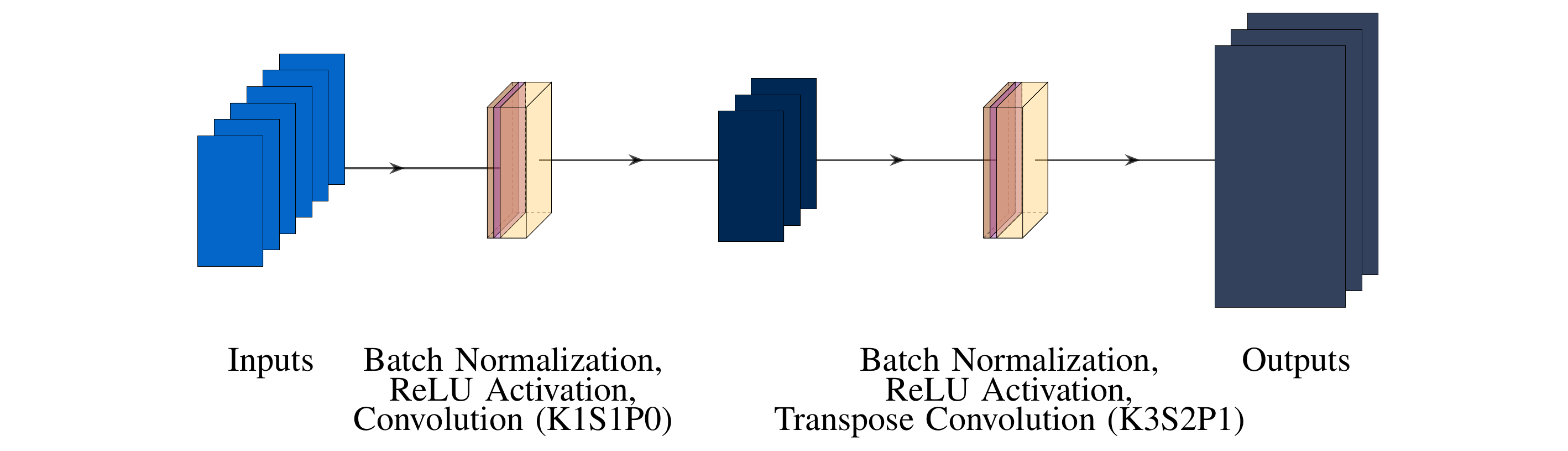}
		\caption{\label{fig:decode_layer}}
	\end{subfigure}
	\caption[Neural network encoding and decoding transition layers]{(\ref{sub@fig:encode_layer}) Encoding layer with convolution and (\ref{sub@fig:decode_layer}) decoding layer with convolution and transpose convolution. The first combination of batch normalization, ReLU activations, and convolution for both encoding and decoding layers reduce the number of feature maps, represented as blue-hued planes in the figure, by half using a kernel with kernel size (K) of 1, stride (S) of 1, and padding (P) of 0. The second combination uses a kernel with K=3, S=2, and P=1 to reduce the feature map sizes by half for encoding with convolution layers and doubles the feature map sizes for decoding with transpose convolution layers.\label{fig:enc_dec_layers}}
\end{figure}

To leverage the advantages of both dense convolutional networks and FCNs, researchers have extended DenseNet to create a fully convolutional architecture \cite{Jegou_2017_CVPR_Workshops}.
To address varying feature map sizes in encoder-decoder architectures, transition layers, illustrated in \cref{fig:encode_layer,fig:decode_layer}, and dense blocks (\cref{fig:dnet_arch}) were defined \cite{Zhu_Zabaras_2018, Jegou_2017_CVPR_Workshops}.
Transition layers, whether down-sampling (encoder) or up-sampling (decoder), serve to connect dense blocks and make the network modular.
The DenseED architecture, developed by Zhu and Zabaras \cite{Zhu_Zabaras_2018}, was adapted from the fully convolutional DenseNet architecture \cite{Jegou_2017_CVPR_Workshops} and tailored for surrogate modeling of PDE-based systems by introducing several key changes: 
(i) the number of feature maps was reduced using the first layer in each transition layer, 
(ii) all feature maps from previous layers were kept in each dense block, 
(iii) no skip connections between the encoder and decoder sections were used, and 
(iv) convolution with stride 2 was used for down-sampling.
The DeepEDH architecture proposed in this work leverages the majority of these changes; however, we introduce several additional enhancements: 
(i) skip connections between encoder and decoder sections due to the stronger correspondence between the input and output for CHT analyses, 
(ii) dropout for regularization, 
(iii) output geometry masks to account for the direct correspondence between the inputs and outputs for flow predictions, and 
(iv) a two-stage model for temperature prediction to take advantage of the coupling between velocity and temperature fields in CHT analyses.

\subsubsection{Output geometry mask}
An output geometry mask is introduced to our architecture to ensure that solid domain regions are correctly represented in the model output for pressure and velocity predictions.
We create the output geometry mask, denoted as $\bm{\delta}$, by inverting the input image.
In the input image, solid regions have a value of 1, while the fluid regions are set to 0.
Hence, we define the mask such that $\bm{\delta} = 1 - \bm{x}$, where $\bm{x}$ is the input image.
Using the output geometry mask, the final pressure ($\hat{P}$) and velocity ($\hat{\bm{u}}$) outputs of the surrogate model $\hat{\bm{y}}_{\text{fluid}}$ are defined using element-wise multiplication with the mask: $\hat{\bm{y}}_{\text{fluid}} = \bm{\delta} \odot \hat{\bm{y}} = \bm{\delta} \odot \hat{f}(\bm{x}, \bm{X_s}, \bm{\theta})$.
Here, $\odot$ is the element-wise multiplication operator. 
This ensures that the predicted pressure and velocity fields are only valid in the fluid domain, improving model accuracy for CHT and flow analyses.

\subsubsection{Two-stage temperature model} 
Due to the significant influence of convection in CHT analyses, temperature exhibits a strong dependence on the velocity field.
In order to leverage this interdependence and improve the temperature predictions, we have developed a two-stage model.
In this approach, the first stage utilizes the output of the velocity model, which is then incorporated into the input of the second-stage model for temperature prediction.
This enables the network to learn the relationship between the temperature and velocity fields, resulting in improved temperature predictions.
The velocity model produces an output denoted as $\hat{\bm{y}}_{\text{vel}} = \hat{f}_{\text{vel}}(\bm{x}, \bm{X_s}, \bm{\theta}_{\text{vel}})$.
This output is used as input for the temperature model, represented as $\hat{\bm{y}}_{\text{temp}} = \hat{f}_{\text{temp}}(\bm{x}, \bm{X_s}, \hat{\bm{y}}_{\text{vel}}, \bm{\theta}_{\text{temp}}) = \hat{f}_{\text{temp}}(\bm{x},\hat{f}_{\text{vel}}(\bm{x}, \bm{X_s}, \bm{\theta}_{\text{vel}}), \bm{\theta}_{\text{temp}})$.

\subsubsection{Final architecture: DeepEDH}
\begin{figure}[H]
	\centering
	\includegraphics[width=0.9\textwidth]{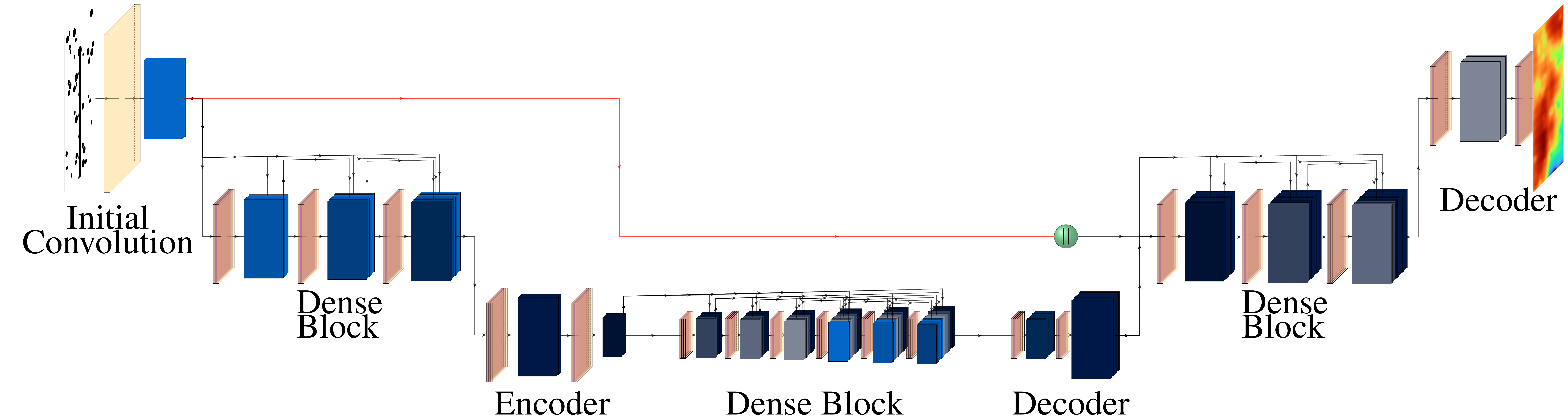}
	\caption[DeepEDH convolutional neural network architecture]{Network architecture: DeepEDH with $L_{\text{dense}} = [3, 4, 3]$ and $K=2$. This example includes 1 encoding dense block with 3 layers, a bottleneck dense block with 4 layers, and 1 decoding dense block with 3 layers. The number of feature maps, represented as blue-hued planes in the figure, grows by two through each dense block layer as defined by the growth rate. \label{fig:final_arch}}
\end{figure}

This work introduces the new neural network architecture called DeepEDH (modular deep encoder-decoder hierarchical convolutional neural network).
DeepEDH uses principles from fully convolutional networks similar to U-Net \cite{Ronneberger_Fischer_Brox_2015}, Segnet \cite{Badrinarayanan_Kendall_Cipolla_2015}, and FPNs \cite{Lin_Dollar_Girshick_He_Hariharan_Belongie_2017}, with dense blocks and skip connections between encoder and decoder sections, similar to fully convolutional DenseNet \cite{Jegou_2017_CVPR_Workshops}.
We use the changes stated above to fully convolutional DenseNet proposed in DenseED \cite{Zhu_Zabaras_2018}. However, we include skip connections and dropout.
DeepEDH is defined by the depth of the network (number of dense blocks), the number of layers within each dense block, the growth rate $K$ through each dense block, and the number of feature maps after the initial convolutional layer.
The depth and the number of layers in each dense block can be defined using a list, $L_{\text{dense}}$.
The length of this list defines the network depth, while the value at each index determines the number of layers in each dense block.
It is important to note that the length of $L_{\text{dense}}$ must be an odd integer, where the middle value defines the number of layers in the bottleneck dense block.
An example of DeepEDH with $L_{\text{dense}} = [3, 4, 3]$ and $K=2$ is shown in \cref{fig:final_arch}.
The encoding section of the network takes the input: $\bm{X_s}$ for velocity and pressure models and $(\bm{X_s}, \hat{\bm{y}}_{\text{vel}})$ for temperature models. 
An initial convolutional layer is applied to reduce the size of the input by half and extract initial feature maps. 
The initial feature maps pass through pairs of dense blocks (\cref{fig:dnet_arch}), and encoding transition layers (\cref{sub@fig:encode_layer}) until reaching the bottleneck block (bottom of \cref{fig:final_arch}), from which code dimension feature maps are extracted.
Subsequently, the code dimension feature maps progress through pairs of decoding transition layers (\cref{sub@fig:decode_layer}) and dense blocks (\cref{fig:dnet_arch}), leading to the final decoding layer that directly predicts the output.
For the pressure and velocity models, the output geometry mask is applied to the network output. In contrast, for temperature predictions, the velocity model output is passed as input in the two-stage architecture. 

\subsection{Network hyperparameters, training loss, and regularization}
To determine the hyperparameters for the DeepEDH architecture, we based our initial values on previous studies such as \cite{Zhu_Zabaras_2018, Jegou_2017_CVPR_Workshops}.
We then performed hyperparameter and architecture optimization with a combination of Bayesian optimization (BO) and Hyperband (HB) with the BOHB algorithm \cite{Falkner_Klein_Hutter_2018}.
The hyperparameters considered for optimization include the dropout rate, the number of layers in the encoding, bottleneck, and decoding dense blocks, the growth rate, the number of feature maps after the initial convolution layer, the initial learning rate, the learning rate weight decay, and the batch size.
It is important to note that hyperparameters were optimized separately for each field.
Full details of the architecture and training hyperparameter optimizations are included with the supplementary materials.

During the training process, we minimized the difference between the structured target field $\bm{y}$, generated by simulation, and the model predicted field $\hat{\bm{y}}$ for all $N_{\text{cells}} = n_x \cdot n_y$ pixels.
We used a regularized mean squared error (MSE) training loss function, as defined in \cref{eqn:mse_loss}.
\begin{equation}
	\label{eqn:mse_loss}
	L(\hat{f}(\bm{x}, \bm{X_s}, \bm{\theta}),\bm{y}) = \frac{1}{N_{\text{cells}}}\sum_{i=1}^{N_{\text{cells}}}\left(\hat{f}(\bm{x}, \bm{X_s}, \bm{\theta})_{i}-\bm{y}_i\right)^{2}+\alpha\Omega(\bm{\theta})
\end{equation}
This loss function incorporates $\pazocal{L}_{2}$ regularization with $\Omega(\bm{\theta}) = \frac{1}{2}\bm{\theta}^{\text{T}}\bm{\theta}$, defined using weight decay ($\alpha$) in PyTorch.
We used both batch normalization \cite{Ioffe_Szegedy_2015} and dropout \cite{JMLR:v15:srivastava14a} for additional regularization.
The network parameters $\bm{\theta}$ include the weights of the convolutional and transpose convolutional layer kernels, as well as the scale and shift parameters in the batch normalization layers.

\section{Surrogate implementation and training}
\label{sec:implementation}
This section details the implementation of the surrogate modeling methodology and DeepEDH architecture presented in \cref{sec:methodology}, considering a liquid-cooled cold-plate-based battery thermal management system.
DeepEDH neural network surrogate models were developed as replacements for computational FEM simulations to predict the pressure, velocity, and temperature fields of the CHT analysis based on the cold plate channel geometry. 
The dataset generation, data transformations, network architecture, and training details are presented for the test case.

\subsection{Problem definition}
\label{sec:implementation_csp}
The surrogate model implementation considered a pin-fin cold plate thermal management system, shown in \cref{sub@fig:case_study_geo_3d}. 
The cold plate was designed to cool a battery module made up of 30 pouch cells, where coolant flows into the cold plate through the inlet tube, around the central wall and pin-fin arrangement, and exits through the outlet tube.
To simulate a transient battery discharge process with non-uniform heat generation from the battery module, a spatially varying heat flux boundary condition was applied to the top surface of the cold plate.
Here, we considered heat flux based on a 3C charge process, where each battery cell in the battery module was charged at a constant current of 3 times its rated capacity. 
This is representative of a battery fast charge process \cite{Al-Zareer_Silva_Amon_2021, Al-Zareer_Ebbs-Picken_Michalak_Escobar_Silva_Davis_Osio_Amon_2023},
with heating rates ranging from 600~W to 900~W, amounting to a total heat generation of 2,693,403~{kJ} considering the 30 pouch cell battery module.
An inlet flow rate of 3~LPM with a temperature of 20~{$^{\circ}$C} was used for the cold plate, leading to laminar flow with a channel Reynolds number of approximately 1000.

The cold plate channel geometry, made up of the pin-fin arrangement, is defined by the position of the pin-fin centers $(x_i, y_i)$ and the pin-fin radii $r_i$ for $i=1,2,\dots,n_{\text{pins}}$ pin-fins.
Using the channel geometry as input to the DeepEDH surrogate models, we predicted pressure and velocity fields at the mid-plane of the cold plate channel and the temperature field at the surface of the cold plate for the final time step of the transient simulation. 
\begin{figure}[H]
	\centering
	\begin{subfigure}{0.6\textwidth}
		\centering
		\includegraphics[height=5cm]{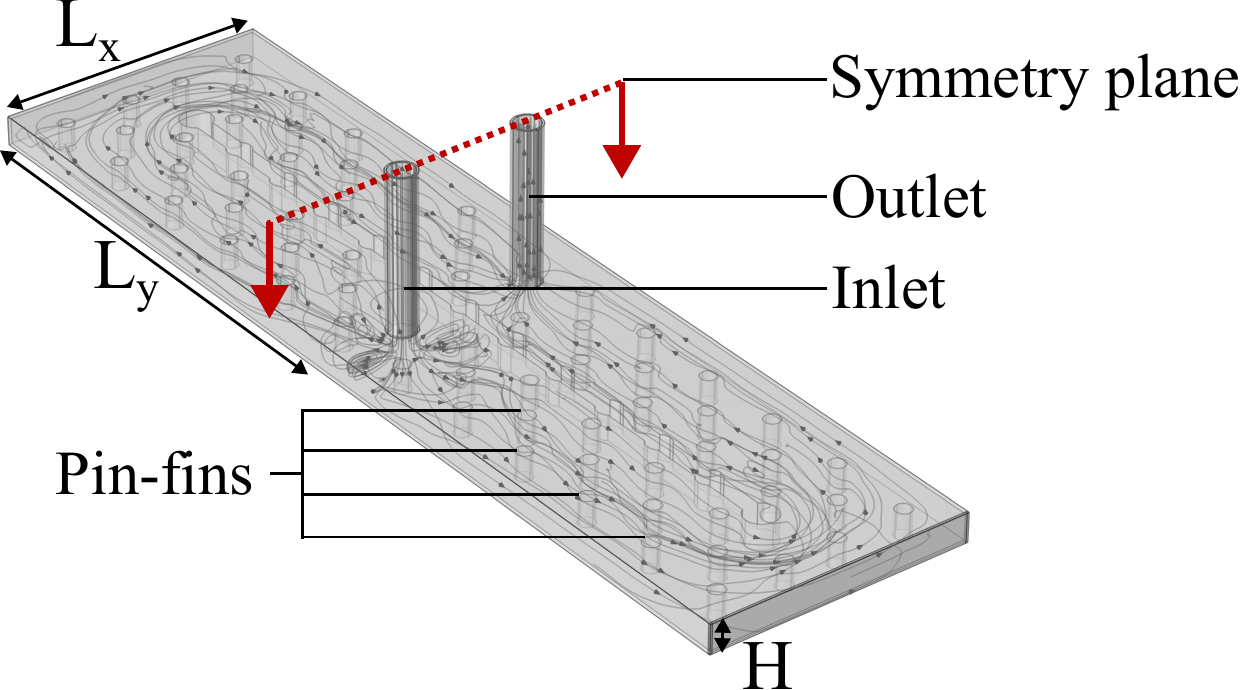}
		\caption{\label{fig:case_study_geo_3d}}
	\end{subfigure}
	\begin{subfigure}{0.35\textwidth}
		\centering
		\includegraphics[height=5cm]{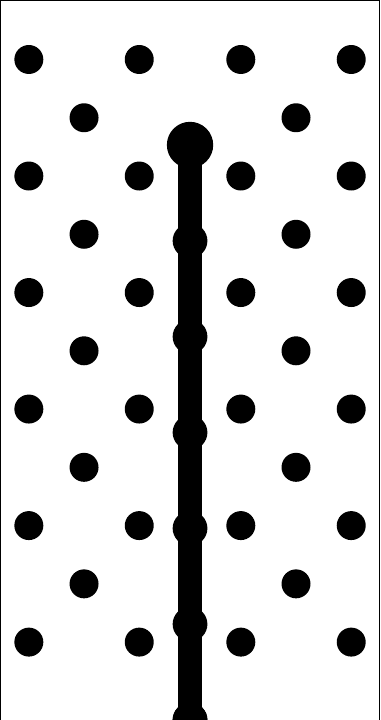}
		\caption{\label{fig:case_study_geo_2d}}
	\end{subfigure}
	\caption[Pin-fin cold plate geometry]{Pin-fin cold plate (\ref{sub@fig:case_study_geo_3d}) three-dimensional geometry and (\ref{sub@fig:case_study_geo_2d}) channel cross-sectional view with the solid metal region shown in black and the flow region shown in white. Symmetry was applied across the width of the cold plate with (\ref{sub@fig:case_study_geo_2d}) showing half of the plate.\label{fig:prob_geometry}}
\end{figure}
The cold plate channel cross-sectional geometry, as shown in \Cref{sub@fig:case_study_geo_2d}, depicts the pin-fin solid regions in black and the fluid region in white.
The domain size was fixed at approximately $L_{x} \approx 0.25$~m and $L_{y} \approx 0.5$~m, with a constant channel height of $H = 0.005$~m.
The design variables included the positions and radii of the pin-fins.
These design variables were constrained to ensure that there was no overlap between solid regions.
The pin-fin radii were further constrained between $r_{\text{min}} = 0.005$~m and $r_{\text{max}} = 0.015$~m, and the system consisted of $n_{\text{pins}}=34$ pin-fins.
A symmetry boundary condition (\cref{eqn:sym}) was applied across the width of the cold plate through the inlet and outlet tubes to reduce the domain size.
Mass flow (\cref{eqn:mass_flow}) and pressure (\cref{eqn:p_out}) boundary conditions were applied to the inlet and outlet respectively. 
No-slip wall (\cref{eqn:no_slip}) boundary conditions were used for the internal channel walls.
The cold plate was thermally insulated on the bottom (\cref{eqn:ht_insulated} with $q_{0} = 0$), while natural convection boundary conditions were applied to the other external cold plate walls using vertical and horizontal wall correlations. 
The top surface used the spatially and time-varying heat flux boundary condition (\cref{eqn:ht_insulated}) determined from a battery discharge process.
Heat inflow and outflow conditions \cref{eqn:ht_inflow,eqn:ht_outflow} were specified at the inlet and outlet.

For the numerical solution of the CHT problem, three-dimensional unstructured meshes were generated and solved using the FEM for each geometry in COMSOL Multiphysics 6.0 with the conjugate heat transfer physics.
From the three-dimensional FEM simulation results the velocity and pressure fields were extracted at the mid-plane of the cold plate channel depth, while the temperature field was extracted at the surface of the cold plate.
The DeepEDH models were developed based on these simulation results to act as surrogates for the FEM models.

\subsubsection{Datasets}
\label{sec:implementation_csp_ds}
\begin{figure}[H]
	\centering
	\begin{subfigure}{0.2\textwidth}
		\centering
		\includegraphics[width=0.9\textwidth]{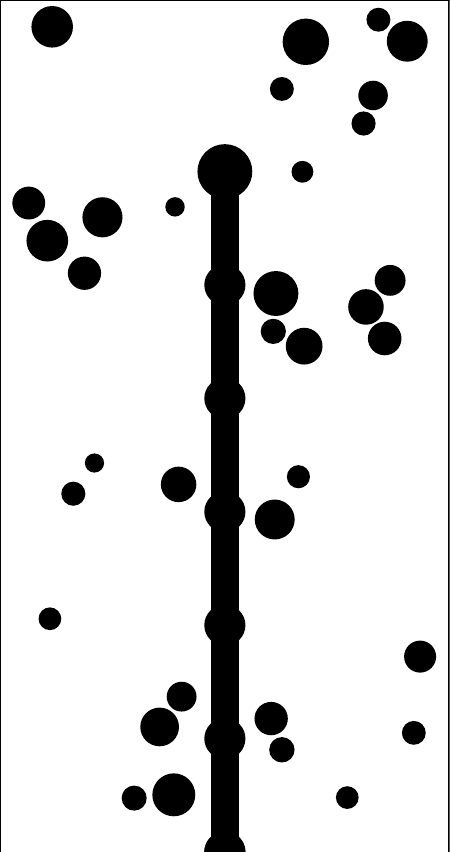}
        \caption{\label{fig:example_samples1}}
	\end{subfigure}
	\qquad
	\begin{subfigure}{0.2\textwidth}
		\centering
		\includegraphics[width=0.9\textwidth]{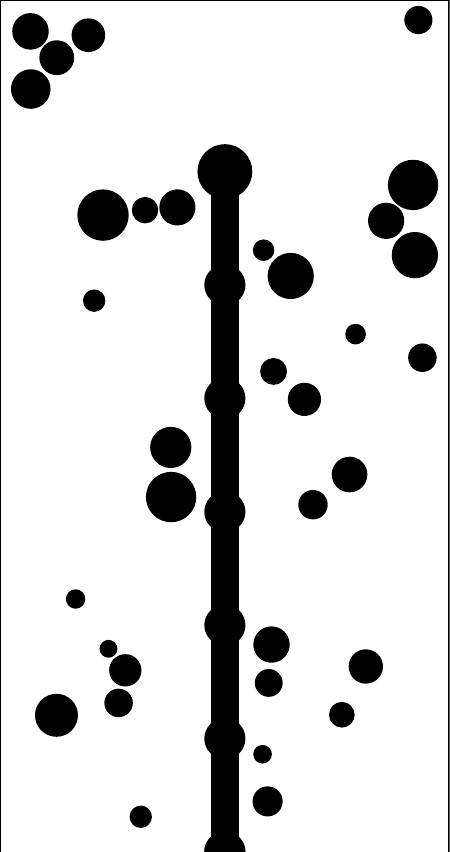}
        \caption{\label{fig:example_samples2}}
	\end{subfigure}
	\qquad
	\begin{subfigure}{0.2\textwidth}
		\centering
		\includegraphics[width=0.9\textwidth]{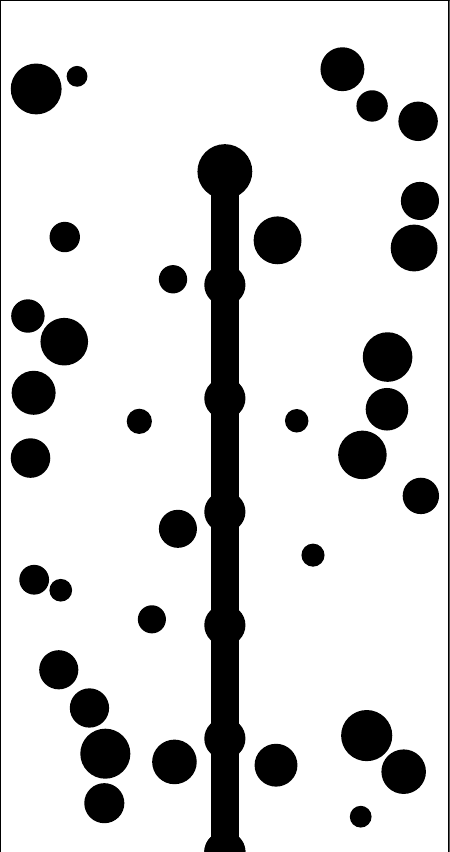}
        \caption{\label{fig:example_samples3}}
	\end{subfigure}
	\qquad
	\begin{subfigure}{0.2\textwidth}
		\centering
		\includegraphics[width=0.9\textwidth]{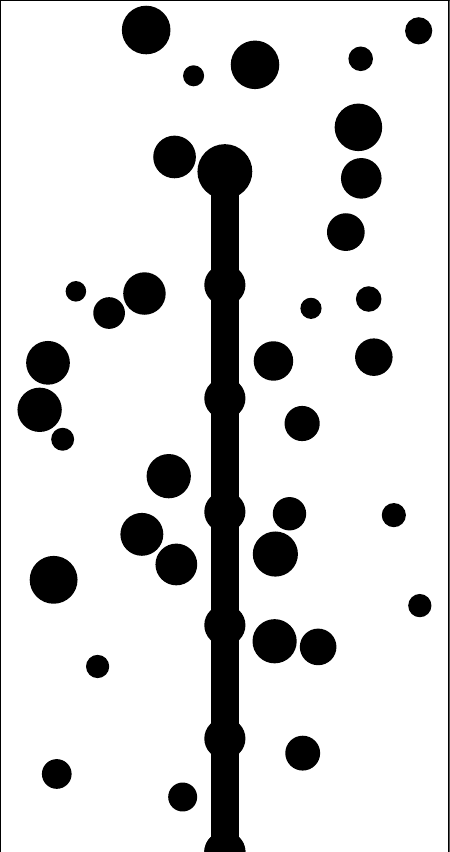}
        \caption{\label{fig:example_samples4}}
	\end{subfigure}
	\caption[Sample pin-fin channel geometries]{Channel cross-sectional views of selected pin-fin cold plate geometry sample points from the Latin Hypercube sample plan. Solid metal regions are shown in black and the flow region is shown in white.\label{fig:example_samples}}
\end{figure}

A set of 1,500 pin-fin arrangements was defined using Latin hypercube sampling, where each arrangement is characterized by the positions and radii of the pin-fins.
Several examples of these pin-fin arrangements are displayed in \cref{fig:example_samples1,fig:example_samples2,fig:example_samples3,fig:example_samples4}.
For each pin-fin arrangement, the FEM CHT simulations determined the output fields at each time step.
These output fields include the pressure ($p$) and velocity components ($u$, $v$, and $w$) at the mid-plane of the cold plate channel, as well as temperature ($T$) at the heated surface of the cold plate.
Parallel computing enabled 50 FEM models to be executed simultaneously, with each simulation taking approximately 0.3 - 0.5~hours, allowing each dataset to be generated in approximately 10 hours.
Each simulation, run in parallel, used 48~GB of memory, 2~GB of storage, and 12 cores on a high-performance computing cluster.

\begin{figure}[H]
	\centering
	\begin{subfigure}{0.125\textwidth}
		\centering
		\includegraphics[width=0.9\textwidth]{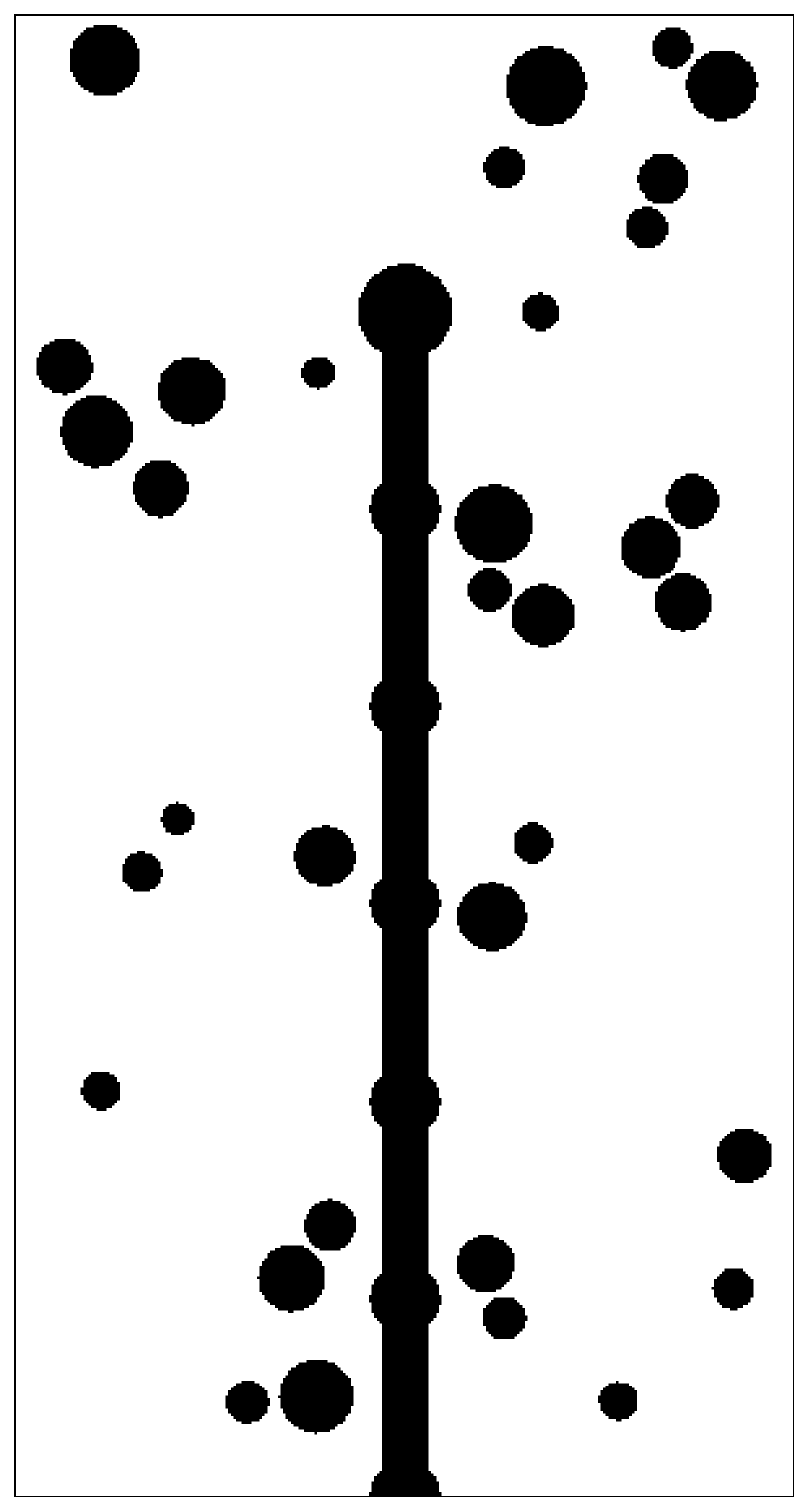}
		\caption{\label{fig:input_ex}}
	\end{subfigure}
	\begin{subfigure}{0.125\textwidth}
		\centering
		\includegraphics[width=0.9\textwidth]{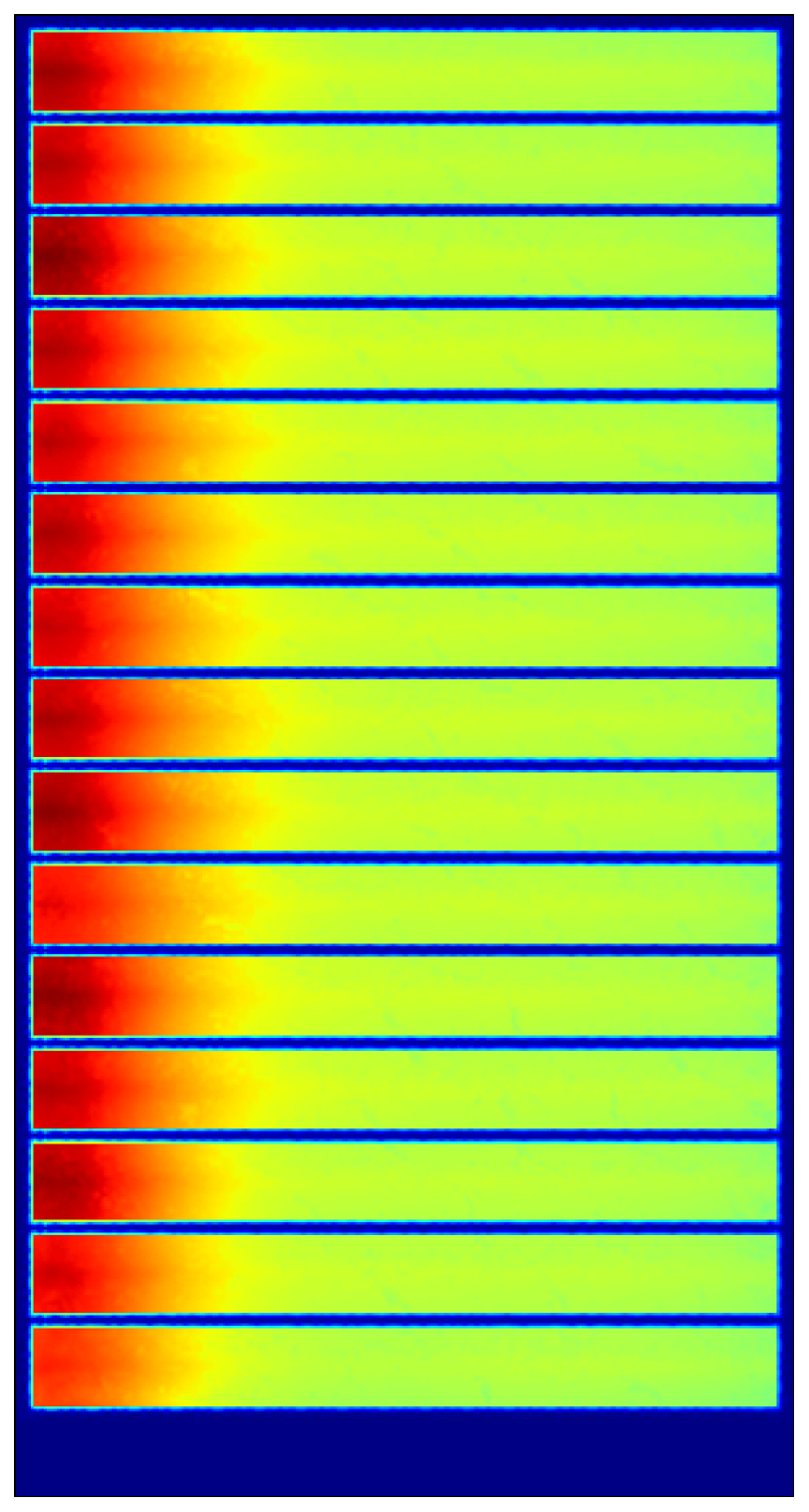}
		\caption{\label{fig:input_hf_ex}}
	\end{subfigure}
	\begin{subfigure}{0.125\textwidth}
		\centering
		\includegraphics[width=0.9\textwidth]{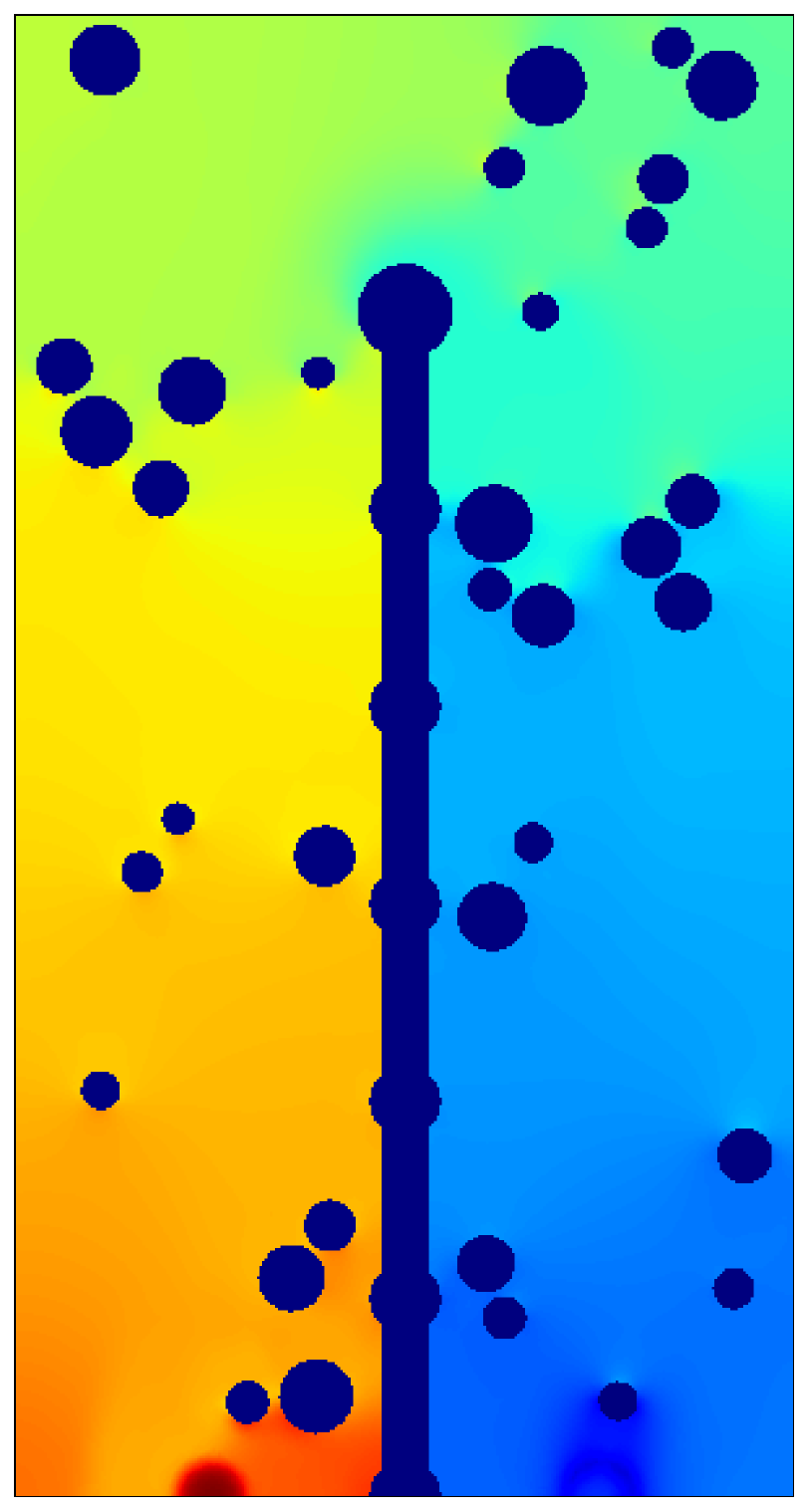}
		\caption{\label{fig:press_ex}}
	\end{subfigure}
	\begin{subfigure}{0.125\textwidth}
		\centering
		\includegraphics[width=0.9\textwidth]{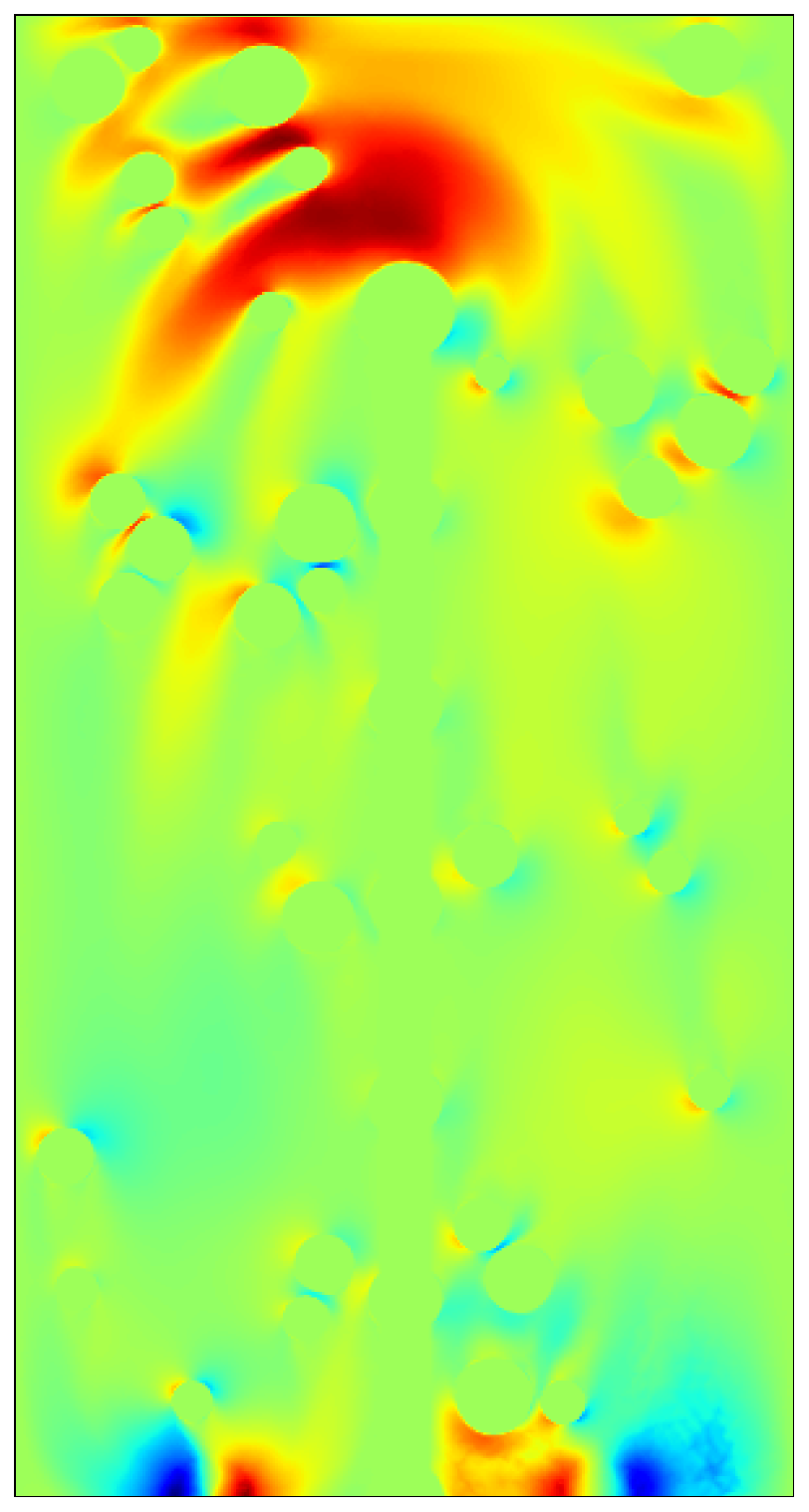}
		\caption{\label{fig:u_ex}}
	\end{subfigure}
	\begin{subfigure}{0.125\textwidth}
		\centering
		\includegraphics[width=0.9\textwidth]{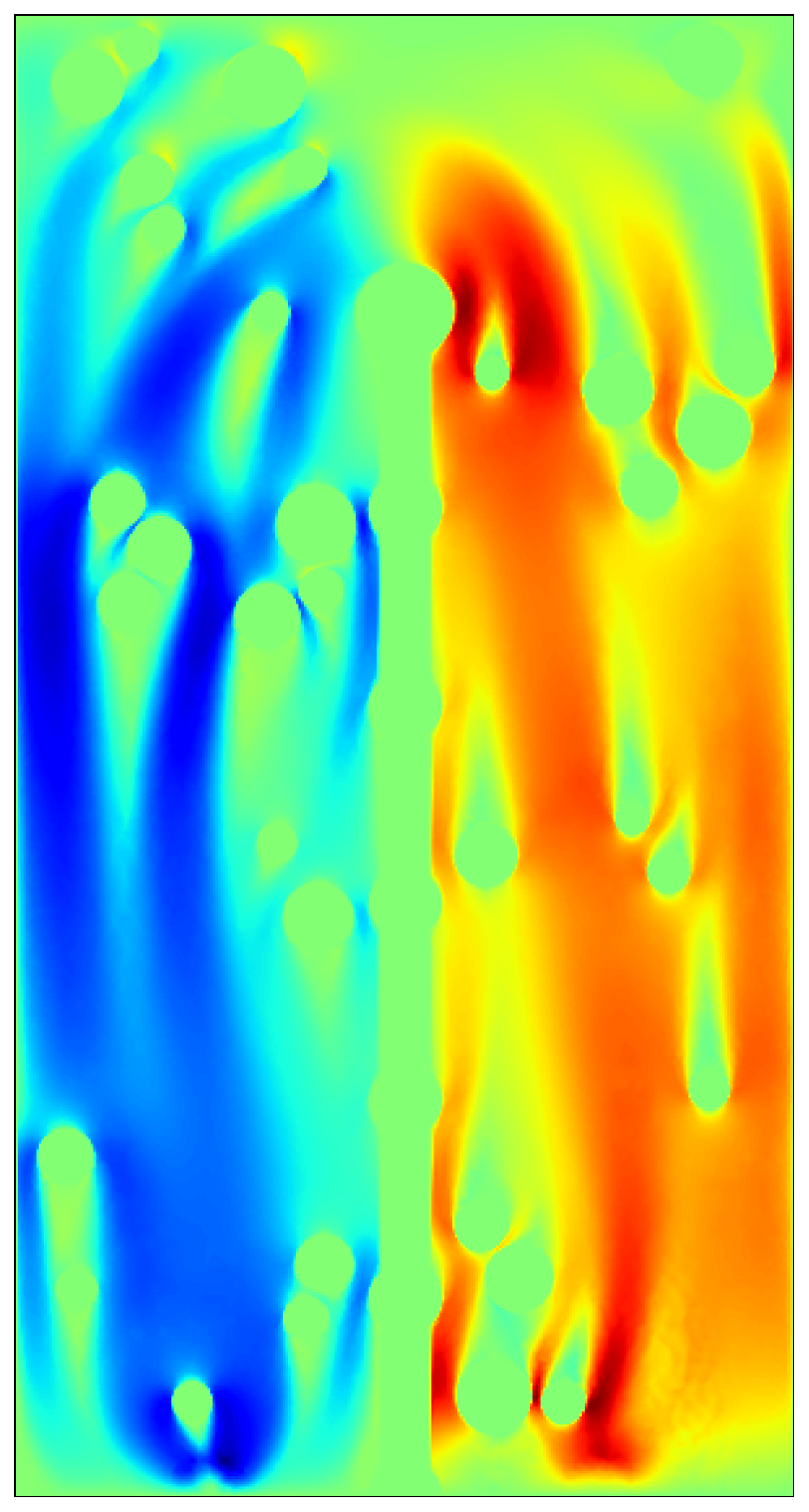}
		\caption{\label{fig:v_ex}}
	\end{subfigure}
	\begin{subfigure}{0.125\textwidth}
		\centering
		\includegraphics[width=0.9\textwidth]{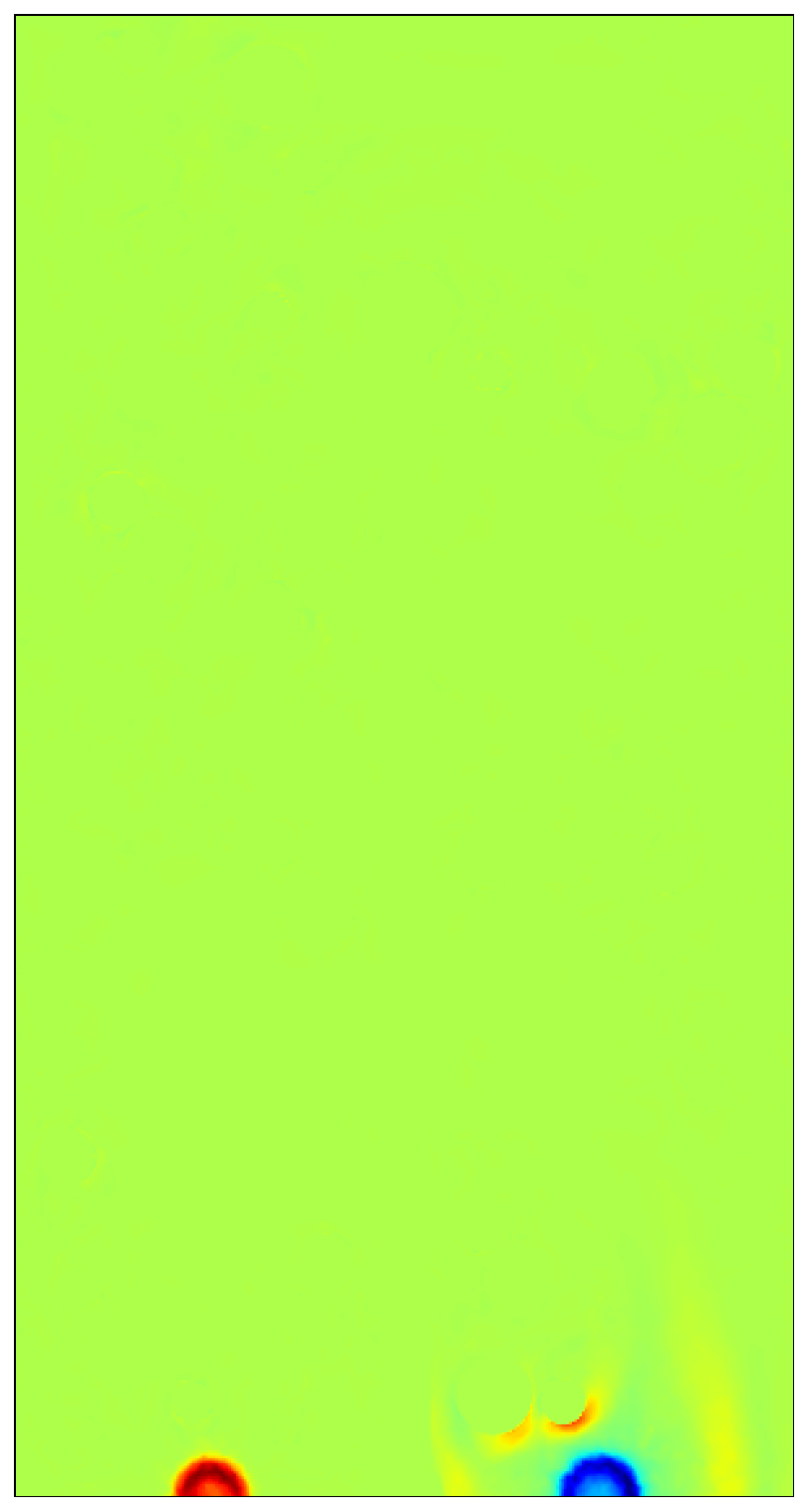}
		\caption{\label{fig:w_ex}}
	\end{subfigure}
	\begin{subfigure}{0.125\textwidth}
		\centering
		\includegraphics[width=0.9\textwidth]{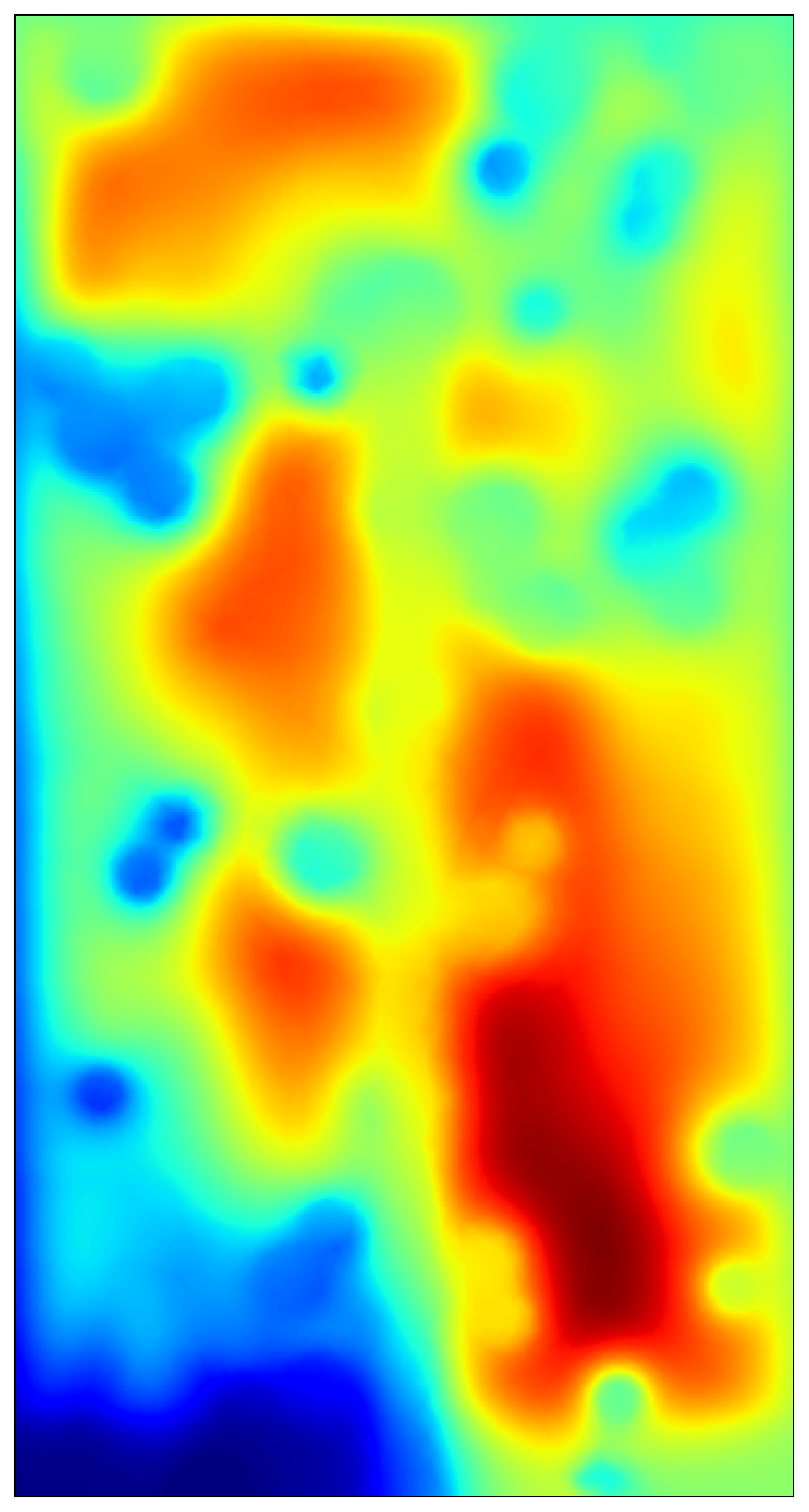}
		\caption{\label{fig:temp_ex}}
	\end{subfigure}
	\caption[Dataset sample point example]{Dataset example: (\ref{sub@fig:input_ex}) geometry input, (\ref{sub@fig:input_hf_ex}) heat flux input, (\ref{sub@fig:press_ex}) pressure output, (\ref{sub@fig:u_ex}) $u$ velocity output, (\ref{sub@fig:v_ex}) $v$ velocity output, (\ref{sub@fig:w_ex}) $w$ velocity output, and (\ref{sub@fig:temp_ex}) temperature output. Pressure and velocity field outputs were extracted at the mid-plane of the cold plate channel depth, while the temperature field was extracted at the surface of the cold plate \label{fig:dataset_example}}
\end{figure}

\begin{figure}[H]
	\centering
	\begin{subfigure}{0.2\textwidth}
		\centering
		\includegraphics[width=0.9\textwidth]{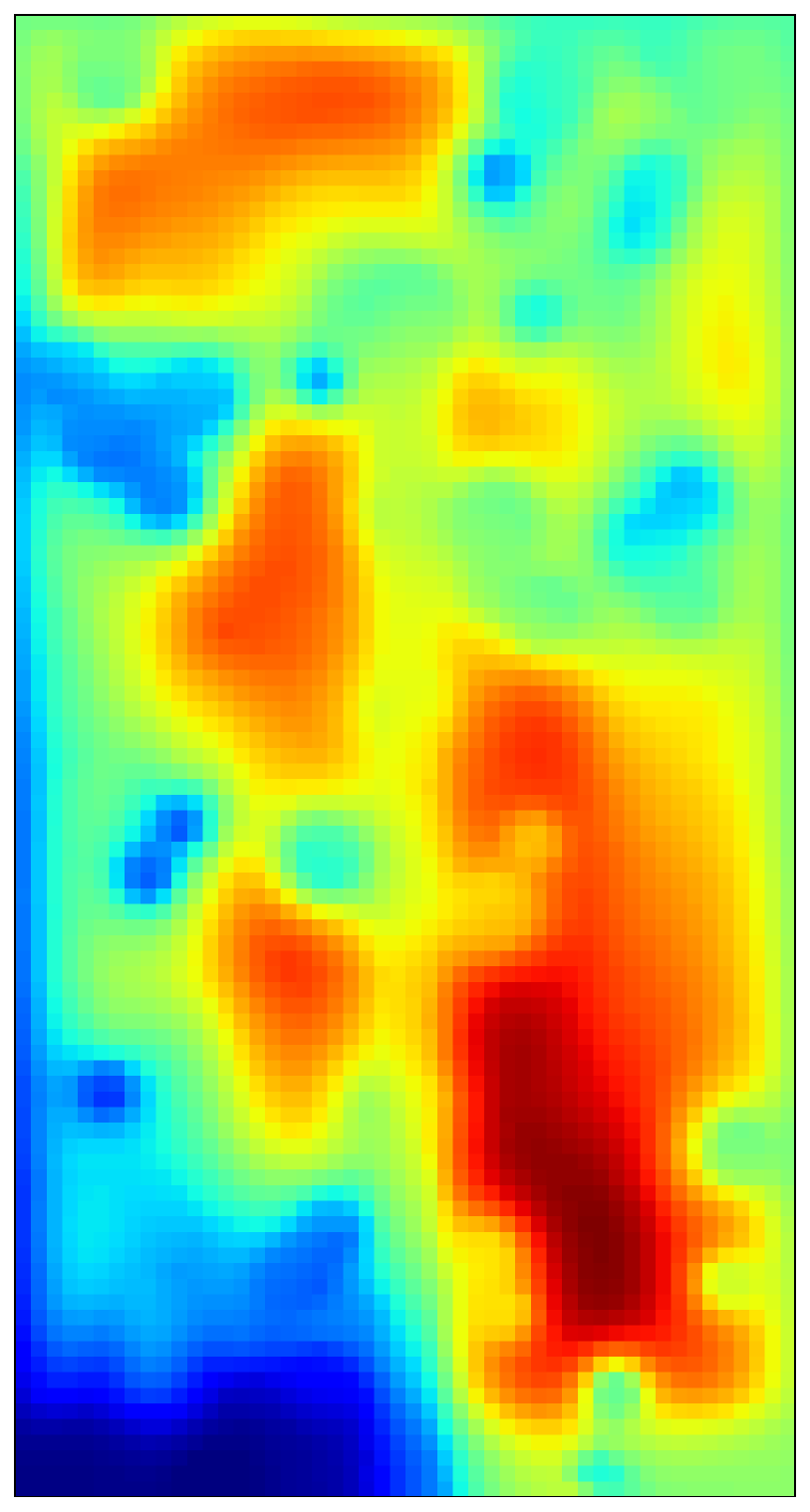}
		\caption{\label{fig:res50_example}}
	\end{subfigure}
	\qquad
	\begin{subfigure}{0.2\textwidth}
		\centering
		\includegraphics[width=0.9\textwidth]{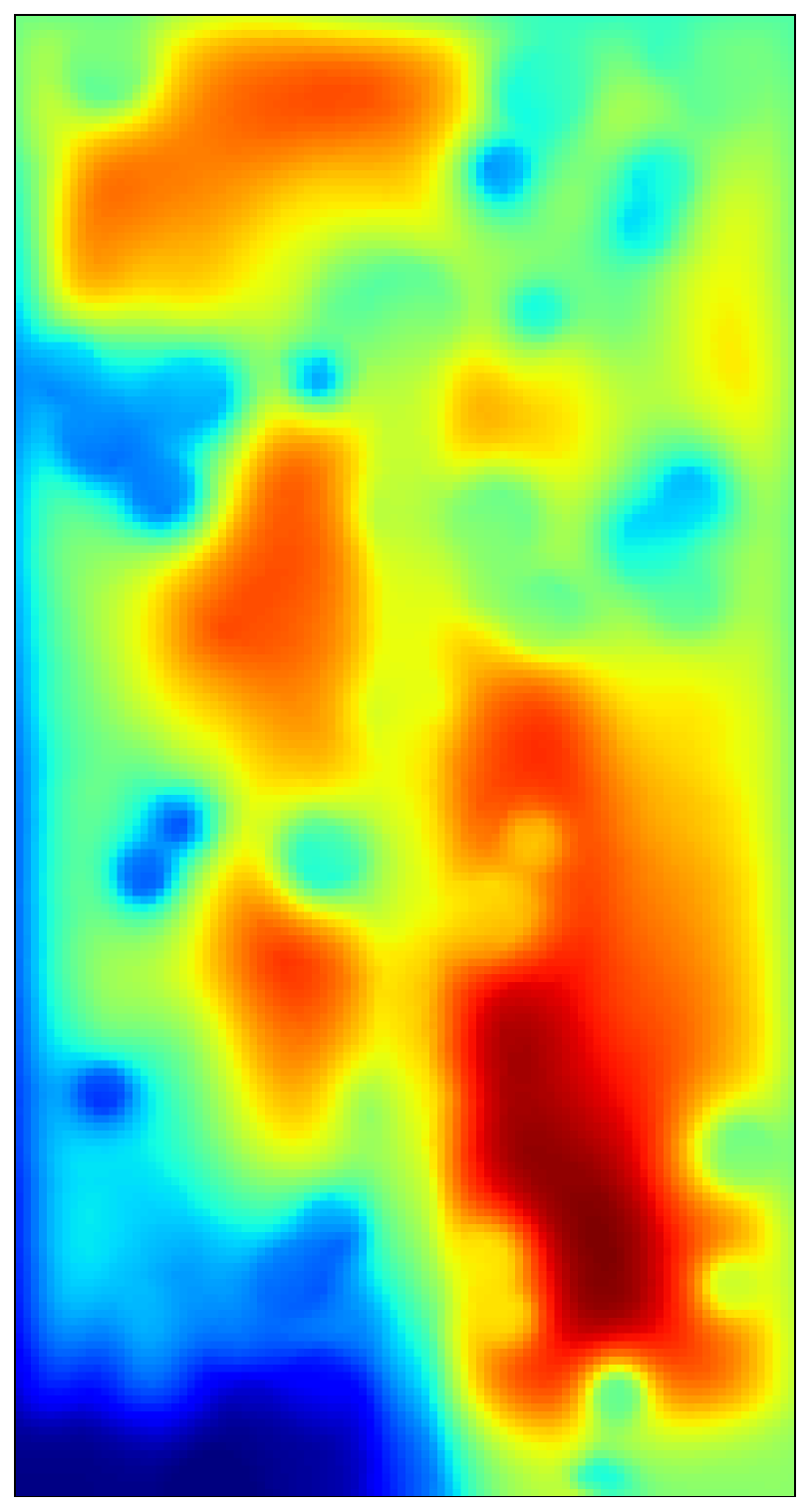}
		\caption{\label{fig:res100_example}}
	\end{subfigure}
	\qquad
	\begin{subfigure}{0.2\textwidth}
		\centering
		\includegraphics[width=0.9\textwidth]{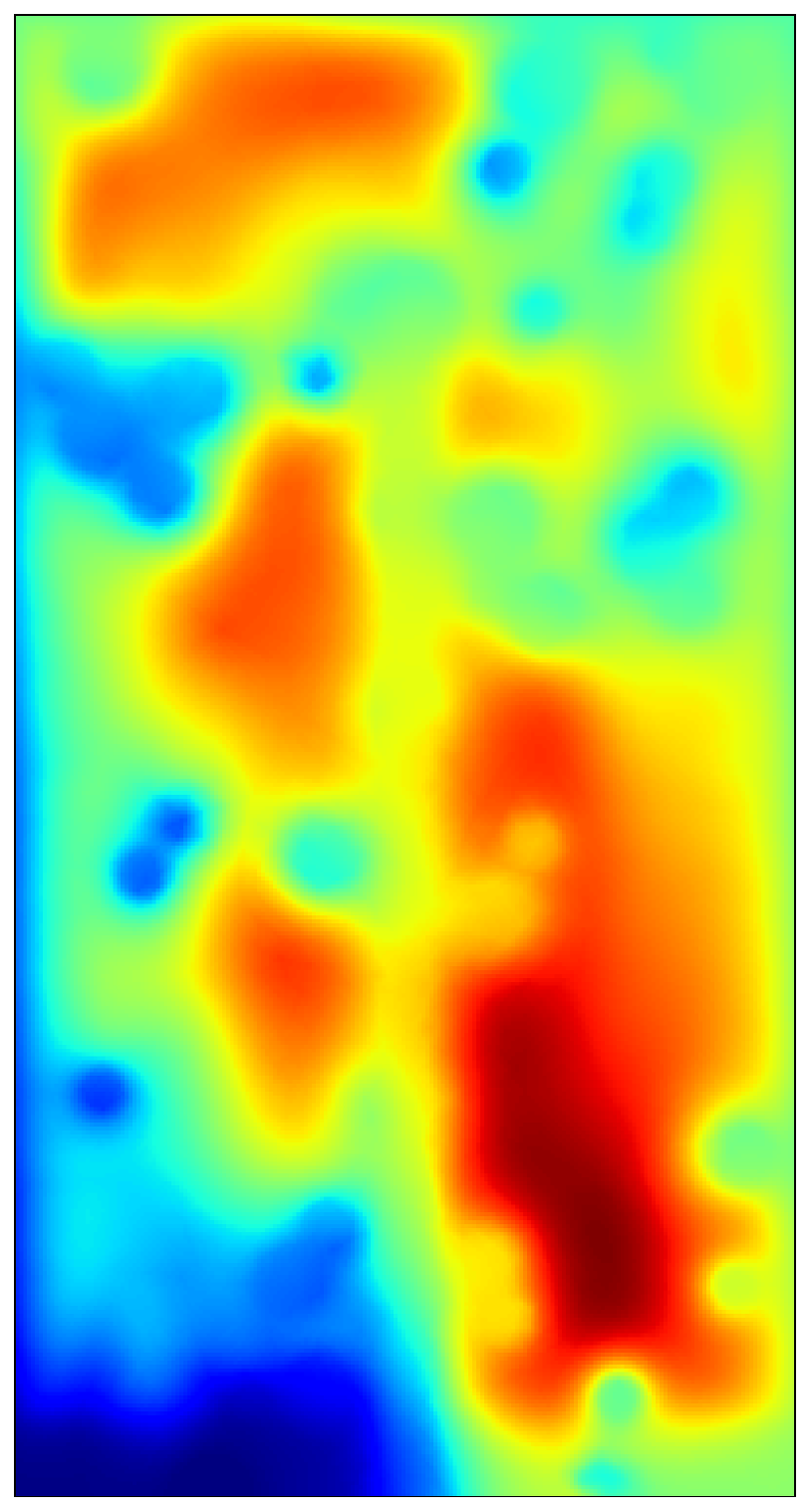}
		\caption{\label{fig:res200_example}}
	\end{subfigure}
	\qquad
	\begin{subfigure}{0.2\textwidth}
		\centering
		\includegraphics[width=0.9\textwidth]{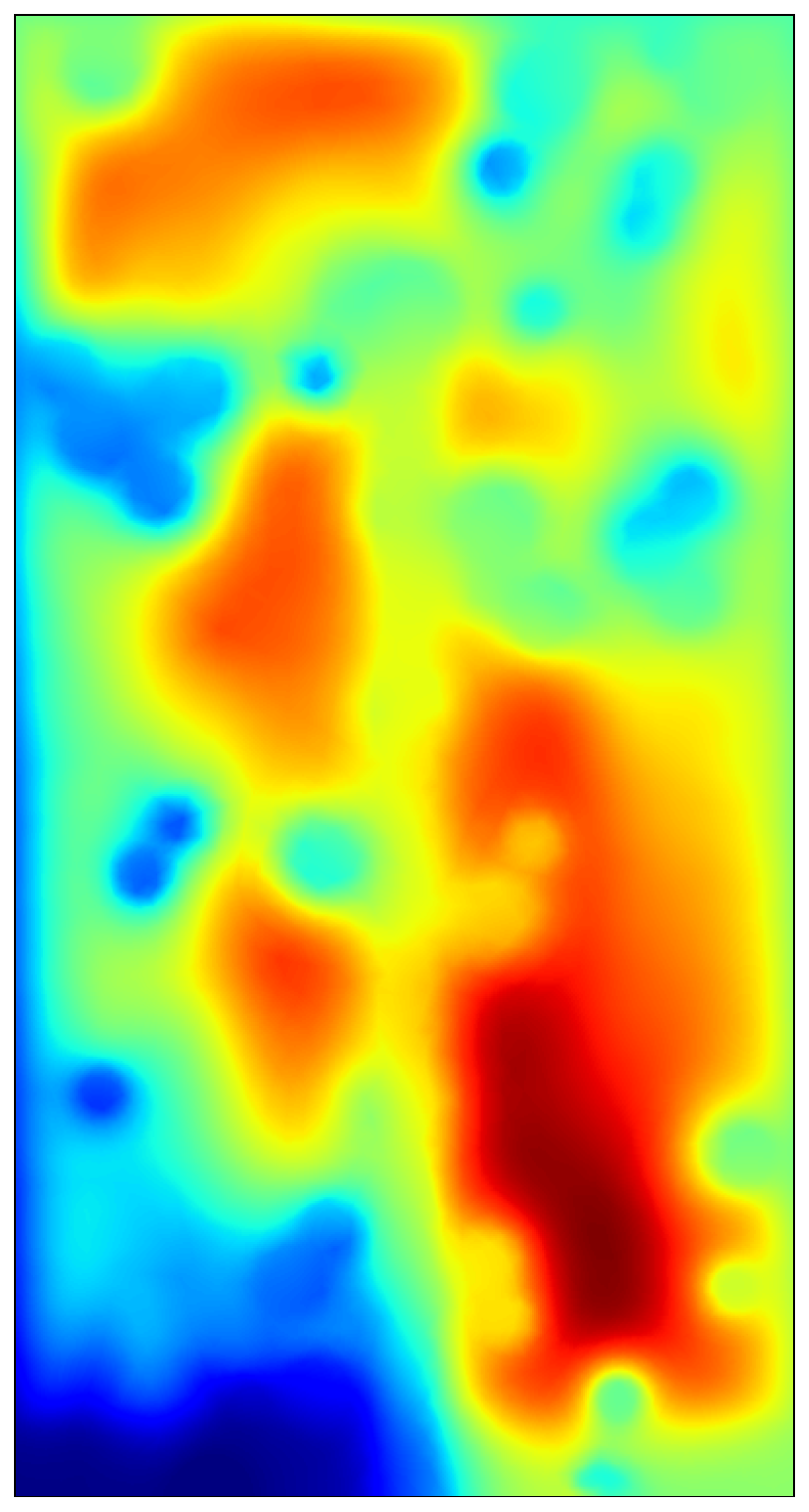}
		\caption{\label{fig:res400_example}}
	\end{subfigure}
	\caption[Data resolutions]{Data resolutions, shown here for a sample temperature field: (\ref{sub@fig:res50_example}) 50$\times$95, (\ref{sub@fig:res100_example}) 100$\times$190, (\ref{sub@fig:res200_example}) 200$\times$380, and (\ref{sub@fig:res400_example}) 400$\times$761.\label{fig:dataset_resolutions}}
\end{figure}

It is important to note that not all of the samples defined by the sampling plan resulted in converged numerical solutions. 
Those non-converged solutions were excluded from the final dataset.
The complete dataset consists of approximately 1,500 simulations, considering only those with converged solutions.
These simulations included various input pin-fin arrangements, heat flux profiles, and output fields for each time step.
An example of a single simulation sample point is presented in \cref{fig:input_ex,fig:input_hf_ex,fig:press_ex,fig:u_ex,fig:v_ex,fig:w_ex,fig:temp_ex}.
The simulation results were based on an unstructured mesh of approximately 3,000,000 cells.
This mesh was composed of approximately 250,000 cells for fluid cross-sections and approximately 200,000 cells for solid cross-sections.
The results were processed using the data transformations described in \cref{sec:data_struct} to structure the results and generate datasets for training and testing the surrogate models.
Four data resolutions were considered, as shown in \cref{fig:res50_example,fig:res100_example,fig:res200_example,fig:res400_example}.
These resolutions were determined by selecting $n_x$ and calculating the appropriate $n_y$ to ensure that the structured cells are physically square. 
The resulting data resolutions were 50$\times$95 with 4,750 cells, 100$\times$190 with 19,000 cells, 200$\times$380 with 76,000 cells, and 400$\times$761 with 304,400 cells, representing cell sizes of approximately 5.2~{mm/px}, 2.6{mm/px}, 1.3~{mm/px}, and 0.65~{mm/px} respectively.
In relation to the original unstructured mesh, these data resolutions represent approximately 2\%, 8\%, 33\%, and 133\% of the original mesh size, providing a range of data scaling.

\subsection{Surrogate model training}
\label{sec:implementation_smt}
DeepEDH models were developed and trained to predict the pressure, velocity, and temperature fields, resulting in three separate DeepEDH field models: DeepEDH-pressure, DeepEDH-velocity, and DeepEDH-temperature.
DeepEDH-temperature used a two-stage architecture, coupling the velocity and temperature fields.
The network architecture for each model was determined through the results of network characterization, hyperparameter optimization, and architecture optimization.
The final architectures for each field model are outlined in \cref{tab:network_layers}.
The outputs of each layer depend on various factors, including the depth of the network, the number of layers in each dense block, and the growth rate $K$.
Each dense block increases the number of feature maps by $K \cdot L$, while the transition layers reduce the number of feature maps by half.
The encoding transition layers halve the feature map size, while decoding transition layers double the feature map size.
\begin{table}[H] 
	\centering
	\caption[Network architecture details for DeepEDH-pressure, DeepEDH-velocity, and DeepEDH-temperature models]{Network architecture details for DeepEDH-pressure, DeepEDH-velocity, and DeepEDH-temperature models.\label{tab:network_layers}}
	\begin{adjustbox}{width=1\columnwidth, center}
		\begin{tabular}{cc|c>{\hspace*{15mm}}c>{\hspace*{15mm}}c|c}
			\toprule
			\multirow{2}{*}{Network section} & \multirow{2}{*}{Layer} & \multicolumn{3}{c|}{Layer parameters} & \multirow{2}{*}{Output} \\
			\cmidrule{3-5}
			& & DeepEDH-pressure & DeepEDH-velocity & DeepEDH-temperature & \\
			\midrule
			\multirow{2}{*}{Initial convolution} & Inputs & 1 & 1 & 2 & ($n_{x}$, $n_{y}$, Inputs) \\ 
			\addlinespace
			& Convolution (K7S2P3) & Output (IC): 16 & Output (IC): 48 & Output (IC): 16 & ($\frac{n_{x}}{2}$, $\frac{n_{y}}{2}$, 1, IC) \\
			\midrule
			\multirow{4}{*}{Encoder} & \multirow{2}{*}{Dense blocks ($n_{enc}$)} & \multirow{2}{*}{K$_{enc}=32$, L$_{enc}=12$} & \multirow{2}{*}{K$_{enc}=16$, L$_{enc}=7$} & \multirow{2}{*}{K$_{enc}=16$, L$_{enc}=5$} & $\left(\frac{n_{x}}{2^{i_{enc}}}, \frac{n_{y}}{2^{i_{enc}}}, \frac{IC+(K_{enc}\times L_{enc})\cdot i_{enc}}{2^{i_{enc}-1}}\right)$ \\
			& & & & & \multicolumn{1}{r}{for $i_{enc} = 1 \ \text{to} \ n_{enc}$}\\ 
			\addlinespace
			& \multirow{2}{*}{Encodings ($n_{enc}$)} & \multicolumn{3}{c|}{\multirow{2}{*}{BN, ReLU, K1S1P0 \& BN, ReLU, K3S2P1}} & $\left(\frac{n_{x}}{2^{i_{enc}+1}}, \frac{n_{y}}{2^{i_{enc}+1}}, \frac{IC+(K_{enc}\times L_{enc})\cdot i_{enc}}{2^{i_{enc}}}\right)$ \\
			& & & & & \multicolumn{1}{r}{for $i_{enc} = 1 \ \text{to} \ n_{enc}$}\\
			\midrule
			\multirow{2}{*}{Bottleneck} & \multirow{2}{*}{Dense block} & \multirow{2}{*}{K$_{bot}=32$, L$_{bot}=5$} & \multirow{2}{*}{K$_{bot}=16$, L$_{bot}=3$} & \multirow{2}{*}{K$_{bot}=16$, L$_{bot}=3$} & $\left(\frac{n_{x}}{2^{n_{enc}+1}}, \frac{n_{y}}{2^{n_{enc}+1}}, FM_{bot}\right)$;\\
			\addlinespace
			& & & & & \multicolumn{1}{r}{$FM_{bot}=\frac{IC+(K_{enc}\times L_{enc})\cdot n_{enc}}{2^{n_{enc}}} + K_{bot}\times L_{bot}$} \\
			\midrule
			\multirow{4}{*}{Decoder} & Decodings ($n_{dec}$) & \multicolumn{3}{c|}{BN, ReLU, K1S1P0 \& BN, ReLU, Transpose K3S2P1} & $\left(\frac{n_{x}}{2^{n_{enc}-i_{dec}+1}}, \frac{n_{y}}{2^{n_{enc}-i_{dec}+1}}, \frac{FM_{bot}+(K_{dec}\times L_{dec})\cdot (i_{dec}-1)}{2^{i_{dec}}}\right)$ \\
			& & & & & \multicolumn{1}{r}{for $i_{dec} = 1 \ \text{to} \ n_{dec}$} \\
			\addlinespace
			& Dense blocks ($n_{dec}$) & K$_{dec}=32$, L$_{dec}=9$ & K$_{dec}=16$, L$_{dec}=10$ & K$_{dec}=16$, L$_{dec}=10$ & ($\frac{n_{x}}{2^{n_{enc}-i_{dec}+1}}$, $\frac{n_{y}}{2^{n_{enc}-i_{dec}+1}}$, $\frac{FM_{bot}+(K_{dec}\times L_{dec})\cdot (i_{dec}-1)}{2^{i_{dec}}}$ \\
			& & & & & \multicolumn{1}{r}{$+ (K_{dec}\times L_{dec})$) for $i_{dec} = 1 \ \text{to} \ n_{dec}$} \\ 
			\midrule
			\multicolumn{2}{c|}{Outputs} & \multicolumn{3}{c|}{1} & ($n_{x}$, $n_{y}$, 1) \\
			\bottomrule
		\end{tabular}
	\end{adjustbox}
\end{table}	

The data was split into three subsets: 80\% for training $\{\bm{x}_i, \bm{y}_i\}^{n_{\text{train}}}_{i=1}$, 10\% for testing $\{\bm{x}_i, \bm{y}_i\}^{n_{\text{test}}}_{i=1}$, and 10\% for validation $\{\bm{x}_i, \bm{y}_i\}^{n_{\text{validation}}}_{i=1}$.
To prepare data for training and testing, data transformations were applied to the datasets, including minimum-maximum scaling and Z-score standardization, as defined in \cref{eqn:min_max_scale,eqn:standardization}.
\begin{equation}
	\label{eqn:min_max_scale}
	\bm{y}_{\text{scaled}} = \frac{\bm{y} - y_{\text{min}}}{y_{\text{max}} - y_{\text{min}}},
\end{equation}
\begin{equation}
	\label{eqn:standardization}
	\bm{y}_{\text{standardized}} = \frac{\bm{y} - \overline{\bm{y}}}{\sigma_{\bm{y}}} = \frac{\bm{y} - \overline{\bm{y}}}{\sqrt{\frac{\sum_{i=1}^{n}\bm{y}_i - \overline{\bm{y}}}{n}}},
\end{equation}
where $y_{\text{min}}$, $y_{\text{max}}$, $\overline{\bm{y}}$, and $\sigma_{\bm{y}}$ are the minimum, maximum, mean, and standard deviation values of the target field $\bm{y}$ based on the training data.
Minimum-maximum scaling bounds the data between 0 and 1, while Z-score standardization results in a distribution with a mean of 0 and a standard deviation of 1.
We found that applying standardization to temperature and velocity outputs and scaling to temperature inputs with no transformations for pressure data gave the best performance for each model.
Before evaluating the performance of the models, inverse data transforms were applied to the predictions to revert them to their original scale.

\begin{figure}[H]
	\centering
	\begin{subfigure}{0.45\textwidth}
		\centering
		\includegraphics[width=0.9\textwidth]{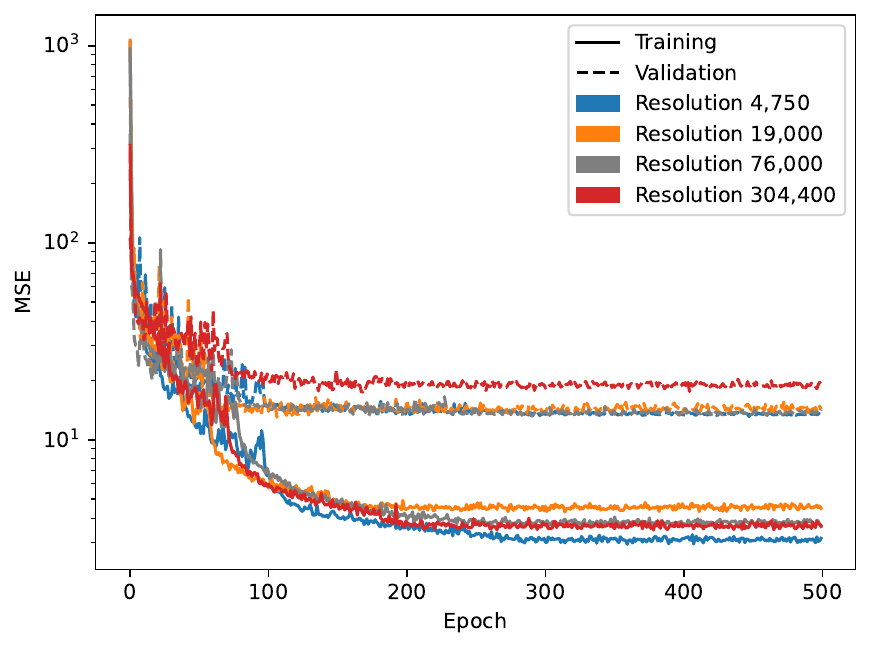}
		\caption{\label{fig:pressure_training}}
	\end{subfigure}
	\begin{subfigure}{0.45\textwidth}
		\centering
		\includegraphics[width=0.9\textwidth]{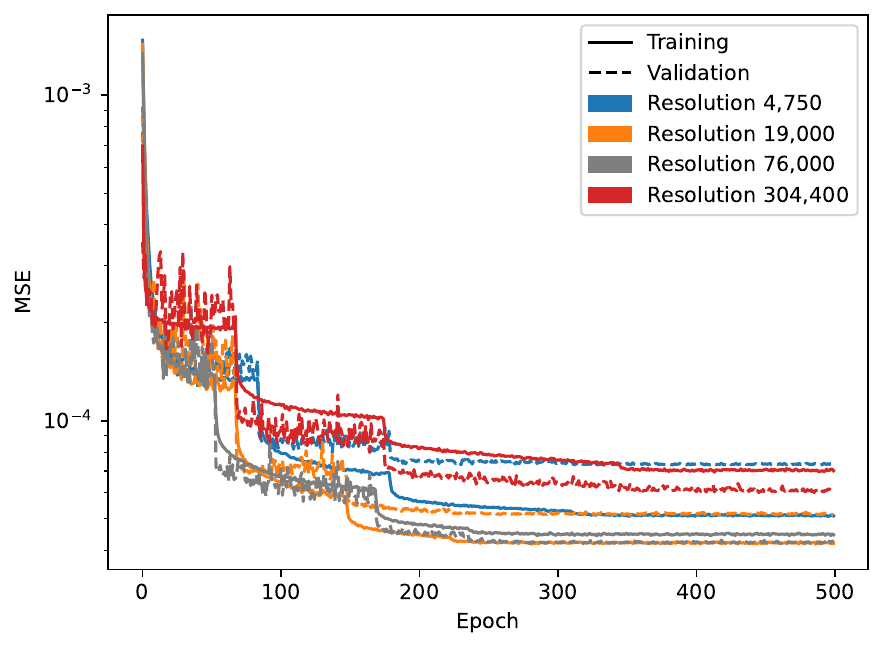}
		\caption{\label{fig:velocity_training}}
	\end{subfigure}
	\begin{subfigure}{0.45\textwidth}
		\centering
		\includegraphics[width=0.9\textwidth]{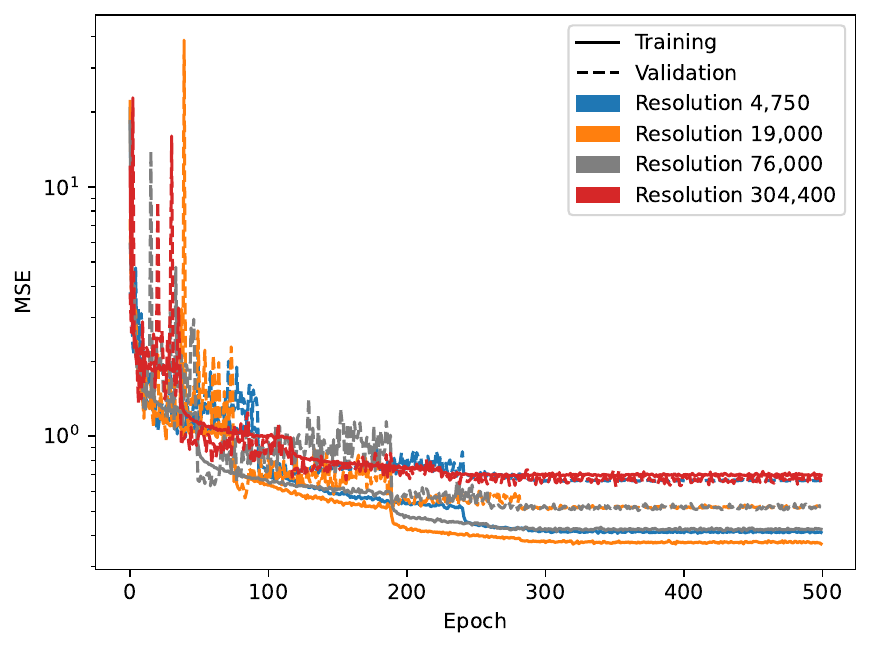}
		\caption{\label{fig:temperature_training}}
	\end{subfigure}
	\caption[Neural network training curves]{Training curves for (\ref{sub@fig:pressure_training}) DeepEDH-pressure [\unit{\pascal\squared}], (\ref{sub@fig:velocity_training}) DeepEDH-velocity [\unit{\meter\squared\per\second\squared}], and (\ref{sub@fig:temperature_training}) DeepEDH-temperature [\unit{\kelvin\squared}] models. These curves show loss computed from the validation dataset.\label{fig:training_curves}}
\end{figure}

\begin{figure}[H]
	\centering
	\includegraphics[width=0.9\textwidth]{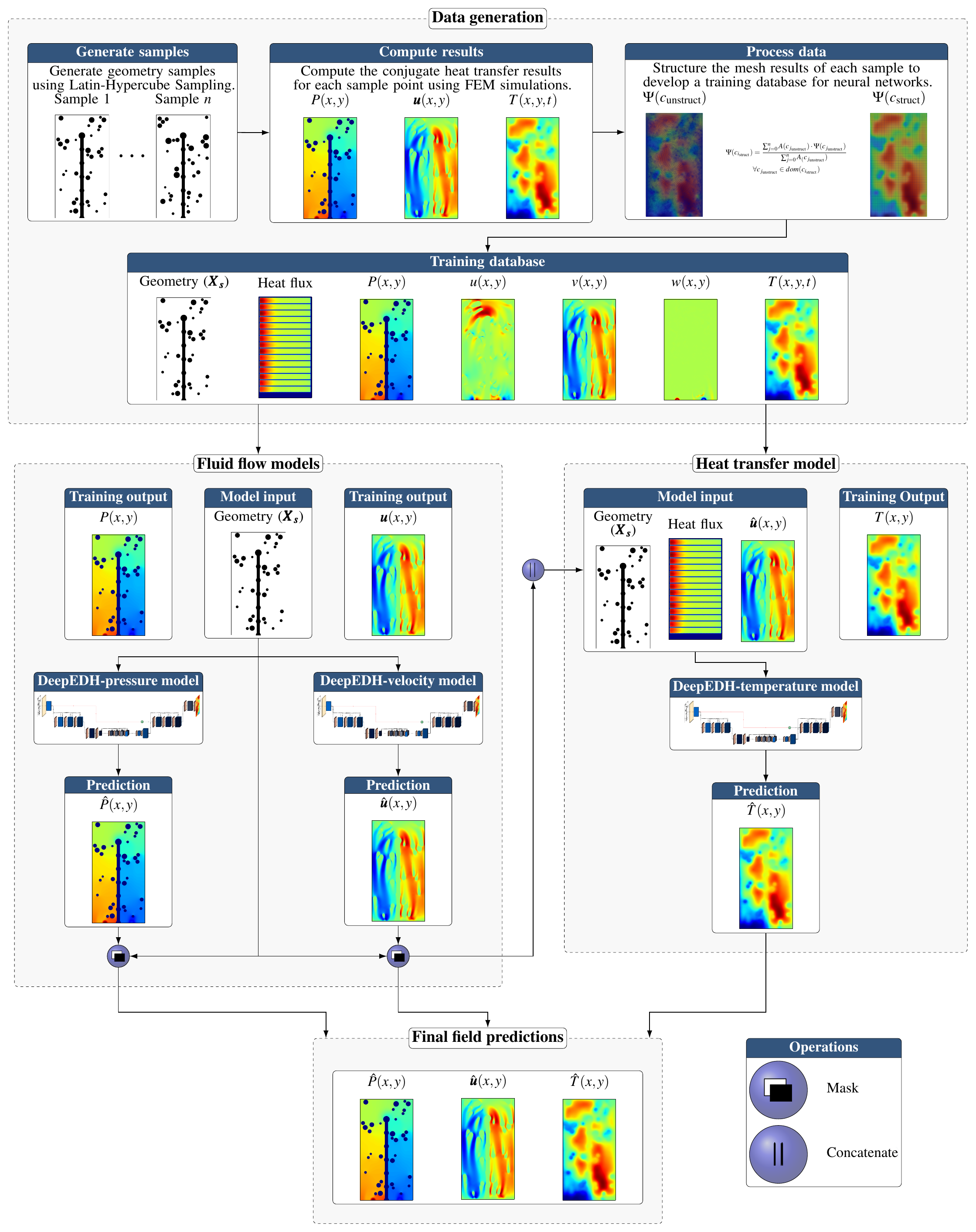}
	\caption[Surrogate methodology overview]{Overview of surrogate methodology from data generation and processing, to final field prediction with the output geometry mask and two-stage temperature architecture.\label{fig:workflow}}
\end{figure}

The training loss function (\cref{eqn:mse_loss}) was minimized by tuning the model parameters $\bm{\theta}$ from \cref{eqn:surr_model}.
Gradients of the training loss function with respect to $\bm{\theta}$ were computed through backpropagation in the DeepEDH models. 
The adaptive moment estimation (ADAM) \cite{Kingma_Ba_2014} optimization algorithm was used with an initial learning rate of $10^{-3}$, weight decay of $10^{-4}$, and batch size of 8.
A learning rate scheduler that reduced the learning rate on a loss plateau was used with a relative tolerance of $1\times10^{-4}$, decay factor of 10, and patience of 10 epochs.
This means that if the loss did not decrease relatively by $1\times10^{-4}$ for 10 epochs, the learning rate was reduced by a factor of 10.
The training process was carried out over 500 epochs, with sample learning curves provided in \cref{fig:pressure_training,fig:velocity_training,fig:temperature_training} for each model and data resolution.
\Cref{fig:pressure_training,fig:velocity_training,fig:temperature_training} indicate good convergence of the training loss for each model and do not show signs of over-fitting.
The full surrogate modeling methodology, from data generation and processing, to final field prediction is outlined in \cref{fig:workflow}.

\section{Surrogate model performance and characterization}
\label{sec:results}
We first characterized the effect of network depth and dataset size on the surrogate model performance before completing hyperparameter and architecture optimization to determine the final architectures of each DeepEDH model.
To demonstrate the advantage of the surrogate methodology proposed in this work, we compared the performance of our DeepEDH models with U-Net \cite{Ronneberger_Fischer_Brox_2015} and DenseED \cite{Zhu_Zabaras_2018} models, specifically examining the impact of each change.
We characterized the effect of the DeepEDH network architectures, output geometry masks, two-stage temperature architecture, and hyperparameter and architecture optimization.
Finally, we present the impact of the heat flux magnitude on the temperature model's performance, demonstrating excellent performance for a range of thermal boundary conditions.

\subsection{Evaluation metrics}
\label{sec:results_em}
To evaluate and characterize the performance of the surrogate models, we considered three metrics based on the test dataset $\{\bm{x}_i, \bm{y}_i\}^{n_{\text{test}}}_{i=1}$. 
The coefficient of determination ($R^{2}$), calculated using \cref{eqn:r2}, quantifies the proportion of variance in the data that the model can explain.
\begin{equation}
	\label{eqn:r2}
	R^{2} = 1 - \frac{\sum_{i=1}^{n_{\text{test}}}\norm[\big]{\bm{y}_{i} - \bm{\hat{y}}_{i}}_{2}^{2}}{\sum_{i=1}^{n_{\text{test}}}\norm[\big]{\bm{y}_{i} - \bm{\overline{y}}}_{2}^{2}}
\end{equation}
The mean of the target fields, $\bm{y}$, is $\bm{\overline{y}}$, while $\bm{\hat{y}}$ are the surrogate model outputs.
The $R^{2}$ value ranges from 0 to 1, with 1 indicating the model can explain 100\% of the variance in data while 0 shows no correlation between the model prediction and the target field.
The root mean square error (RMSE), defined by \cref{eqn:rmse}, measures the average error between the model predictions and target fields.
\begin{equation}
	\label{eqn:rmse}
	RMSE = \sqrt{\frac{1}{n_{\text{test}}}\sum_{i=1}^{n_{\text{test}}}\norm[\big]{\bm{\hat{y}}_{i}-\bm{y}_{i}}_{2}^{2}}
\end{equation}
The RMSE metric is expressed in the same units as the target field, and a lower RMSE indicates a model with a lower average error.
Spearman's rank correlation coefficient (SCC), \cref{eqn:scc}, assesses the ordered correlation between the model predictions and the target field.
\begin{equation}
	\label{eqn:scc}
	SCC = \frac{\sum_{i=1}^{n_{\text{test}}}\left(R(\bm{y}_{i}) - \overline{R(\bm{y})}\right) \cdot \left(R(\bm{\hat{y}}_{i}) - \overline{R(\bm{\hat{y}}_{i})}\right)}{\sqrt{\sum_{i=1}^{n_{\text{test}}}\left(R(\bm{y}_{i}) - \overline{R(\bm{y})}\right)^{2}} \cdot \sqrt{\sum_{i=1}^{n_{\text{test}}}\left(R(\bm{\hat{y}}_{i}) - \overline{R(\bm{\hat{y}}_{i})}\right)^{2}}}
\end{equation}
here $\bm{y}_{i}$ and $\bm{\hat{y}}_{i}$ are converted to their corresponding rank variables $R(\bm{y}_{i})$ and $R(\bm{\hat{y}}_{i})$, while the means of the rank variables are $\overline{R(\bm{y})}$ and $\overline{R(\bm{\hat{y}})}$.
The SCC value ranges from -1 to 1 and measures the monotonic relationship, with 1 indicating a positive correlation and -1 indicating an inverse correlation.
The SCC is particularly useful in applications where preserving the relative rank between alternatives is crucial, such as design optimization.
These three metrics collectively offer a comprehensive evaluation of the surrogate model's performance. 
The $R^{2}$ and SCC values provide insights into the overall regression performance and correlation, while RMSE quantifies the average prediction error.

\subsection{Characterization results}
\label{sec:results_cr}
To characterize the model performance, we examined the effect of the network depth, dataset size, data resolution, and thermal boundary condition magnitude.
All three metrics introduced in \cref{sec:results_em} show similar trends. Hence, characterization results in this section are presented with the $R^{2}$ metric, while the RMSE and SCC results are included in the supplementary materials.
Additionally, we include model field predictions for the extreme values of each characterization parameter and tabulated data for all results presented in this section in the supplementary materials.

\subsubsection{Code dimension}
\label{sec:results_cr:cd}
The term code dimension, in the context of the DeepEDH network, refers to the spatial dimension or number of pixels in the feature maps at the network's bottleneck layer.
The number of dense blocks used in the DeepEDH network and data resolution define the code dimension.
Unlike fully connected networks, where the entire input affects each unit in the network, units within a convolutional network are only dependent on a region of the input, while only a region of the output impacts each unit in the network during backpropagation. 
This region is referred to as the receptive field or field of view \cite{Luo_Li_Urtasun_Zemel_2017} and is related to the code dimension, which represents a mapping of the input geometry to the output field.

Smaller code dimensions correspond to deeper networks, mapping the input field to a lower dimension space, resulting in more information loss.
However, by implementing skip connections, this information can be retained, allowing smaller code dimensions to generate more features and capture fine-grained information from the input geometry. This results in less smooth outputs.
Conversely, these deeper networks require more parameters, which increases their computational cost and makes them more difficult to train.
In comparison, shallower networks with larger code dimensions and fewer parameters can capture only more high-level or coarse information, resulting in smoother outputs.
For the dense prediction task considered here, the code dimension must be a suitable size such that the receptive field can capture the required information from the input and output images, while limiting the network size for computational efficiency and ease of training.
The code dimension enables some physical interpretations, corresponding to the spatial resolution of input features that impact the output field. 
With a code dimension that is too large, the model cannot capture the small features from the input geometry that impact the output fields.
Hence, investigating model performance across different code dimensions allows the minimum geometry size that impacts output fields to be determined.

\begin{figure}[H] 
	\centering
	\begin{subfigure}{0.925\textwidth}
		\centering
		\includegraphics[width=0.95\textwidth]{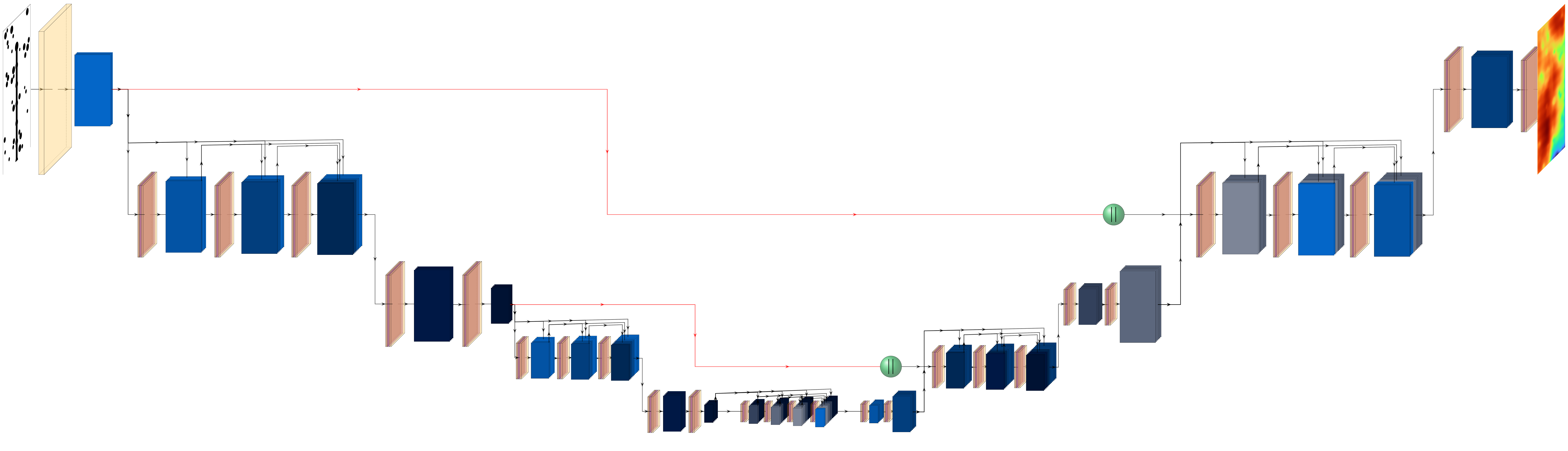}
		\caption{\label{fig:arch_5db}}
	\end{subfigure}
	\begin{subfigure}{0.925\textwidth}
		\centering
		\includegraphics[width=0.95\textwidth]{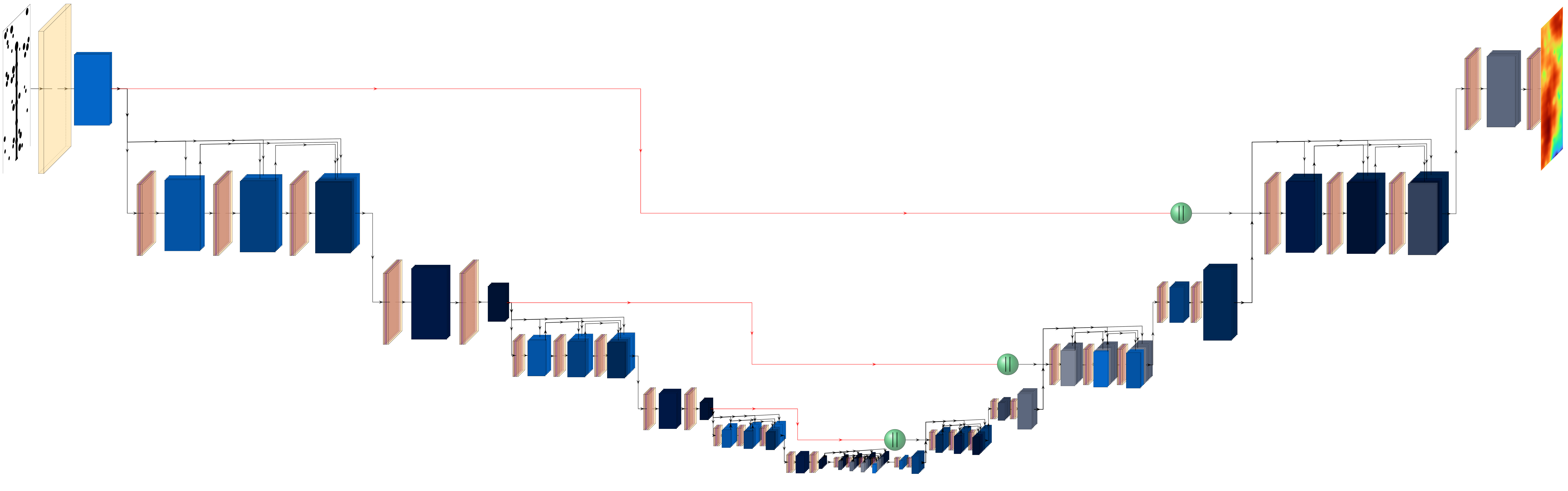}
		\caption{\label{fig:arch_7db}}
	\end{subfigure}
	\caption[Varying DeepEDH architecture depths]{DeepEDH architecture with varying depths: (\ref{sub@fig:arch_5db}) 5 dense blocks with $L_{\text{dense}} = [3, 3, 4, 3, 3]$ and $K=2$, and (\ref{sub@fig:arch_7db}) 7 dense blocks with $L_{\text{dense}} = [3, 3, 3, 4, 3, 3, 3]$ and $K=2$.\label{fig:net_arch_cd}}
\end{figure}

\begin{figure}[H]
	\centering
	\begin{subfigure}{0.35\textwidth}
		\centering
		\includegraphics[width=0.95\textwidth]{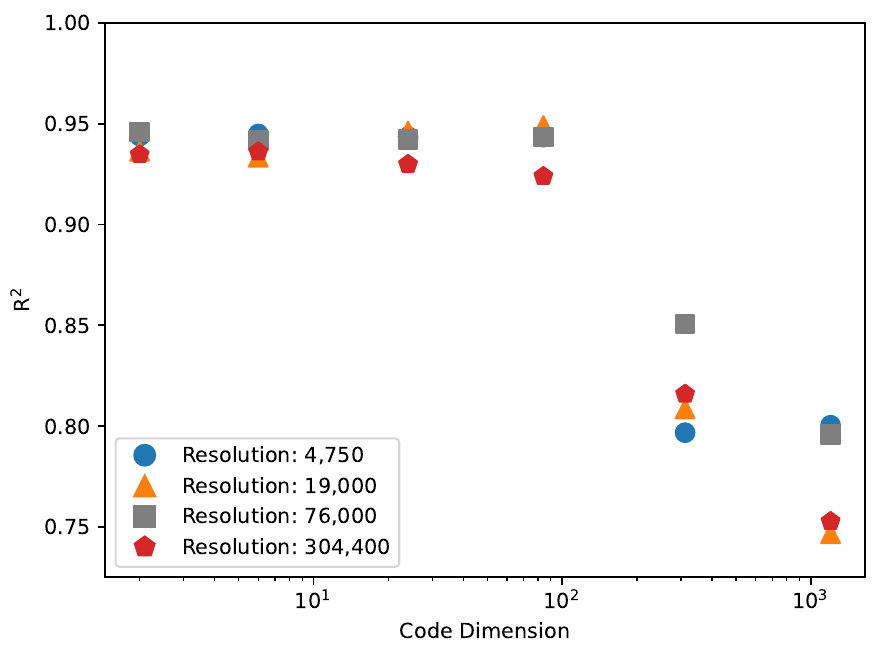}
		\caption{\label{fig:pressure_cd_r2}}
	\end{subfigure}
	\begin{subfigure}{0.35\textwidth}
		\centering
		\includegraphics[width=0.95\textwidth]{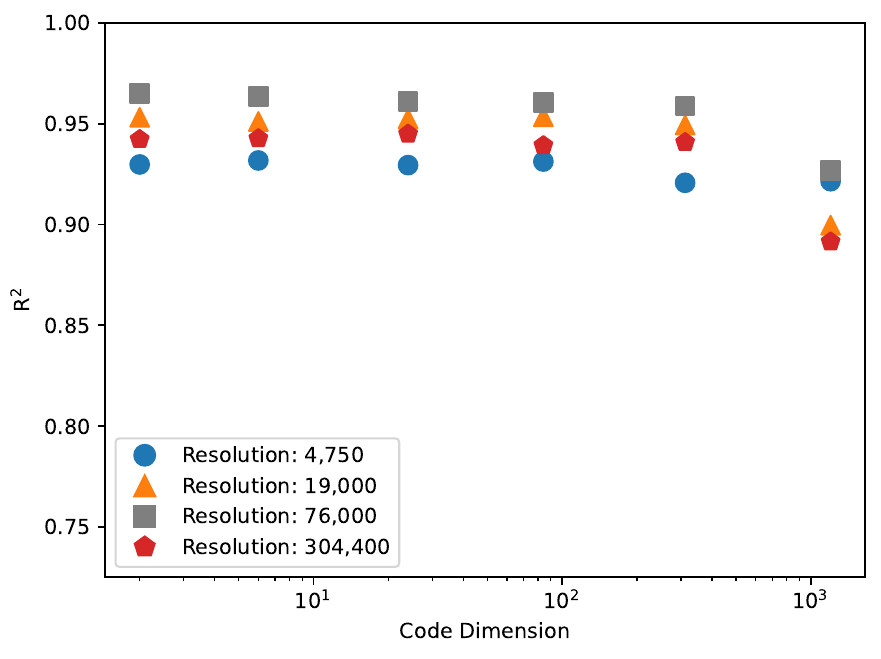}
		\caption{\label{fig:velocity_cd_r2}}
	\end{subfigure}
	\begin{subfigure}{0.35\textwidth}
		\centering
		\includegraphics[width=0.95\textwidth]{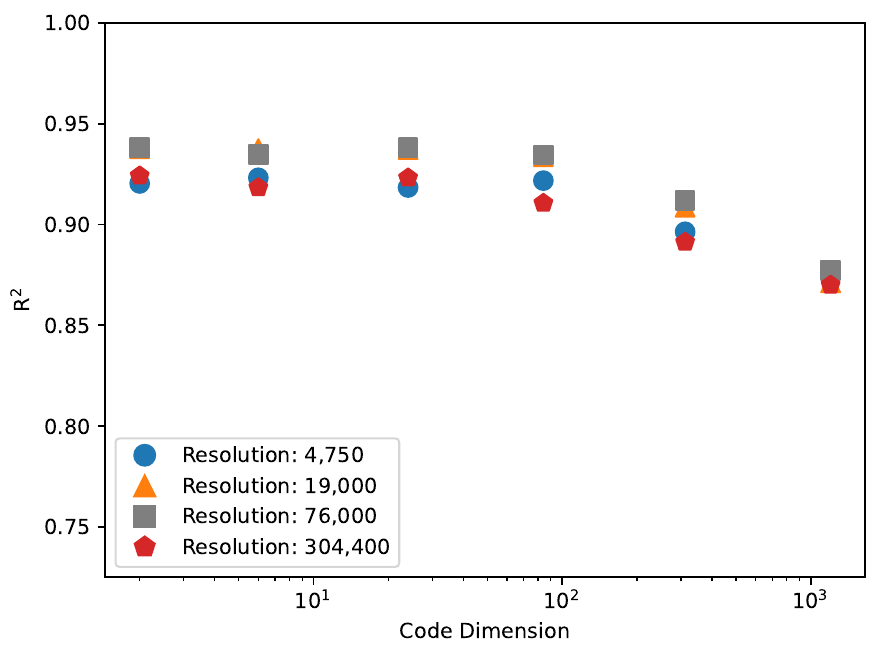}
		\caption{\label{fig:temperature_cd_r2}}
	\end{subfigure}
	\caption[Code dimension R$^{2}$ characterization]{Model code dimension characterization R$^{2}$ results for: (\ref{sub@fig:pressure_cd_r2}) DeepEDH-pressure, (\ref{sub@fig:velocity_cd_r2}) DeepEDH-velocity, and (\ref{sub@fig:temperature_cd_r2}) DeepEDH-temperature models.\label{fig:model_cd_r2}}
\end{figure}

When comparing results across different data resolutions, it is important to maintain a consistent code dimension. 
Therefore, different network depths are required for each resolution.
For example, for a 50$\times$95 data resolution with a DeepEDH architecture defined as $L_{\text{dense}} = [3, 4, 3]$, shown in \cref{fig:final_arch}, the resulting code dimension is 312 pixels with dimensions 13$\times$24.
The input size of 50$\times$95 is halved by the initial convolution layer, then halved again by the encoding layer, resulting in the corresponding code dimension of 13$\times$24.
\Cref{fig:net_arch_cd} shows DeepEDH architectures with varying depths of $L_{\text{dense}} = [3, 3, 4, 3, 3]$ in \cref{sub@fig:arch_5db} and $L_{\text{dense}} = [3, 3, 3, 4, 3, 3, 3]$ in \cref{sub@fig:arch_7db}. 
The network architectures in \cref{sub@fig:arch_5db,sub@fig:arch_7db} result in code dimensions of 84 and 24 pixels, respectively, for the 50$\times$95 data resolution.
To achieve code dimensions of 312, 84, and 24, the dataset with 400$\times$761 resolution would require 9, 11, and 13 dense blocks.
We considered code dimensions ranging from 1200 pixels, based on a single dense block for the 50$\times$95 dataset, to 2 pixels.
A summary of the code dimensions for each data resolution with the corresponding number of dense blocks is shown in \cref{tab:network_code_level}.
\begin{table}[H]
	\centering
	\caption[Network depth, data resolution, and corresponding code dimension]{Network depth, data resolution, and corresponding code dimension.\label{tab:network_code_level}}
		\begin{tabular}{c|cccc}
			\toprule
			\multirow{2}{*}{Dense blocks} & \multicolumn{4}{c}{Data resolution (cells [W$\times$H])} \\
			& 4,750 [50$\times$95] & 19,000 [100$\times$190] & 76,000 [200$\times$380] & 304,400 [400$\times$761] \\
			\midrule
			1 & 1,200 [25$\times$48] & - & - & - \\
			3 & 312 [13$\times$24] & 1,200 [25$\times$48] & - & - \\
			5 & 84 [7$\times$12] & 312 [13$\times$24] & 1,200 [25$\times$48] & - \\
			7 & 24 [4$\times$6] & 84 [7$\times$12] & 312 [13$\times$24] & 1,200 [5$\times$48] \\
			9 & 6 [2$\times$3] & 24 [4$\times$6] & 84 [7$\times$12] & 312 [13$\times$24] \\
			11 & 2 [1$\times$2] & 6 [2$\times$3] & 24 [4$\times$6] & 84 [7$\times$12] \\
			13 & - & 2 [1$\times$2] & 6 [2$\times$3] & 24 [4$\times$6] \\
			15 & - & - & 2 [1$\times$2] & 6 [2$\times$3] \\
			17 & - & - & - & 2 [1$\times$2]\\
			\bottomrule
		\end{tabular}
\end{table}	
To investigate the impact of code dimension, we used the same network architecture for each DeepEDH model, with the code dimension defined by the number of dense blocks. 
Here, architectures used encoding and decoding dense blocks with 3 layers and a bottleneck block with 6 layers; all blocks had a growth rate of 16. 
The models considered 48 initial feature maps and a dropout of 0.05.
The networks were trained for 500 epochs with a batch size of 16, learning rate of $3\times10^{-3}$, and learning rate decay of $5\times10^{-4}$.
For each data resolution, models were trained with the appropriate number of dense blocks to achieve code dimensions ranging from 1200 to 2 pixels as defined in \cref{tab:network_code_level}.
\Cref{fig:pressure_cd_r2,fig:velocity_cd_r2,fig:temperature_cd_r2} show the $R^{2}$ results for each field and data resolution.
\Cref{fig:pressure_cd_r2,fig:velocity_cd_r2,fig:temperature_cd_r2} show that the code dimension has a significant impact on the model performance - specifically at larger code dimensions, which result in poor performance for all fields and data resolutions.
This is due to the inability of the higher code dimension models to capture small input features and the resulting smooth nature of the output fields.
However, as the code dimension is reduced, model performance improves across all fields and data resolutions before plateauing at a code dimension of 24 pixels.
Based on the physical interpretation of the code dimension, this is expected: as the code dimension decreases the minimum geometry size that impacts the output fields is captured. 
Further refinement past this point does not improve model performance, as the model already captures the necessary information from the input and output fields.
For optimal performance, a code dimension small enough to capture the minimum information scale is required. However, this must be balanced against the higher computational training cost and training difficulty associated with the increased number of network parameters from smaller code dimensions.
Considering these results, future experiments use a code dimension of 24 pixels to limit the number of network parameters while maintaining model performance.

\subsubsection{Dataset size}
\label{sec:results_cr:ds}
\begin{figure}[H]
	\centering
	\begin{subfigure}{0.35\textwidth}
		\centering
		\includegraphics[width=0.95\textwidth]{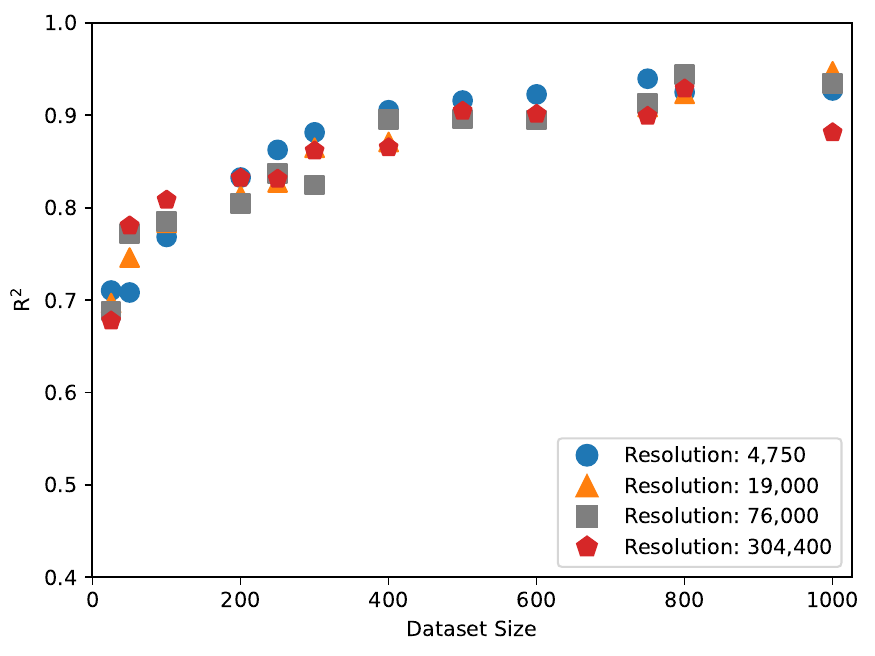}
		\caption{\label{fig:pressure_characterization_r2}}
	\end{subfigure}
	\begin{subfigure}{0.35\textwidth}
		\centering
		\includegraphics[width=0.95\textwidth]{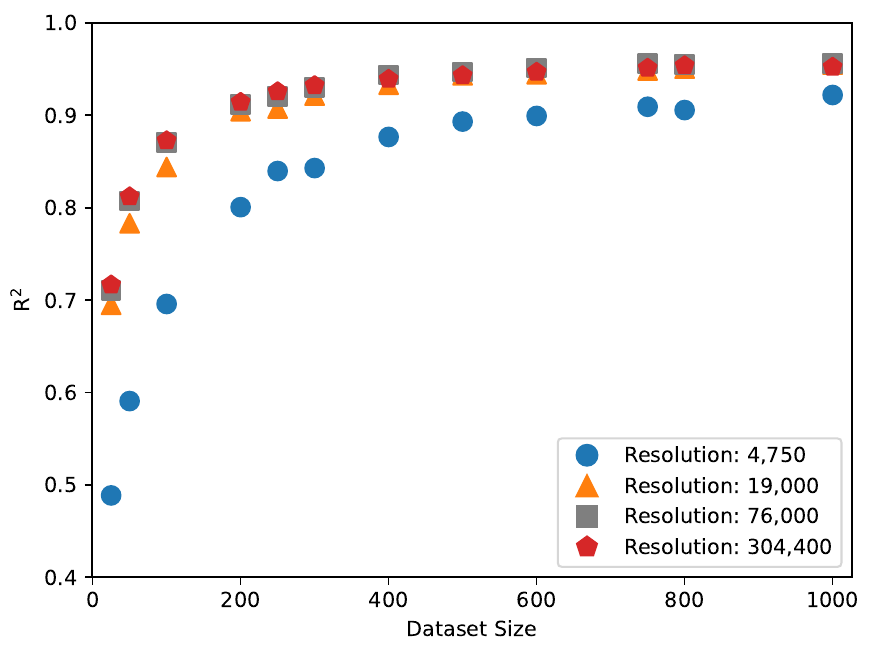}
		\caption{\label{fig:velocity_characterization_r2}}
	\end{subfigure}
	\begin{subfigure}{0.35\textwidth}
		\centering
		\includegraphics[width=0.95\textwidth]{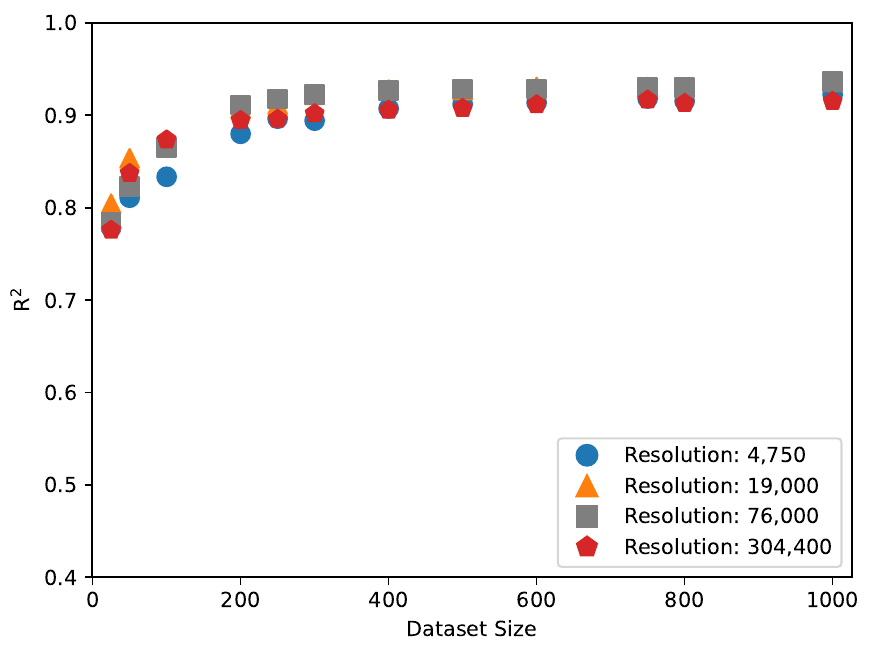}
		\caption{\label{fig:temperature_characterization_r2}}
	\end{subfigure}
	\caption[Dataset size and dimension R$^{2}$ characterization]{Model training dataset size and dimension characterization R$^{2}$ results for: (\ref{sub@fig:pressure_characterization_r2}) DeepEDH-pressure, (\ref{sub@fig:velocity_characterization_r2}) DeepEDH-velocity, and (\ref{sub@fig:temperature_characterization_r2}) DeepEDH-temperature models.\label{fig:model_characterization_r2}}
\end{figure}

Dataset size refers to the number of simulations in the dataset $\{\bm{x}_i, \bm{y}_i\}^{n_{\text{train}}}_{i=1}$ used to train the surrogate models.
For surrogate modeling, generating the dataset is typically the most computationally expensive step.
Increased dataset size also results in longer training times. However, this is generally less significant than the simulation time required to generate the dataset.
By selecting an appropriate dataset size we can reduce the computational cost of generating the dataset while maintaining good model performance.
We considered dataset sizes ranging from 25 to 1000 simulations, corresponding to approximately 1.5\% and 66\% of the total pre-generated dataset size.
\Cref{fig:pressure_characterization_r2,fig:velocity_characterization_r2,fig:temperature_characterization_r2} show the $R^{2}$ results for each field, data resolution, and dataset size.
\Cref{fig:pressure_characterization_r2,fig:velocity_characterization_r2,fig:temperature_characterization_r2} show that increasing the dataset size results in improved DeepEDH model performance with diminishing returns.
Increases from 25 to approximately 200 simulations significantly improve model performance, while performance plateaus at approximately 500 simulations.

\subsubsection{Data resolution}
\label{sec:results_cr:dr}
\begin{figure}[H]
	\centering
	\begin{subfigure}{0.35\textwidth}
		\centering
		\includegraphics[width=0.95\textwidth]{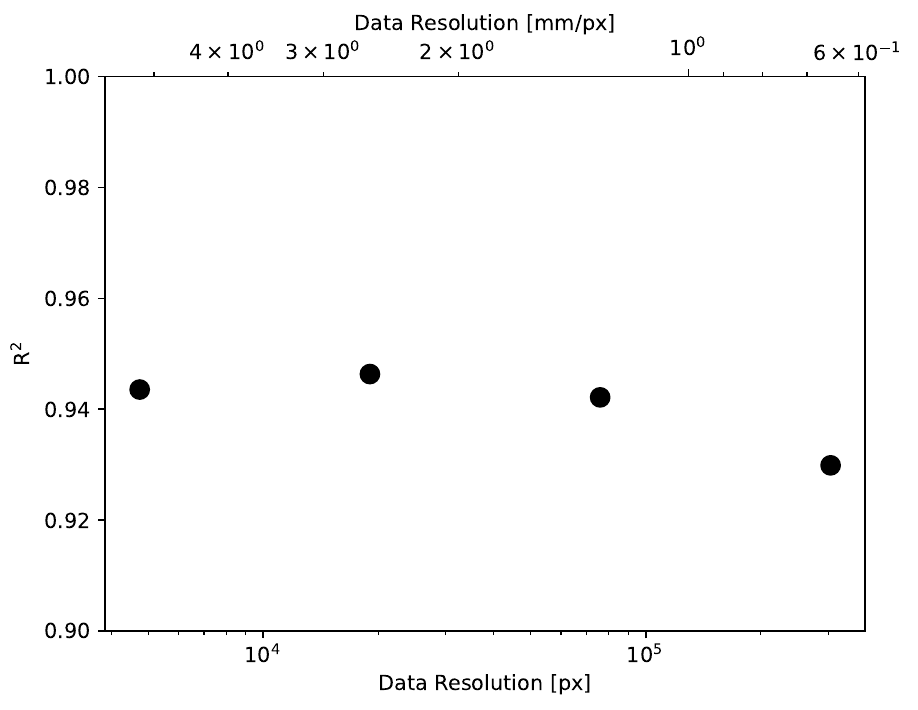}
		\caption{\label{fig:pressure_res_r2}}
	\end{subfigure}
	\begin{subfigure}{0.35\textwidth}
		\centering
		\includegraphics[width=0.95\textwidth]{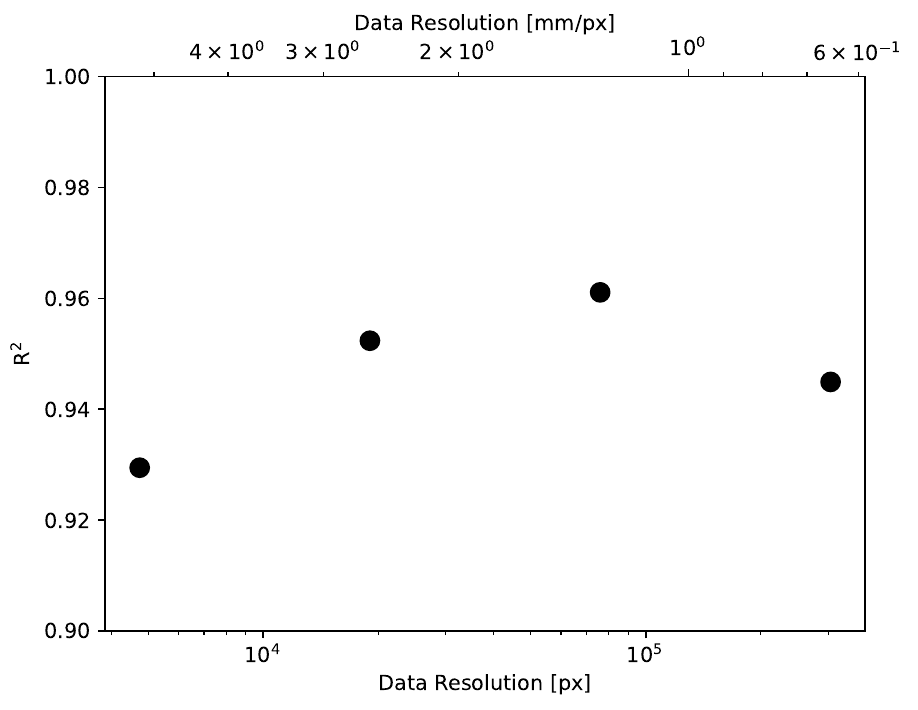}
		\caption{\label{fig:velocity_res_r2}}
	\end{subfigure}
	\begin{subfigure}{0.35\textwidth}
		\centering
		\includegraphics[width=0.95\textwidth]{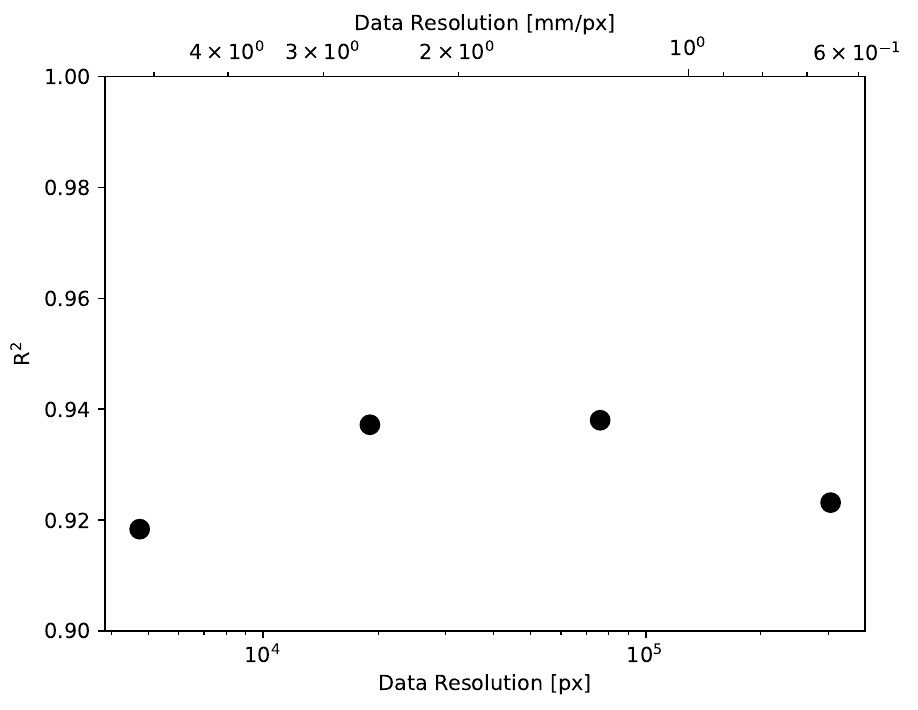}
		\caption{\label{fig:temperature_res_r2}}
	\end{subfigure}
	\caption[Data resolution R$^{2}$ characterization]{Model data resolution characterization R$^{2}$ results for: (\ref{sub@fig:pressure_res_r2}) DeepEDH-pressure, (\ref{sub@fig:velocity_res_r2}) DeepEDH-velocity, and (\ref{sub@fig:temperature_res_r2}) DeepEDH-temperature models. The bottom axis shows results for the total number of pixels, while the top shows the corresponding pixel resolutions.\label{fig:model_resolution_r2}}
\end{figure}

From the results presented in \cref{sec:results_cr:cd,sec:results_cr:ds}, we determined that a code dimension of 24 pixels and dataset size of 500 simulations resulted in good model performance while limiting the computational cost of training. 
We have considered data resolutions ranging from 50$\times$95 to 400$\times$761, corresponding to 4,750 and 304,400 pixels with dimensions of approximately 5.2~{mm/px} and 0.65~{mm/px}.
Physically, the data resolution indicates the required spatial resolution of the dataset representation necessary to accurately capture the input geometry and output field features. 
Lower resolutions can result in the loss of finer geometry and field features, whereas higher resolutions can overrefine the input and output fields, adding redundant information.
\Cref{fig:pressure_res_r2,fig:velocity_res_r2,fig:temperature_res_r2} show the $R^{2}$ results for each field and data resolution.
As the data resolution increases the computational cost of model training increases significantly and does not improve model performance.
From \cref{fig:pressure_res_r2,fig:velocity_res_r2,fig:temperature_res_r2} resolutions of 19,000 or 76,000 pixels, corresponding to 2.6~{mm/px} and 1.3~{mm/px}, show the best performance across all fields.
Lower resolutions result in poor model performance compared to full-resolution data due to the reduction of fidelity in the input and output fields, leading to unpredictable input-to-output behavior.
Particularly for the circular geometry of the pin-fins and central wall, at low structured resolutions, the geometry becomes significantly distorted.
At the other extreme, higher resolutions require many more pixel values to be predicted, increasing the difficulty of the regression task and reducing model performance.

\subsection{Overall DeepEDH model performance}
\label{sec:results_mp}
\begin{figure}[H]
	\centering
	\includegraphics[width=0.8\textwidth]{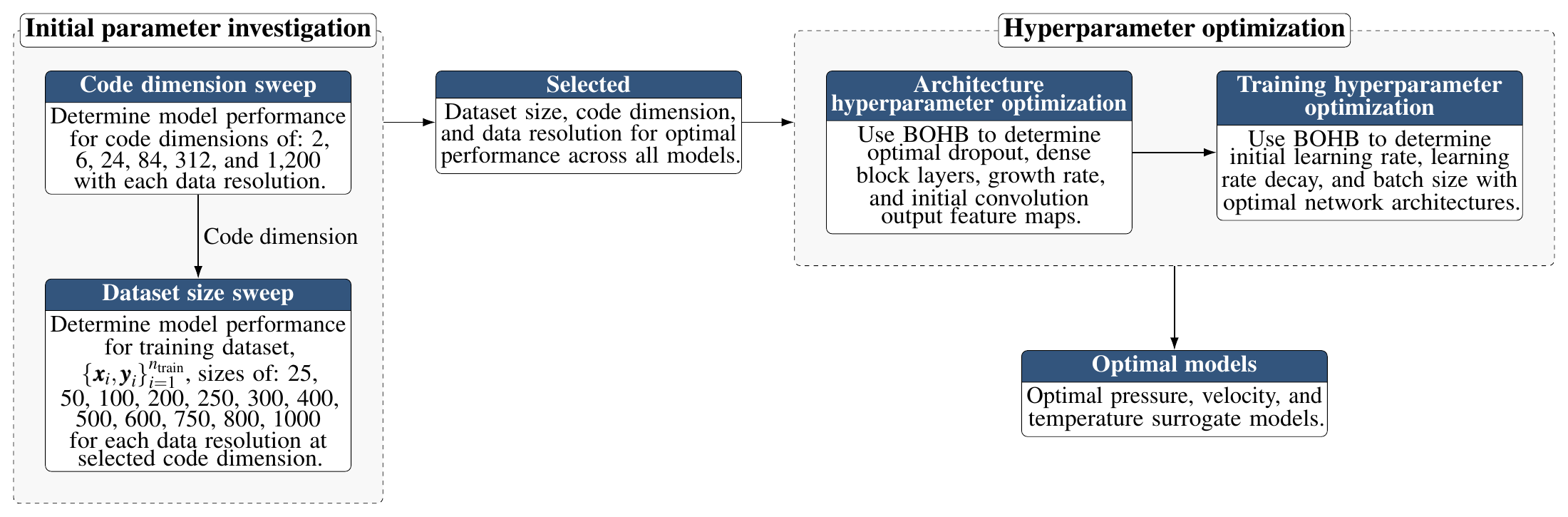}
	\caption[DeepEDH surrogate model characterization and optimization process]{Overview of DeepEDH surrogate model characterization and optimization process.\label{fig:char_workflow}}
\end{figure}

Through the DeepEDH model characterization, we determined the optimal code dimension, dataset size, and data resolution for the surrogate models, which allowed for good model performance while limiting computational cost. 
After selecting a code dimension of 24 pixels and a resolution of 19,000 pixels based on the characterization results presented in \cref{sec:results_cr}, a new hyperparameter and architecture optimization effort was conducted to determine the final architectures. 
We employed Bayesian optimization (BO) and Hyperband (HB) techniques using the BOHB algorithm \cite{Falkner_Klein_Hutter_2018} to determine the final network architectures, which are detailed in \cref{tab:network_layers}.
An overview of the characterization and optimization process is illustrated in \cref{fig:char_workflow}.

We evaluated our proposed DeepEDH surrogate models against other convolutional neural network models, including U-Net \mbox{\cite{Ronneberger_Fischer_Brox_2015}} and DenseED \mbox{\cite{Zhu_Zabaras_2018}}. 
Compared to other deep learning methods, such as fully connected neural networks, U-Net and DenseED are more suitable architectures for this work's CHT physical field predictions, allowing for more relevant benchmarking results.
For this comparison, we maintained the same training dataset size of 80\%, a code dimension of 24 pixels for both DenseED and DeepEDH, and data resolution of 19,000. 
These comparison results are included in \cref{tab:network_feature_perf}.
\begin{table}[H]
	\caption[Model performance comparison]{DeepEDH model performance with various features compared to U-Net \cite{Ronneberger_Fischer_Brox_2015} and DenseED \cite{Zhu_Zabaras_2018}. Each column considers an added feature, where features in previous columns are also included. For example the \textit{DeepEDH optimized} column includes output geometry masking for flow fields, multi-stage temperature architecture, and optimized hyperparameters.\label{tab:network_feature_perf}}
	\begin{adjustbox}{width=1\columnwidth, center}
		\centering
		\begin{tabular}{cc|c|c|cccc}
			\toprule
			\multirow{3}{*}{Model} & \multirow{3}{*}{Metric} & \multicolumn{6}{c}{Architecture} \\ 
			\cmidrule{3-8}
			& & \multirow{2}{*}{U-Net \cite{Ronneberger_Fischer_Brox_2015}} & \multirow{2}{*}{DenseED \cite{Zhu_Zabaras_2018}} & \multicolumn{4}{c}{DeepEDH} \\
			\cmidrule{5-8}
			& & & & Base & Output geometry mask & Multi-stage & Hyperparameter optimized \\
			\midrule
			\multirow{3}{*}{Pressure} & RMSE [\unit{\pascal}] & 4.9571 & 5.6956 & 3.6105 & 3.5472 & - & 3.3031 \\
			& R$^2$ & 0.8578 & 0.8123 & 0.9464 & 0.9465 & - & 0.9550 \\
			& SCC & 0.9929 & 0.9855 & 0.9941 & 0.9951 & - & 0.9958 \\
			\midrule
			\multirow{3}{*}{Velocity} & RMSE [\unit{\meter\per\second}] & 0.01932 & 0.01443 & 0.007154 & 0.007115 & - & 0.006054 \\
			& R$^2$ & 0.6302 & 0.7938 & 0.9524 & 0.9524 & - & 0.9657 \\
			& SCC & 0.8591 & 0.9250 & 0.9807 & 0.9808 & - & 0.9845 \\
			\midrule
			\multirow{3}{*}{Temperature} & RMSE [\unit{\kelvin}] & 1.8649 & 1.4476 & 0.8922 & - & 0.7127 & 0.6889 \\
			& R$^2$ & 0.5701 & 0.7409 & 0.9015 & - & 0.9372 & 0.9413 \\
			& SCC & 0.9185 & 0.9506 & 0.9823 & - & 0.9888 & 0.9891 \\
			\bottomrule
		\end{tabular}
	\end{adjustbox}
\end{table}
The performance comparison presented in \cref{tab:network_feature_perf} demonstrates that DeepEDH outperforms U-Net and DenseED for all models.
Compared to DenseED, DeepEDH results in 18\%, 22\%, and 27\% improvements for R$^2$ of pressure, velocity, and temperature predictions respectively.
Similarly, R$^2$ improvements of 11\%, 53\%, and 65\% for pressure, velocity, and temperature predictions were achieved compared to U-Net.
The largest improvements are for the temperature predictions due to the coupled model architecture.
Model parameters and training time for each network are summarized in \cref{tab:network_train_perf}. 
The training times are based on training completed using a Nvidia GeForce RTX 3080 GPU with 10~GB of GPU memory and 64~GB of RAM.
From \cref{tab:network_train_perf}, training time of DeepEDH is approximately 26\% less than U-Net and requires 91\% fewer parameters. 
Compared to DenseED, DeepEDH uses individual networks for each model and requires sequential training for the temperature model, resulting in roughly doubled training time and tripled parameter count.
\begin{table}[H]
	\caption[Model parameters and training time]{Model parameters and training time for U-Net \cite{Ronneberger_Fischer_Brox_2015}, DenseED \cite{Zhu_Zabaras_2018}, and DeepEDH.\label{tab:network_train_perf}}
		\centering
		\begin{tabular}{lcc}
			\toprule
			Architecture & Parameters & Training time [s] \\ 
			\midrule
			U-Net \cite{Ronneberger_Fischer_Brox_2015} & 31,037,187 & 4,872 \\
			DenseED \cite{Zhu_Zabaras_2018} & 593,489 & 1,886 \\
			DeepEDH & 2,684,391 & 3,596 \\ 
			\bottomrule
		\end{tabular}
\end{table}

\begin{figure}[H]
	\centering
    \begin{subfigure}[T]{0.075\textwidth}
		\centering
		\includegraphics[height=10cm]{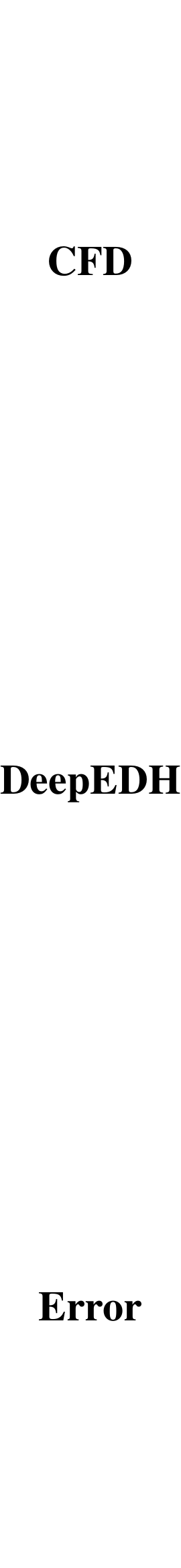}
	\end{subfigure}%
	\begin{subfigure}[T]{0.3\textwidth}
		\centering
		\includegraphics[height=10cm]{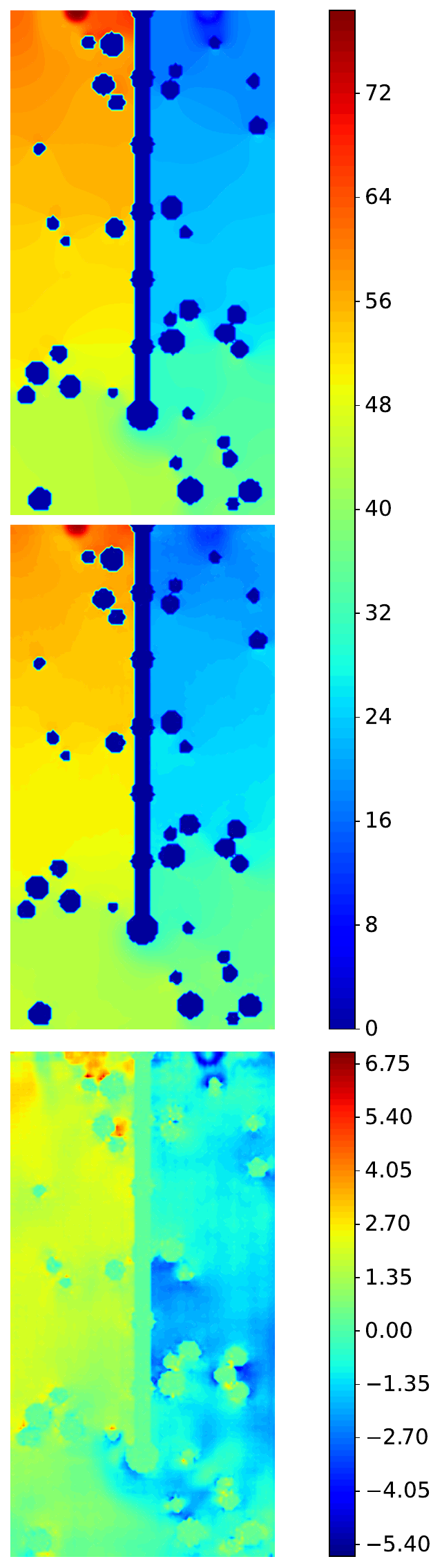}
		\caption{\label{fig:pressure_results}}
	\end{subfigure}
	\begin{subfigure}[T]{0.3\textwidth}
		\centering
		\includegraphics[height=10cm]{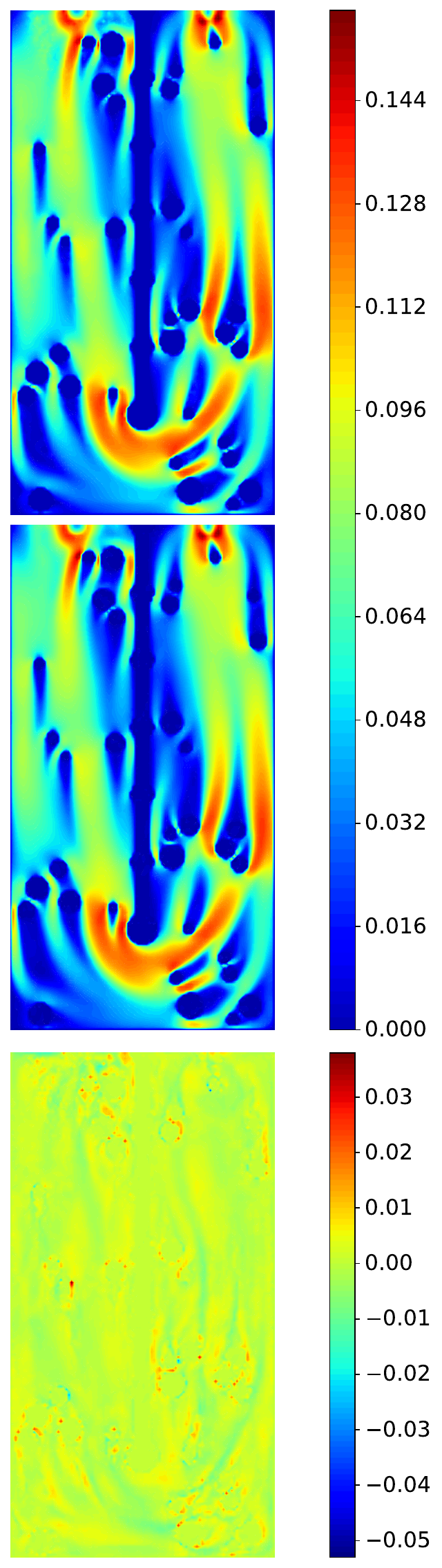}
		\caption{\label{fig:velocity_results}}
	\end{subfigure}
	\begin{subfigure}[T]{0.3\textwidth}
		\centering
		\includegraphics[height=10cm]{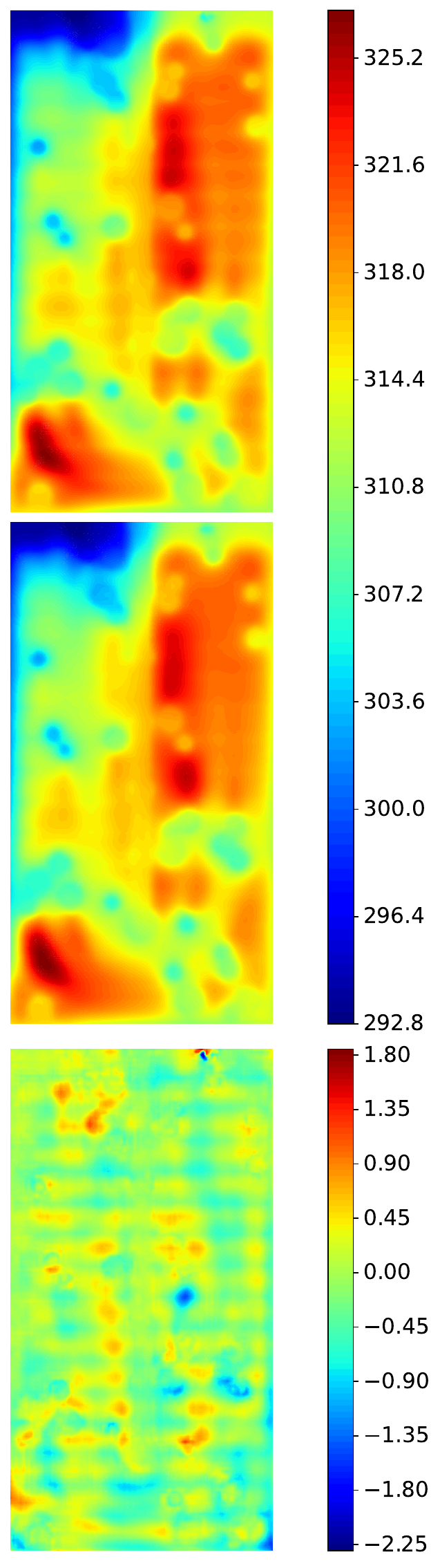}
		\caption{\label{fig:temperature_results}}
	\end{subfigure}
	\caption[Optimized model predictions]{Optimized model predictions: (\ref{sub@fig:pressure_results}) DeepEDH-pressure [\unit{\pascal}], (\ref{sub@fig:velocity_results}) DeepEDH-velocity [\unit{\meter\per\second}], and (\ref{sub@fig:temperature_results}) DeepEDH-temperature [\unit{\kelvin}] fields. Within each subfigure, the first row shows the target (CFD) fields, while the second row shows the model (DeepEDH) predictions, and the third row shows the pixel-level error.\label{fig:result_fields}}
\end{figure}

\mbox{\Cref{fig:pressure_results,fig:velocity_results,fig:temperature_results}} show the target (CFD) fields, surrogate model (DeepEDH) predictions, and pixel-level error fields for the best-performing DeepEDH models, while \mbox{\cref{tab:best_model_perf}} shows the tabulated performance metrics.
Qualitatively the results shown in \cref{fig:pressure_results,fig:velocity_results,fig:temperature_results} show that the model is able to capture the general trends of the field, while the metrics presented in \cref{tab:network_feature_perf} show that the model is able to capture the field with high accuracy and low error.

\begin{table}[H]
	\caption[Best model performance]{Best DeepEDH model performance metrics.\label{tab:best_model_perf}}
		\centering
		\begin{tabular}{cccc}
			\toprule
			Model & R$^{2}$ & RMSE & SCC\\
			\midrule
			DeepEDH-pressure & 0.9550 & 3.3073~\unit{\pascal} & 0.9958 \\
			\addlinespace
			DeepEDH-velocity & 0.9657 & 0.006071~\unit{\meter\per\second} & 0.9845 \\
			\addlinespace
			DeepEDH-temperature & 0.9413 & 0.6889~\unit{\kelvin} & 0.9892 \\
			\bottomrule
		\end{tabular}
\end{table}

\subsection{DeepEDH-temperature model performance at varying heat fluxes}
\label{sec:results_hf}
This section quantifies the impact of the heat flux boundary condition on the performance of the DeepEDH-temperature model.
We varied the heat flux by scaling the input heat flux field by factors of 1/10, 1/5, 2/5, 2, 5, and 10.
The accumulated heat was calculated using \cref{eqn:accumulated_heat} for each heat flux scaling factor.
DeepEDH-temperature performance is summarized in \cref{fig:hf_perf}, with tabulated results included in the supplementary materials.
\begin{equation}
	\label{eqn:accumulated_heat}
	Q_{\text{accumulated}} = \int_{0}^{t_{\text{final}}} \left[\int_{\partial \Omega} q_{0}(x,y,t) dS\right] dt - \int_{0}^{t_{\text{final}}} \dot{m}C_{p}(T_{\text{outlet}}(t)-T_{\text{inlet}}(t)) dt
\end{equation}
\begin{figure}[H]
	\centering
	\includegraphics[width=0.35\textwidth]{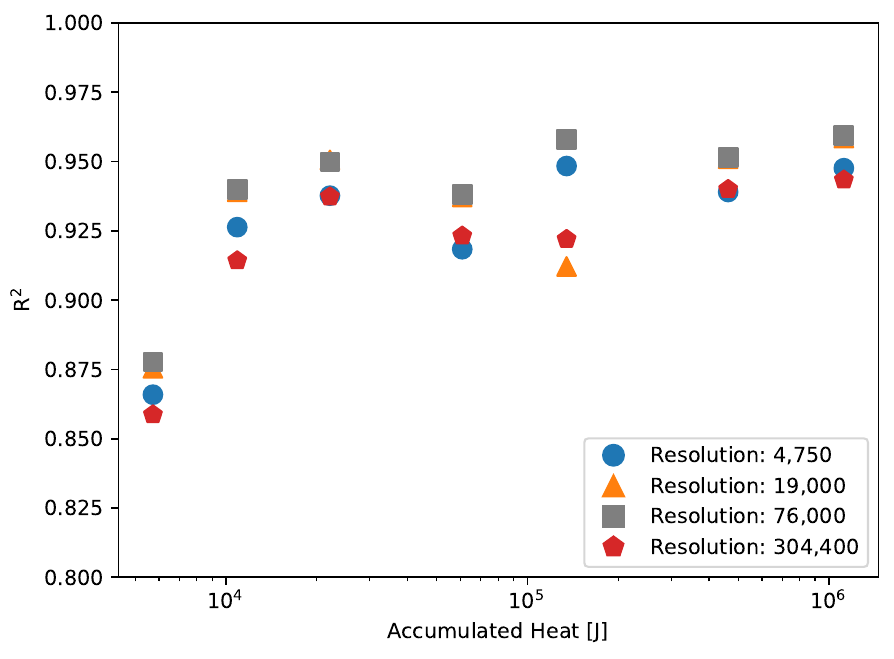}
	\caption[Temperature model performance for a varying heat flux boundary condition]{DeepEDH-temperature model R$^{2}$ performance for varying heat flux boundary conditions.\label{fig:hf_perf}}
\end{figure}

\begin{figure}[H]
	\centering
    \begin{subfigure}[T]{0.075\textwidth}
		\centering
		\includegraphics[height=10cm]{figure_labels.pdf}
	\end{subfigure}
	\begin{subfigure}[T]{0.18\textwidth}
		\centering
		\includegraphics[height=10cm]{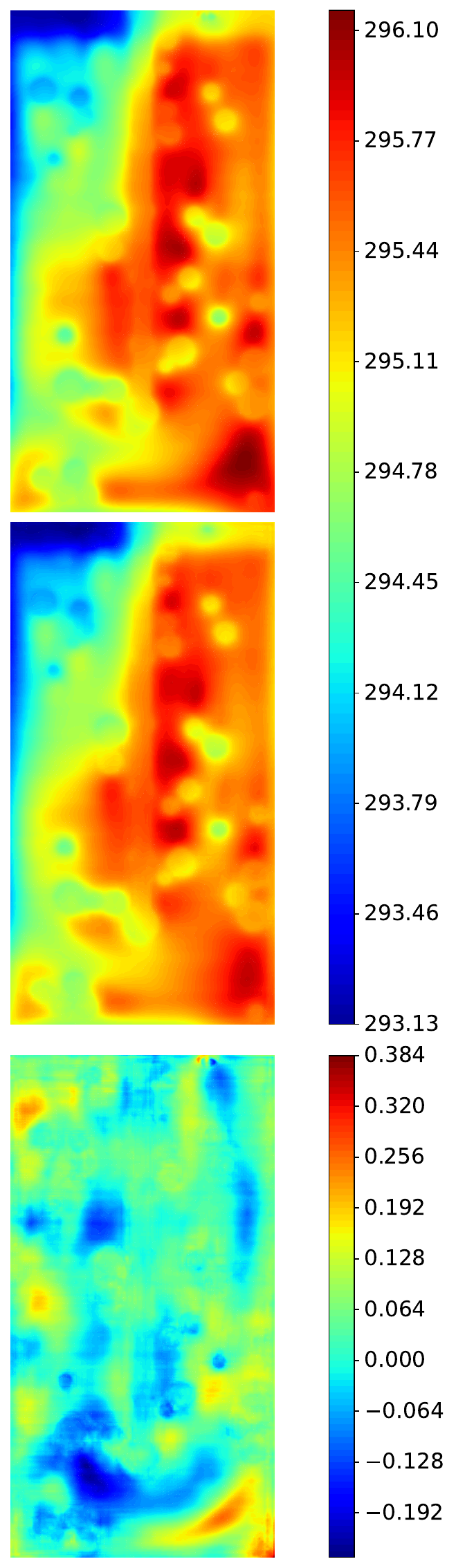}
		\caption{\label{fig:temp_05hf}}
	\end{subfigure}
	\begin{subfigure}[T]{0.18\textwidth}
		\centering
		\includegraphics[height=10cm]{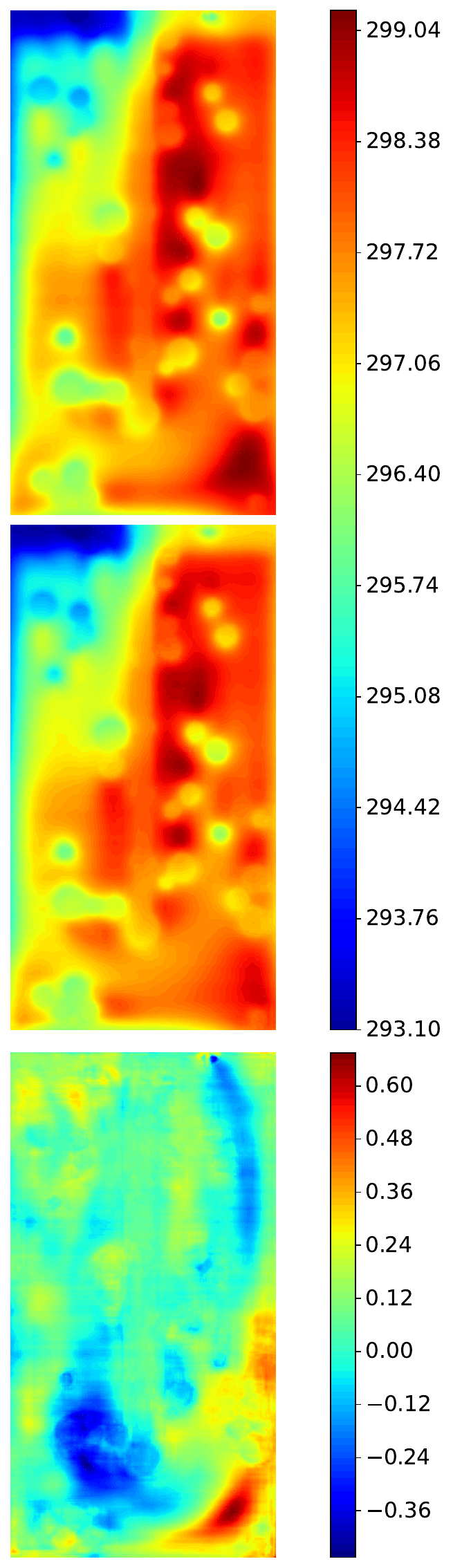}
		\caption{\label{fig:temp_1hf}}
	\end{subfigure}
	\begin{subfigure}[T]{0.18\textwidth}
		\centering
		\includegraphics[height=10cm]{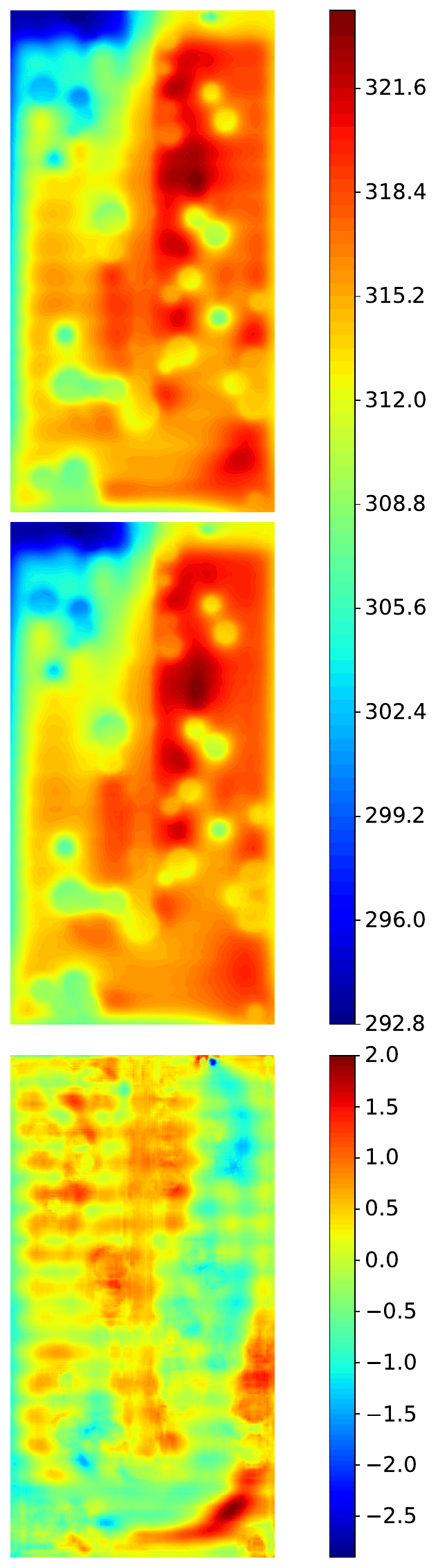}
		\caption{\label{fig:temp_5hf}}
	\end{subfigure}
	\begin{subfigure}[T]{0.18\textwidth}
		\centering
		\includegraphics[height=10cm]{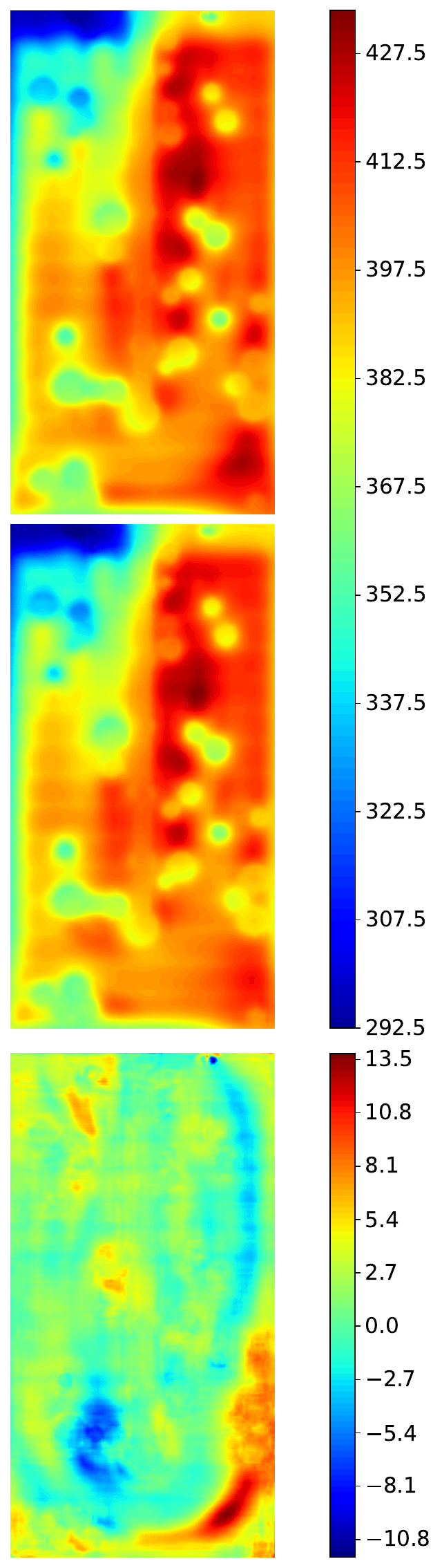}
		\caption{\label{fig:temp_25hf}}
	\end{subfigure}
	\begin{subfigure}[T]{0.18\textwidth}
		\centering
		\includegraphics[height=10cm]{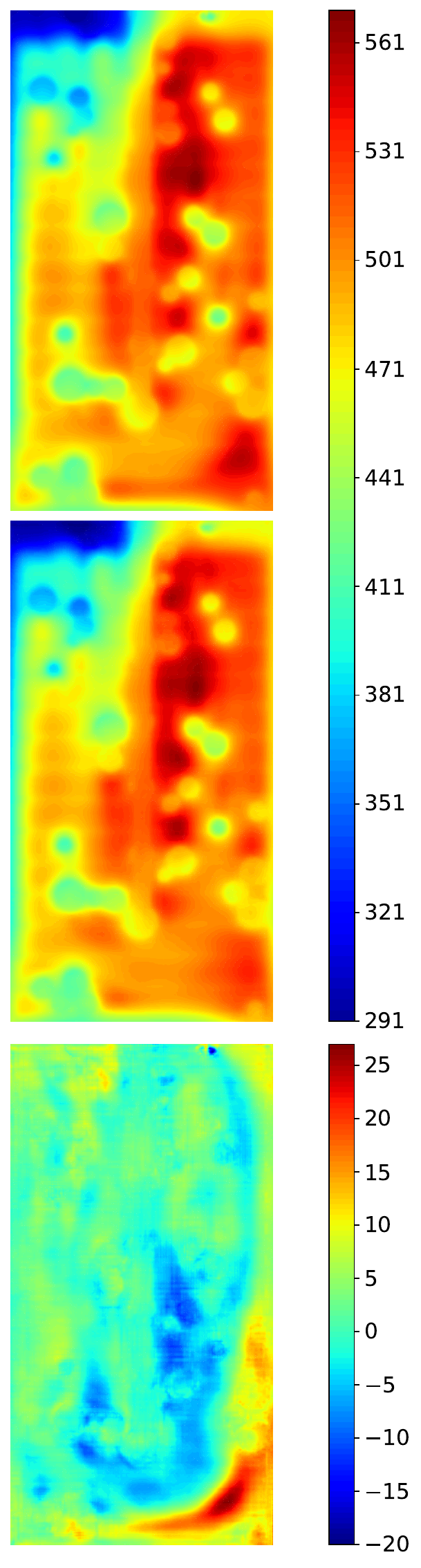}
		\caption{\label{fig:temp_50hf}}
	\end{subfigure}
	\caption[Temperature field predictions for a varying heat flux boundary condition]{Temperature [\unit{\kelvin}] field predictions with heat flux factors of: (\ref{sub@fig:temp_05hf}) 1/10, (\ref{sub@fig:temp_1hf}) 1/5, (\ref{sub@fig:temp_5hf}) 1, (\ref{sub@fig:temp_25hf}) 5, and (\ref{sub@fig:temp_50hf}) 10. Within each subfigure, the first row shows the target (CFD) fields, while the second row shows the model (DeepEDH) predictions, and the third row shows the pixel-level error. Note the difference in scale for the pixel-level error results. \label{fig:hf_characterization}}
\end{figure}

\Cref{fig:hf_perf} shows that the DeepEDH-temperature model performance improves and then plateaus as heat flux increases.
At lower heat fluxes, the accumulated heat is limited, resulting in a temperature profile with lower variation. 
For example, consider \cref{fig:temp_05hf,fig:temp_1hf,fig:temp_5hf,fig:temp_25hf,fig:temp_50hf} that show the target (CFD) fields, surrogate model (DeepEDH) predictions, and the pixel-level error fields for a sample point in the test dataset.
\Cref{fig:temp_05hf,fig:temp_1hf,fig:temp_5hf,fig:temp_25hf,fig:temp_50hf} show the limited variation of the temperature field at low heat fluxes, while at higher values the effect of geometry is more clear. 
It is important to note the difference in scale for the pixel-level error results in \cref{fig:temp_05hf,fig:temp_1hf,fig:temp_5hf,fig:temp_25hf,fig:temp_50hf}.
Physically, this indicates that the cold plate geometry has a limited impact on the temperature field at lower heat fluxes; in contrast, the temperature field is more sensitive to the cold plate geometry at higher heat fluxes. 
Interpreting these results allows for the identification of the level of heating required for changes in the cold plate geometry to significantly impact the temperature field and, hence, the cold plate's thermal performance. 
With heat fluxes below this limit, changes in the cold plate geometry will have a more limited impact.

\subsection{Application feasibility}
\label{sec:results_af}
The results presented in this work demonstrate the effectiveness of the DeepEDH surrogate modeling methodology for CHT analyses. 
This section examines the feasibility of applying these models to common surrogate modeling tasks, including design optimization, real-time control systems, and real-time digital twin applications. 
For design optimization, thousands of objective evaluations are required to evaluate design candidates during the optimization process. 
By using DeepEDH models to replace the FEM numerical models, the optimization computational time can be reduced by approximately 97\% as summarized in \cref{tab:opt_times}.
For real-time control or digital twin applications, the computational resources and execution time required to run the FEM CHT simulations in real time are often infeasible.
DeepEDH requires limited computational resources and can be executed in less than 1 s, as shown in \cref{tab:opt_times}.
Compared to the FEM models, the storage and memory requirements for the DeepEDH models are also significantly reduced by approximately 98\% and 92\%, respectively. 
Overall, these results demonstrate the feasibility of using DeepEDH surrogate models for these typical surrogate applications.

\begin{table}[H]
	\caption[Optimization computational time comparison]{Application feasibility summary based on the models used in the work, comparing computational time and memory requirements. The computational time estimates are based on executing 50 FEM simulations in parallel. Memory estimates are based on disk space, COMSOL memory, and PyTorch memory usage.\label{tab:opt_times}}
	\centering
	\begin{tabular}{lcc}
		\toprule
		& Finite element model & DeepEDH model \\
		\midrule
		Single model evaluation [s] & 1,250 & 0.5 \\
		Dataset generation (\num{1500} samples) [s] & - & 37,500 \\
		DeepEDH training [s] & - & 3,600 \\
		Optimization (1000 evaluations) [s] & 1,250,000 & 500 \\
		Complete optimization [s] & 1,250,000 ($\approx$14 days) & 41,600 ($\approx$0.5 days) \\
		Model storage [GB] & 2 & 0.05 \\
		Execution memory [GB] & 48 & 4 \\
		\bottomrule
	\end{tabular}
\end{table}
The values presented in \cref{tab:opt_times} are approximate based on the FEM simulations, DeepEDH training, and DeepEDH execution times.
Further, the times included in \cref{tab:opt_times} are based on executing 50 numerical models in parallel, as completed in this work.
The database generation for surrogate model training can be executed in parallel as the simulations are independent of each other. 
In contrast, for optimization, future simulations depend on the previous simulations’ results and must be executed sequentially.

\section{Summary and conclusions}
\label{sec:concl_sm}
In this work, we developed a data-driven surrogate modeling methodology for conjugate heat transfer (CHT) analyses, featuring a novel image-to-image regression convolutional neural network architecture called DeepEDH.
This architecture is a novel modular deep encoder-decoder hierarchical convolutional neural network that uses dense blocks at each layer, along with skip connections between the encoding and decoding sections.
We introduced output geometry masks, multi-stage temperature architectures, and separate models for each field to enhance the surrogate model performance for CHT analyses.

We demonstrated the effectiveness of the proposed DeepEDH methodology modeling a liquid-cooled pin-fin cold plate for a battery thermal management system application.
The field-specific DeepEDH surrogate models, DeepEDH-pressure, DeepEDH-velocity, and DeepEDH-temperature, were trained and tested using a dataset of 1,500 transient CHT finite element method simulations with varying pin-fin geometries.
These numerical results were processed from unstructured meshes into image-like data with resolutions ranging from 50$\times$95 to 400$\times$761 pixels.
The DeepEDH model performances were characterized for varying code dimensions, dataset sizes, and data resolutions.
The results highlighted the significant impact of code dimension, with larger code dimensions resulting in poor performance due to the smoothness of the output fields.
As the code dimension was reduced, the models' performances improved significantly before plateauing at a code dimension of 4$\times$6 (24) pixels.
Similarly, increasing the dataset size also improved model performances, but with diminishing returns and performance plateauing at approximately 500 simulations.
Data resolution had a limited impact on the model performances compared to the code dimension and dataset size.
Regardless, a compressed resolution of 100$\times$190 (19,000) or 200$\times$380 (76,000) pixels, corresponding to physical sizes of 2.6{mm/px} and 1.3~{mm/px}, had the best performance across all models.
Optimal network hyperparameters and architectures were determined through Bayesian optimization (BO) and Hyperband (HB) using the BOHB algorithm.

Our proposed DeepEDH convolutional neural network was demonstrated to be an effective surrogate modeling methodology for CHT analyses, outperforming U-Net and DenseED for all pressure, velocity, and temperature field predictions. 
The methodology and architecture components introduced in this work, including the output geometry mask, multi-stage temperature architecture, and hyperparameter optimization, all contributed to the improved performance of the DeepEDH models. 
Finally, we quantified the impact of the heat flux boundary condition on the temperature model performance. 
Decreased temperature field prediction performance for low heat flux values suggests that the temperature field is less sensitive to cold plate geometry under such conditions. 
The coupling methodology introduced here for the temperature model yielded excellent results, improving performance by up to 65\% compared to other architectures, offering the potential for application in other multiphysics surrogate modeling tasks. 

The methodology presented here could enable advancements in the design of systems governed by CHT, with applications in the aerospace, automotive, and electronics industries. 
The most significant challenges facing surrogate modeling for CHT problems are the computational cost of generating the dataset, the complexity of the models required to capture the relevant physics, and obtaining accurate full-field predictions.
The proposed DeepEDH methodology addresses these challenges by reducing the required dataset size and improving model performance and accuracy of the full-field predictions. 
Future works can use the DeepEDH methodology and the characterization and hyperparameter optimization techniques presented here to develop optimal surrogate models for other CHT systems. 
The optimal hyperparameters and architectures determined in this work can be used as a starting point for future surrogate modeling tasks, reducing the computational cost of developing new models.
Finally, the modular blocks and hierarchical architecture of the DeepEDH architecture can inform other neural network architecture designs, with the potential to extend to other systems governed by coupled physics, including electro-thermal or electro-magnetic surrogate modeling tasks.

Future directions for this work include expanding DeepEDH models to predict field results at all time steps and different cross-sections of three-dimensional geometries. 
Examining different system conditions, such as turbulent flow or conjugate radiative heat transfer, could demonstrate the DeepEDH models' effectiveness for different applications.
In addition, the introduction of more advanced neural network architecture features and techniques such as physics-based constraints, cross-domain generalization, or adaptive loss functions could further improve model performance and reduce the required dataset size.

%% file: backmatter.tex


\section*{Data availability}
\addcontentsline{toc}{section}{Data availability}
The code developed and used for this study is available in the GitHub repository:
\\
\href{https://github.com/takiahebbspicken/conjugate-heat-transfer-surrogate-modeling}{takiahebbspicken/conjugate-heat-transfer-surrogate-modeling}

\section*{Acknowledgments}
\addcontentsline{toc}{section}{Acknowledgments}
This work was supported by the Natural Sciences and Engineering Research Council of Canada (NSERC), MITACS, and the Ford Motor Company of Canada Ltd.
Simulations were facilitated by CMC Microsystems software licences and performed on the SciNet high performance computing clusters supported by the Canada Foundation for Innovation.

%% file: main-edsm.bbl
\begin{thebibliography}{10}

\bibitem{John_Senthilkumar_Sadasivan_2019}
B.~John, P.~Senthilkumar, and S.~Sadasivan, ``Applied and theoretical aspects of conjugate heat transfer analysis: A review,'' {\em Archives of Computational Methods in Engineering}, vol.~26, no.~2, p.~475–489, 2019.

\bibitem{EbbsPicken_Silva_Amon_2023}
T.~Ebbs-Picken, C.~M.~D. Silva, and C.~H. Amon, ``Design optimization methodologies applied to battery thermal management systems: A review,'' {\em Journal of Energy Storage}, vol.~67, p.~107460, 2023.

\bibitem{Hachem_Ghraieb_Viquerat_Larcher_Meliga_2021}
E.~Hachem, H.~Ghraieb, J.~Viquerat, A.~Larcher, and P.~Meliga, ``Deep reinforcement learning for the control of conjugate heat transfer,'' {\em Journal of Computational Physics}, vol.~436, p.~110317, 2021.

\bibitem{Naseri_Gil_Barbu_Cetkin_Yarimca_Jensen_Larsen_Gomes_2023}
F.~Naseri, S.~Gil, C.~Barbu, E.~Cetkin, G.~Yarimca, A.~Jensen, P.~Larsen, and C.~Gomes, ``Digital twin of electric vehicle battery systems: Comprehensive review of the use cases, requirements, and platforms,'' {\em Renewable and Sustainable Energy Reviews}, vol.~179, p.~113280, 2023.

\bibitem{Raissi_Perdikaris_Karniadakis_2019}
M.~Raissi, P.~Perdikaris, and G.~Karniadakis, ``Physics-informed neural networks: A deep learning framework for solving forward and inverse problems involving nonlinear partial differential equations,'' {\em Journal of Computational Physics}, vol.~378, p.~686–707, 2019.

\bibitem{Meng_Karniadakis_2020}
X.~Meng and G.~E. Karniadakis, ``A composite neural network that learns from multi-fidelity data: Application to function approximation and inverse pde problems,'' {\em Journal of Computational Physics}, vol.~401, p.~109020, 2020.

\bibitem{Jin_Cai_Li_Karniadakis_2021}
X.~Jin, S.~Cai, H.~Li, and G.~E. Karniadakis, ``Nsfnets (navier-stokes flow nets): Physics-informed neural networks for the incompressible navier-stokes equations,'' {\em Journal of Computational Physics}, vol.~426, p.~109951, 2021.

\bibitem{Natale_Svetozarevic_Heer_Jones_2022}
L.~D. Natale, B.~Svetozarevic, P.~Heer, and C.~Jones, ``Physically consistent neural networks for building thermal modeling: Theory and analysis,'' {\em Applied Energy}, vol.~325, p.~119806, 2022.

\bibitem{Yu_Yang_Yue_2011}
K.~Yu, X.~Yang, and Z.~Yue, ``Aerodynamic and heat transfer design optimization of internally cooling turbine blade based different surrogate models,'' {\em Structural and Multidisciplinary Optimization}, vol.~44, no.~1, p.~75–83, 2011.

\bibitem{Maakala_Jarvinen_Vuorinen_2018}
V.~Maakala, M.~J\"arvinen, and V.~Vuorinen, ``Optimizing the heat transfer performance of the recovery boiler superheaters using simulated annealing, surrogate modeling, and computational fluid dynamics,'' {\em Energy}, vol.~160, p.~361–377, 2018.

\bibitem{Ren_Gao_Ma_Zhang_Sun_2024}
H.~Ren, D.-c. Gao, Z.~Ma, S.~Zhang, and Y.~Sun, ``Data-driven surrogate optimization for deploying heterogeneous multi-energy storage to improve demand response performance at building cluster level,'' {\em Applied Energy}, vol.~356, p.~122312, 2024.

\bibitem{Fiore_Koloszar_Mendez_Duponcheel_Bartosiewicz_2022}
M.~Fiore, L.~Koloszar, M.~A. Mendez, M.~Duponcheel, and Y.~Bartosiewicz, ``Turbulent heat flux modelling in forced convection flows using artificial neural networks,'' {\em Nuclear Engineering and Design}, vol.~399, p.~112005, 2022.

\bibitem{Antonio_Afonso_2011}
C.~C.~a. Ant\'onio and C.~Afonso, ``Air temperature fields inside refrigeration cabins: A comparison of results from cfd and ann modelling,'' {\em Applied Thermal Engineering}, vol.~31, no.~6–7, p.~1244–1251, 2011.

\bibitem{Ozsunar_Arcaklioglu_Dur_2009}
A.~Ozsunar, E.~Arcakl{\i}oglu, and F.~N. Dur, ``The prediction of maximum temperature for single chips’ cooling using artificial neural networks,'' {\em Heat and Mass Transfer}, vol.~45, no.~4, p.~443–450, 2009.

\bibitem{Kargar_Ghasemi_Aminossadati_2011}
A.~Kargar, B.~Ghasemi, and S.~M. Aminossadati, ``An artificial neural network approach to cooling analysis of electronic components in enclosures filled with nanofluids,'' {\em Journal of Electronic Packaging}, vol.~133, no.~1, p.~011010, 2011.

\bibitem{Ben-Nakhi_Mahmoud_Mahmoud_2008}
A.~Ben-Nakhi, M.~A. Mahmoud, and A.~M. Mahmoud, ``Inter-model comparison of cfd and neural network analysis of natural convection heat transfer in a partitioned enclosure,'' {\em Applied Mathematical Modelling}, vol.~32, no.~9, p.~1834–1847, 2008.

\bibitem{Varol_Avci_Koca_Oztop_2007}
Y.~Varol, E.~Avci, A.~Koca, and H.~F. Oztop, ``Prediction of flow fields and temperature distributions due to natural convection in a triangular enclosure using adaptive-network-based fuzzy inference system (anfis) and artificial neural network (ann),'' {\em International Communications in Heat and Mass Transfer}, vol.~34, no.~7, p.~887–896, 2007.

\bibitem{Mahmoud_Ben-Nakhi_2007}
M.~A. Mahmoud and A.~E. Ben-Nakhi, ``Neural networks analysis of free laminar convection heat transfer in a partitioned enclosure,'' {\em Communications in Nonlinear Science and Numerical Simulation}, vol.~12, no.~7, p.~1265–1276, 2007.

\bibitem{Wang_Zhou_Yang_Huang_2022}
Q.~Wang, W.~Zhou, L.~Yang, and K.~Huang, ``Comparison between conventional and deep learning-based surrogate models in predicting convective heat transfer performance of u-bend channels,'' {\em Energy and AI}, vol.~8, p.~100140, 2022.

\bibitem{Jiang_Durlofsky_2023}
S.~Jiang and L.~J. Durlofsky, ``Use of multifidelity training data and transfer learning for efficient construction of subsurface flow surrogate models,'' {\em Journal of Computational Physics}, vol.~474, p.~111800, 2023.

\bibitem{ZHANG2020115552}
J.~Zhang and X.~Zhao, ``A novel dynamic wind farm wake model based on deep learning,'' {\em Applied Energy}, vol.~277, p.~115552, 2020.

\bibitem{ZHANG2022121747}
J.~Zhang and X.~Zhao, ``Wind farm wake modeling based on deep convolutional conditional generative adversarial network,'' {\em Energy}, vol.~238, p.~121747, 2022.

\bibitem{Bai_Kolter_Koltun_2018}
S.~Bai, J.~Z. Kolter, and V.~Koltun, ``An empirical evaluation of generic convolutional and recurrent networks for sequence modeling,'' {\em arXiv}, 2018.

\bibitem{tc-16-1447-2022}
F.~Madaeni, K.~Chokmani, R.~Lhissou, S.~Homayouni, Y.~Gauthier, and S.~Tolszczuk-Leclerc, ``Convolutional neural network and long short-term memory models for ice-jam predictions,'' {\em The Cryosphere}, vol.~16, no.~4, pp.~1447--1468, 2022.

\bibitem{DEHGHANI2023102119}
A.~Dehghani, H.~M. Z.~H. Moazam, F.~Mortazavizadeh, V.~Ranjbar, M.~Mirzaei, S.~Mortezavi, J.~L. Ng, and A.~Dehghani, ``Comparative evaluation of lstm, cnn, and convlstm for hourly short-term streamflow forecasting using deep learning approaches,'' {\em Ecological Informatics}, vol.~75, p.~102119, 2023.

\bibitem{Yang_Min_Yue_Rao_Chyu_2019}
L.~Yang, Z.~Min, T.~Yue, Y.~Rao, and M.~K. Chyu, ``High resolution cooling effectiveness reconstruction of transpiration cooling using convolution modeling method,'' {\em International Journal of Heat and Mass Transfer}, vol.~133, p.~1134–1144, 2019.

\bibitem{Jin_Cheng_Chen_Li_2018}
X.~Jin, P.~Cheng, W.-L. Chen, and H.~Li, ``Prediction model of velocity field around circular cylinder over various reynolds numbers by fusion convolutional neural networks based on pressure on the cylinder,'' {\em Physics of Fluids}, vol.~30, no.~4, p.~047105, 2018.

\bibitem{He_Wang_Lu_Chai_Yang_2024}
N.~He, Q.~Wang, Z.~Lu, Y.~Chai, and F.~Yang, ``Early prediction of battery lifetime based on graphical features and convolutional neural networks,'' {\em Applied Energy}, vol.~353, p.~122048, 2024.

\bibitem{Tikka_Haapaniemi_Raisanen_Honkapuro_2022}
V.~Tikka, J.~Haapaniemi, O.~R\"ais\"anen, and S.~Honkapuro, ``Convolutional neural networks in estimating the spatial distribution of electric vehicles to support electricity grid planning,'' {\em Applied Energy}, vol.~328, p.~120124, 2022.

\bibitem{Guo_Li_Iorio_2016}
X.~Guo, W.~Li, and F.~Iorio, ``Convolutional neural networks for steady flow approximation,'' {\em Proceedings of the 22nd ACM SIGKDD International Conference on Knowledge Discovery and Data Mining}, p.~481–490, 2016.

\bibitem{Bhatnagar_Afshar_Pan_Duraisamy_Kaushik_2019}
S.~Bhatnagar, Y.~Afshar, S.~Pan, K.~Duraisamy, and S.~Kaushik, ``Prediction of aerodynamic flow fields using convolutional neural networks,'' {\em Computational Mechanics}, vol.~64, no.~2, p.~525–545, 2019.

\bibitem{Zhou_Li_Qi_Zhao_Yi_2024}
W.~Zhou, X.~Li, Z.~Qi, H.~Zhao, and J.~Yi, ``A shale gas production prediction model based on masked convolutional neural network,'' {\em Applied Energy}, vol.~353, p.~122092, 2024.

\bibitem{Peng_Liu_Xia_Aubry_Chen_Wu_2021}
J.-Z. Peng, X.~Liu, Z.-D. Xia, N.~Aubry, Z.~Chen, and W.-T. Wu, ``Data-driven modeling of geometry-adaptive steady heat convection based on convolutional neural networks,'' {\em Fluids}, vol.~6, no.~12, p.~436, 2021.

\bibitem{Ronneberger_Fischer_Brox_2015}
O.~Ronneberger, P.~Fischer, and T.~Brox, ``U-net: Convolutional networks for biomedical image segmentation,'' {\em arXiv}, 2015.

\bibitem{Hua_Yu_Peng_Wu_He_Zhou_2022}
Y.~Hua, C.-H. Yu, J.-Z. Peng, W.-T. Wu, Y.~He, and Z.-F. Zhou, ``Thermal performance estimation of nanofluid-filled finned absorber tube using deep convolutional neural network,'' {\em Applied Sciences}, vol.~12, no.~21, p.~10883, 2022.

\bibitem{Ma_Zhang_Haidn_Thuerey_Hu_2020}
H.~Ma, Y.-x. Zhang, O.~J. Haidn, N.~Thuerey, and X.-y. Hu, ``Supervised learning mixing characteristics of film cooling in a rocket combustor using convolutional neural networks,'' {\em Acta Astronautica}, vol.~175, p.~11–18, 2020.

\bibitem{He_Zhang_Ren_Sun_2016}
K.~He, X.~Zhang, S.~Ren, and J.~Sun, ``Deep residual learning for image recognition,'' {\em 2016 IEEE Conference on Computer Vision and Pattern Recognition (CVPR)}, p.~770–778, 2016.

\bibitem{NIPS2015_215a71a1}
R.~K. Srivastava, K.~Greff, and J.~Schmidhuber, ``Training very deep networks,'' in {\em Advances in Neural Information Processing Systems} (C.~Cortes, N.~Lawrence, D.~Lee, M.~Sugiyama, and R.~Garnett, eds.), vol.~28, Curran Associates, Inc., 2015.

\bibitem{Huang_Liu_Maaten_Weinberger_2016}
G.~Huang, Z.~Liu, L.~v.~d. Maaten, and K.~Q. Weinberger, ``Densely connected convolutional networks,'' {\em arXiv}, 2016.

\bibitem{Hua_Yu_Zhao_Li_Wu_Wu_2023}
Y.~Hua, C.-H. Yu, Q.~Zhao, M.-G. Li, W.-T. Wu, and P.~Wu, ``Surrogate modeling of heat transfers of nanofluids in absorbent tubes with fins based on deep convolutional neural network,'' {\em International Journal of Heat and Mass Transfer}, vol.~202, p.~123736, 2023.

\bibitem{Tang_Liu_Durlofsky_2020}
M.~Tang, Y.~Liu, and L.~J. Durlofsky, ``A deep-learning-based surrogate model for data assimilation in dynamic subsurface flow problems,'' {\em Journal of Computational Physics}, vol.~413, p.~109456, 2020.

\bibitem{10.2118/203924-MS}
{\em {History Matching Complex 3D Systems Using Deep-Learning-Based Surrogate Flow Modeling and CNN-PCA Geological Parameterization}}, vol.~Day 1 Tue, October 26, 2021 of {\em SPE Reservoir Simulation Conference}, 10 2021.

\bibitem{Wen_Tang_Benson_2021}
G.~Wen, M.~Tang, and S.~M. Benson, ``Towards a predictor for co2 plume migration using deep neural networks,'' {\em International Journal of Greenhouse Gas Control}, vol.~105, p.~103223, 2021.

\bibitem{Zhu_Zabaras_2018}
Y.~Zhu and N.~Zabaras, ``Bayesian deep convolutional encoder–decoder networks for surrogate modeling and uncertainty quantification,'' {\em Journal of Computational Physics}, vol.~366, p.~415–447, 2018.

\bibitem{Jegou_2017_CVPR_Workshops}
S.~Jegou, M.~Drozdzal, D.~Vazquez, A.~Romero, and Y.~Bengio, ``The one hundred layers tiramisu: Fully convolutional densenets for semantic segmentation,'' in {\em Proceedings of the IEEE Conference on Computer Vision and Pattern Recognition (CVPR) Workshops}, July 2017.

\bibitem{Mo_Zhu_Zabaras_Shi_Wu_2019}
S.~Mo, Y.~Zhu, N.~Zabaras, X.~Shi, and J.~Wu, ``Deep convolutional encoder-decoder networks for uncertainty quantification of dynamic multiphase flow in heterogeneous media,'' {\em Water Resources Research}, vol.~55, no.~1, p.~703–728, 2019.

\bibitem{Romero_Hasanpoor_Antonini_Amon_2024}
D.~A. Romero, S.~Hasanpoor, E.~G.~A. Antonini, and C.~H. Amon, ``Predicting wind farm wake losses with deep convolutional hierarchical encoder–decoder neural networks,'' {\em APL Machine Learning}, vol.~2, no.~1, p.~016111, 2024.

\bibitem{EbbsPicken_Silva_Amon_2024}
T.~Ebbs-Picken, C.~M.~D. Silva, and C.~H. Amon, ``Multi-objective design optimization of pin-fin cold plates for electric vehicle battery packs using convolutional neural networks and genetic algorithms,'' in {\em 2024 23nd IEEE Intersociety Conference on Thermal and Thermomechanical Phenomena in Electronic Systems (ITherm)}, pp.~1--10, 2024, to appear.

\bibitem{EbbsPicken_Silva_Amon_2024_ate}
T.~Ebbs-Picken, C.~M.~D. Silva, and C.~H. Amon, ``Hierarchical thermal modeling and surrogate-model-based design optimization framework for cold plates used in battery thermal management systems,'' {\em Applied Thermal Engineering}, 2024, accepted.

\bibitem{Al-Zareer_Silva_Amon_2021}
M.~Al-Zareer, C.~D. Silva, and C.~H. Amon, ``Predicting anisotropic thermophysical properties and spatially distributed heat generation rates in pouch lithium-ion batteries,'' {\em Journal of Power Sources}, vol.~510, p.~230362, 2021.

\bibitem{Hochreiter_1998}
S.~Hochreiter, ``The vanishing gradient problem during learning recurrent neural nets and problem solutions,'' {\em International Journal of Uncertainty, Fuzziness and Knowledge-Based Systems}, vol.~6, no.~02, p.~107–116, 1998.

\bibitem{Ioffe_Szegedy_2015}
S.~Ioffe and C.~Szegedy, ``Batch normalization: Accelerating deep network training by reducing internal covariate shift,'' {\em arXiv}, 2015.

\bibitem{Glorot2010UnderstandingTD}
X.~Glorot and Y.~Bengio, ``Understanding the difficulty of training deep feedforward neural networks,'' in {\em International Conference on Artificial Intelligence and Statistics}, 2010.

\bibitem{Long_Shelhamer_Darrell_2015}
J.~Long, E.~Shelhamer, and T.~Darrell, ``Fully convolutional networks for semantic segmentation,'' {\em 2015 IEEE Conference on Computer Vision and Pattern Recognition (CVPR)}, p.~431–440, 2015.

\bibitem{Badrinarayanan_Kendall_Cipolla_2015}
V.~Badrinarayanan, A.~Kendall, and R.~Cipolla, ``Segnet: A deep convolutional encoder-decoder architecture for image segmentation,'' {\em IEEE Transactions on Pattern Analysis and Machine Intelligence}, vol.~39, no.~12, p.~2481–2495, 2015.

\bibitem{Lin_Dollar_Girshick_He_Hariharan_Belongie_2017}
T.-Y. Lin, P.~Doll\'ar, R.~Girshick, K.~He, B.~Hariharan, and S.~Belongie, ``Feature pyramid networks for object detection,'' {\em 2017 IEEE Conference on Computer Vision and Pattern Recognition (CVPR)}, p.~936–944, 2017.

\bibitem{Falkner_Klein_Hutter_2018}
S.~Falkner, A.~Klein, and F.~Hutter, ``Bohb: Robust and efficient hyperparameter optimization at scale,'' {\em arXiv}, 2018.

\bibitem{JMLR:v15:srivastava14a}
N.~Srivastava, G.~Hinton, A.~Krizhevsky, I.~Sutskever, and R.~Salakhutdinov, ``Dropout: A simple way to prevent neural networks from overfitting,'' {\em Journal of Machine Learning Research}, vol.~15, no.~56, pp.~1929--1958, 2014.

\bibitem{Al-Zareer_Ebbs-Picken_Michalak_Escobar_Silva_Davis_Osio_Amon_2023}
M.~Al-Zareer, T.~Ebbs-Picken, A.~Michalak, C.~Escobar, C.~M.~D. Silva, T.~Davis, I.~Osio, and C.~H. Amon, ``Heat generation rates and anisotropic thermophysical properties of cylindrical lithium-ion battery cells with different terminal mounting configurations,'' {\em Applied Thermal Engineering}, vol.~223, p.~119990, 2023.

\bibitem{Kingma_Ba_2014}
D.~P. Kingma and J.~Ba, ``Adam: A method for stochastic optimization,'' {\em arXiv}, 2014.

\bibitem{Luo_Li_Urtasun_Zemel_2017}
W.~Luo, Y.~Li, R.~Urtasun, and R.~Zemel, ``Understanding the effective receptive field in deep convolutional neural networks,'' {\em arXiv}, 2017.

\end{thebibliography}
